\documentclass{aa} 

\usepackage{graphicx}
\usepackage{txfonts}
\usepackage{placeins}
\usepackage{subfigure}
\usepackage{float}
\usepackage{natbib}
\usepackage{color}

\begin{document}

   \title{Interpreting the photometry and spectroscopy\\
    of directly imaged planets:\\
    a new atmospheric model applied to $\beta$ Pictoris b \\
    and SPHERE observations}
   \titlerunning {Interpreting the photometry and spectroscopy of 
   directly imaged planets}

   \subtitle{}

   \author{J.-L. Baudino
          \inst{1}
          \and
          B. B\'ezard\inst{1}
          \and
          A. Boccaletti\inst{1}
          \and
          M. Bonnefoy\inst{2}
          \and
          A.-M. Lagrange\inst{2}
           \and
           R. Galicher\inst{1}
          }
      \authorrunning{JL Baudino et al.}

   \institute{LESIA, Observatoire de Paris, PSL Research 
   University, CNRS, Sorbonne Universités, UPMC Univ. Paris 06, 
   Univ. Paris Diderot, Sorbonne Paris Cité, 5 place Jules 
   Janssen, 92195 Meudon, France~ 
   \email{jean-loup.baudino@obspm.fr}
         \and
              Univ. Grenoble Alpes, IPAG, F-38000 Grenoble, France 
              CNRS, IPAG, F-38000 Grenoble, France  }

   \date{Received 16/04/2015; accepted 31/07/2015}


 \abstract
   {Since the end of 2013 a new generation of instruments optimized to  image young 
   giant planets  around nearby stars directly is becoming available on 8-m class telescopes, both at 
   Very Large Telescope and Gemini in the southern hemisphere. Beyond the achievement of 
   high contrast and the discovery capability, these instruments are designed to obtain 
   photometric and spectral information to characterize the atmospheres of these planets. }
   {We aim to interpret future photometric and spectral measurements from these instruments, 
   in terms of physical parameters of the planets, with an atmospheric model using a minimal 
   number of assumptions and parameters.}
   {We developed  
    Exoplanet Radiative-convective Equilibrium Model (Exo-REM) to analyze the photometric and 
   spectroscopic data of directly imaged planets. The input parameters are a planet's surface 
   gravity ($g$), effective temperature ($T_\mathrm{eff}$), and elemental composition. The 
   model predicts the equilibrium temperature profile and mixing ratio profiles of the most 
   important gases. Opacity sources include the H$_2$-He collision-induced absorption and 
   molecular lines from eight compounds (including CH$_4$ updated with the Exomol line list). 
   Absorption by iron and silicate cloud particles is added above the expected condensation 
   levels with a fixed scale height and a given optical depth at some reference wavelength. 
   Scattering was not included at this stage.}
   {We applied Exo-REM to photometric and spectral observations of the planet $\beta$ 
   Pictoris b obtained in a series of near-IR filters. We derived $T_\mathrm{eff}$ = 1550 
   $\pm$ 150 K, $\log(g)$ = 3.5 $\pm$ 1, and  radius $R$ = 1.76 $\pm$ 0.24 $R_\mathrm{Jup}$ 
   (2-$\sigma$ error bars from photometric measurements). These values are comparable to those
   found in the literature, although with more conservative error bars, consistent with the model 
   accuracy. We were able to reproduce, within error bars, the J- and H-band spectra
   of $\beta$ Pictoris b.
   We finally investigated the precision to which the above parameters can be constrained 
   from SPHERE measurements using different sets of near-IR filters as well as low-resolution spectroscopy.}
   {}


   \maketitle
%

\section{Introduction}
          
   Following the detection of 51 Peg $b$ by \citet{Mayor1995b} using 
   velocimetry, almost 2000 exoplanets\footnote{\url{exoplanet.eu}} are known as of today 
   (March 10, 2015) and many more candidates are awaiting confirmation 
   \citep{Rowe2015a}. Among them, only a few were detected with direct imaging.\\
  
   The first image of a planetary mass object orbiting a star, 2M1207 
   $b$, was obtained by \cite{Chauvin2004b} with NaCo at the Very Large 
   Telescope (VLT) and this result has inspired several other discoveries in the 
   last decade \citep{Marois2008c, Lagrange2009, Rameau2013d}. In the 
   first case, the mass ratio was highly favorable as the central star is a brown dwarf 
   (BD). The detection was enabled by the use of an adaptive optics (AO) system 
   in the L' band and no specific 
   device to attenuate the star, such as a coronagraph, was needed.
   Later, larger mass ratios became feasible with the improvement 
   of high contrast imaging techniques. For now, the planet with the largest 
   mass ratio with respect to its host star, and for which we have an image, is HD95086 $b$
    with a mass of 5 $\pm$ 2 $M_\mathrm{Jup}$ around a star of 1.6 $M_\mathrm{Sun}$ 
    \citep{Rameau2013d}. Conveniently, direct imaging also allows us to collect 
    spectroscopic data if one is able to attenuate the starlight at the location of the 
   planet. \citet{Janson2010c} presented the first spatially resolved spectra of HR8799 
   $c$ still with NaCo. The spectrum covered the 3.88-4.10 $\mu$m interval 
   but with a low signal-to-noise ratio. The OSIRIS instrument at Keck II allowed 
   \cite{Bowler2010} and \cite{Barman2011c} to obtain spectra of the exoplanet HR8799 b in the 
   K band, sensitive to methane opacity, as well as a spectrum in the H band \citep{Barman2011c}. 
   Later, the same instrument provided spectra of the same planet \cite{Barman2015e} and of 
   HR8799 c \citep{Konopacky2013} in the K band at higher resolution (R $\sim$ 4000).
   Near-infrared, low-resolution spectra of the planet \object{$\beta$ 
   Pictoris $b$} \citep{Lagrange2010e} were obtained with GPI \citep{Macintosh2014c} in the 
   $J$ and $H$ bands \citep{Bonnefoy2014, Chilcote2015}, an instrument tailored for the 
   search of young giant planets.\\
 
    \begin{figure}[htb!]
      \centering
      \includegraphics[trim = 1cm 0cm 2cm 1cm, clip,width=0.5\textwidth]{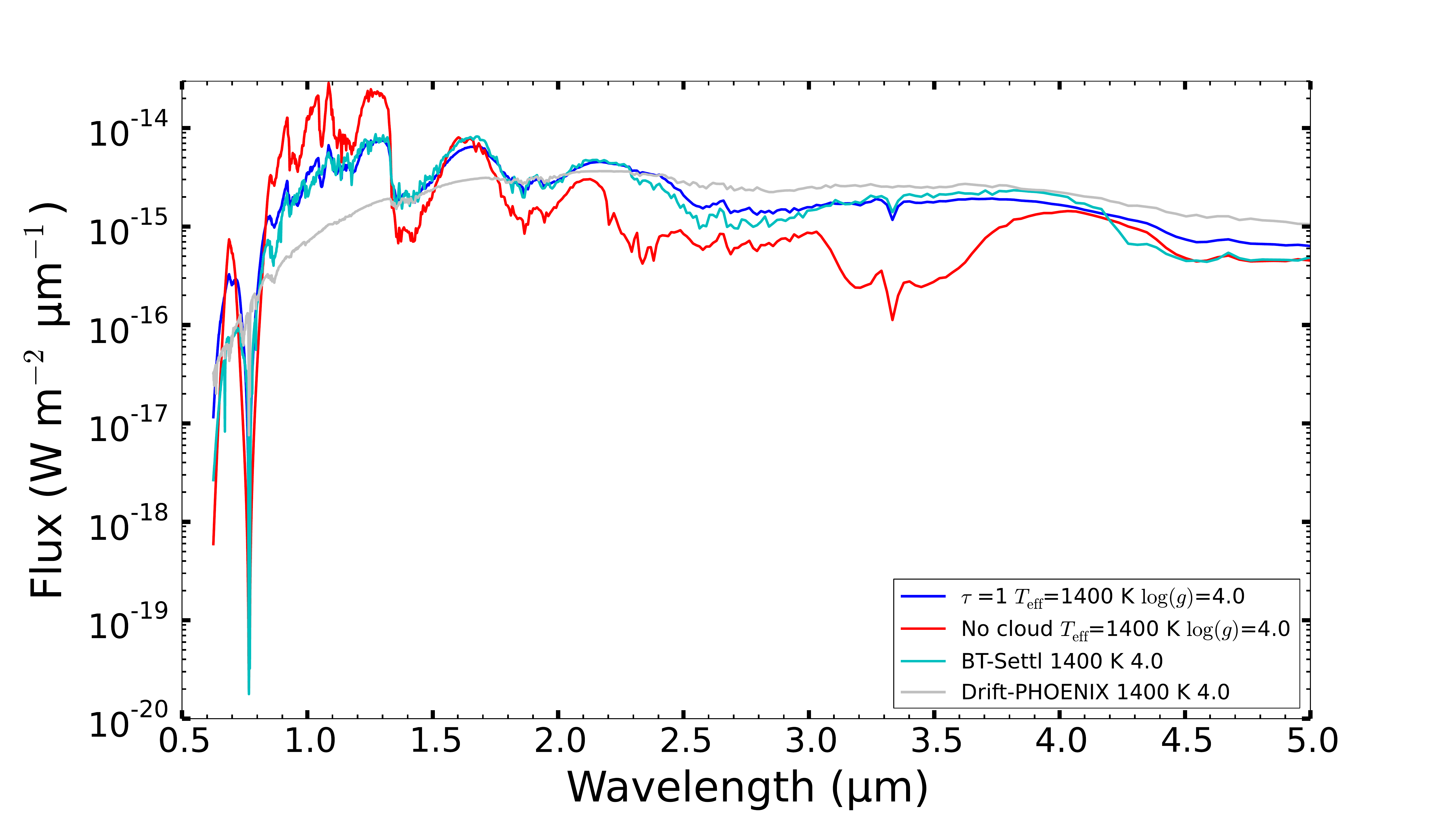} 
      \caption{Spectra calculated for a cloudy (blue) and cloud-free (red) 
      atmospheres at a distance of 10 pc, for a radius of 1 $R_\mathrm{Jup}$, $\log (g)$
      = 4 and $T_\mathrm{eff}$ = 1400 K . 
      Examples of BT-Settl (cyan) and Drift-PHOENIX (gray) models with the same parameters
      are also shown for comparison.}
      \label{CompareModel}
   \end{figure} 
 
   At the present time,  instruments for direct imaging combine AO, 
   coronography, and differential imaging to detect faint planets. The Spectro-Polarimetric 
   High-contrast Exoplanet Research, SPHERE \citep{Beuzit2008}, installed 
   at the VLT, is designed to  perform high contrast imaging for detecting young 
   giant planets and for characterizing their atmospheres. SPHERE provides broad-
   and narrowband photometry and spectroscopy in the near-infrared (NIR) range with the
   InfraRed Dual-band Imager and Spectrograph \citep[IRDIS, ][]{Langlois2010g} and the
   Integral Field Spectrograph \citep[IFS, ][]{Claudi2008a}, and photometry and 
   polarimetry in the visible range with the Zurich Imaging Polarimeter 
   \citep[ZIMPOL, ][]{Schmid2010v}.\\  
   
   To directly detect the light  from a planet around a star other than 
   the sun, the following conditions have to be met:
    \begin{itemize}
     \item the star-to-planet angular separation must be larger than the 
     angular resolution offered by an 8-m telescope in the NIR (25--50\,mas 
     for SPHERE). This restrains the sample of targets to less than 100\,pc 
     as well as the minimal physical separation to $\geq$1 AU.
     \item the star-to-planet brightness ratio must be smaller than the 
     achievable instrumental contrast, which is typically 
     $10^{5}$--$10^{7}$ at less than 1$''$. Only giant planets can be warm 
     enough at young ages to produce a detectable emission 
     \citep{Burrows1995h, Chabrier2000b}. These young extrasolar giant planets (YEGP) are 
     $\sim$10--100 millions years old.
    \end{itemize}     
    
   One theoretical challenge is to understand planetary formation 
   mechanisms. Brown dwarf and young extrasolar giant planet are two types of objects almost impossible to 
   differentiate, which are formed in different ways. Proposed mechanisms are 
   gravitational instabilities \citep{Boss2001b}, such as stars,  
   without a sufficient mass to start to burn hydrogen or core accretion 
   \citep{Lin1997}. A precise determination of the luminosity, mass, and age can
   supply information about the initial entropy of the planet and allows us to 
   identify the formation mechanism \citep{Bonnefoy2014,Marleau2014}.

   \begin{figure}[htb!]
      \centering
      \includegraphics[trim = 1cm 0cm 2cm 1cm, clip,width=0.5\textwidth]{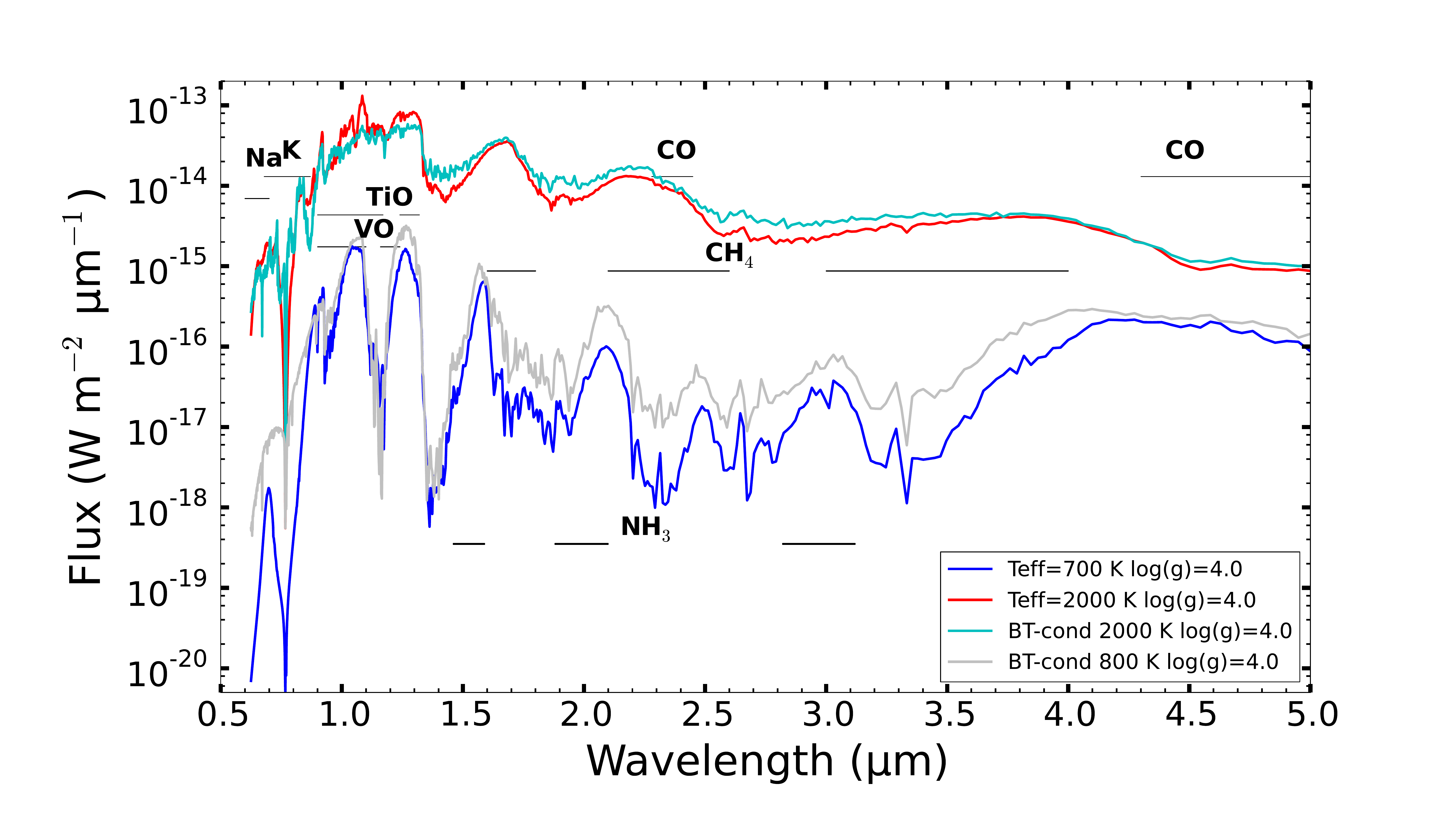} 
      \caption{Spectra calculated for a cloud-free atmosphere at a distance of 10 pc,
       for a radius of 1 $R_\mathrm{Jup}$, $\log (g)$ = 4.0 and two different 
       effective temperatures ($T_\mathrm{eff}$): 700 K (blue) and 2000 K (red).
        Molecular absorptions other than H$_2$O are indicated. Examples of 
        BT-cond models (cyan and gray) are also shown for comparison.}
      \label{Identification}
   \end{figure} 

   In parallel, atmospheric models for objects with mass and 
   temperature lower than an M dwarf were developed since the end of the 
   1990's. The basic idea is to include some chemistry
   and other physical processes in a H-He atmosphere  to account for the range of pressure temperature
   expected in such low-mass objects. The models differ in particular by their treatment of 
   dust opacity. This is a crucial component of the models since it was noticed very 
   soon that these atmospheres must contain clouds below $T_\mathrm{eff} \simeq$ 2600 K to 
   account for the spectroscopic observations \citep{Tsuji1996}.\\

   The model of Tsuji and collaborators is a direct adaptation of their M dwarf 
   models to BD. Dust grains in the atmosphere are treated through a parametrized model 
   \citep{Tsuji2002g}, with three cases. In Case B, the part of the atmosphere where the 
   temperature is lower than the condensation temperature is full of dust. In an opposite 
   case (Case C), dust forms but is immediately removed by precipitation so that it does not 
   contribute to the opacity. The third case, called the Unified Cloud Model, is 
   intermediate between these two extremes with dust present only between the condensation 
   level and a level at a slightly lower temperature $T_\mathrm{c}$.

         \begin{table}[htb!]
            \caption{Atom and molecular opacity sources}
            \begin{center}
               \begin{tabular}{c|c|c}
               \hline\hline
               {Opacities} & {Intensity cutoff} & {References}\\
               {} & {(cm molecule$^{-1}$)} & {}\\
               {} & {} & {}\\\hline
               {H$_2$O} & {$10^{-27}$ at 2500 K} & {\tiny HITEMP line list}\\
               {CO} & {$10^{-27}$ at 3000 K} & {\tiny \citep{Rothman2010h}}\\\hline
               {CH$_4$} & {$10^{-27}$ at 1500 K} & {\tiny\cite{Yurchenko2014a}}\\\hline
               {NH$_3$} & {$5 \times 10^{-27}$ at 4000 K} & {\tiny\cite{Yurchenko2011}}\\\hline
               {TiO, VO} & {$10^{-22}$ at 4000 K} & {\tiny\cite{Plez1998} (with update)}\\\hline
               {Na, K} & {} & {\tiny \cite{NIST_ASD}}\\
               {} & {} & {\tiny \cite{Burrows2003m}}\\\hline
               {H$_2$-H$_2$,} & {} & {\tiny \cite{Borysow1988, Borysow1989a}}\\
               {H$_2$-He} & {} & {\tiny \cite{Borysow1989d}}\\
               {} & {} & {\tiny \cite{BorysowU.G.Jorgensen2001} }\\
               {} & {} & {\tiny \cite{Borysow2002} }\\\hline
               \end{tabular}
            \end{center}
            \label{tabRef}
         \end{table}     
   
  \cite{2000AAS...19712705M} proposed a model adapted from solar system planets and 
  using the \cite{Ackerman2001g} cloud model, which is parametrized with a factor 
  $f_\mathrm{sed}$ representing the sedimentation efficiency.
  
  The Lyon's group atmospheric models, DUSTY and COND \citep{Allard2001a} are two 
  cases similar to Tsuji's B and C cases. BT-Settl \citep{Allard2003j} is a more complex 
  model, which compares condensation, sedimentation, and mixing timescales of  dust to define cloud 
  parameters.
   
   Finally, Drift-PHOENIX \citep{Helling2008h} considers microphysical processes 
   (nucleation, condensation, particle growth, sedimentation, and evaporation) to calculate 
   composition, number density, and size distribution of dust particles as a function of 
   atmospheric pressure level.

         \begin{figure*}[htb!]
            \begin{center}
               $\begin{array}{ccc}
                  \includegraphics[width=0.332\textwidth, trim = 0cm 1.9cm 5cm 3cm, clip]{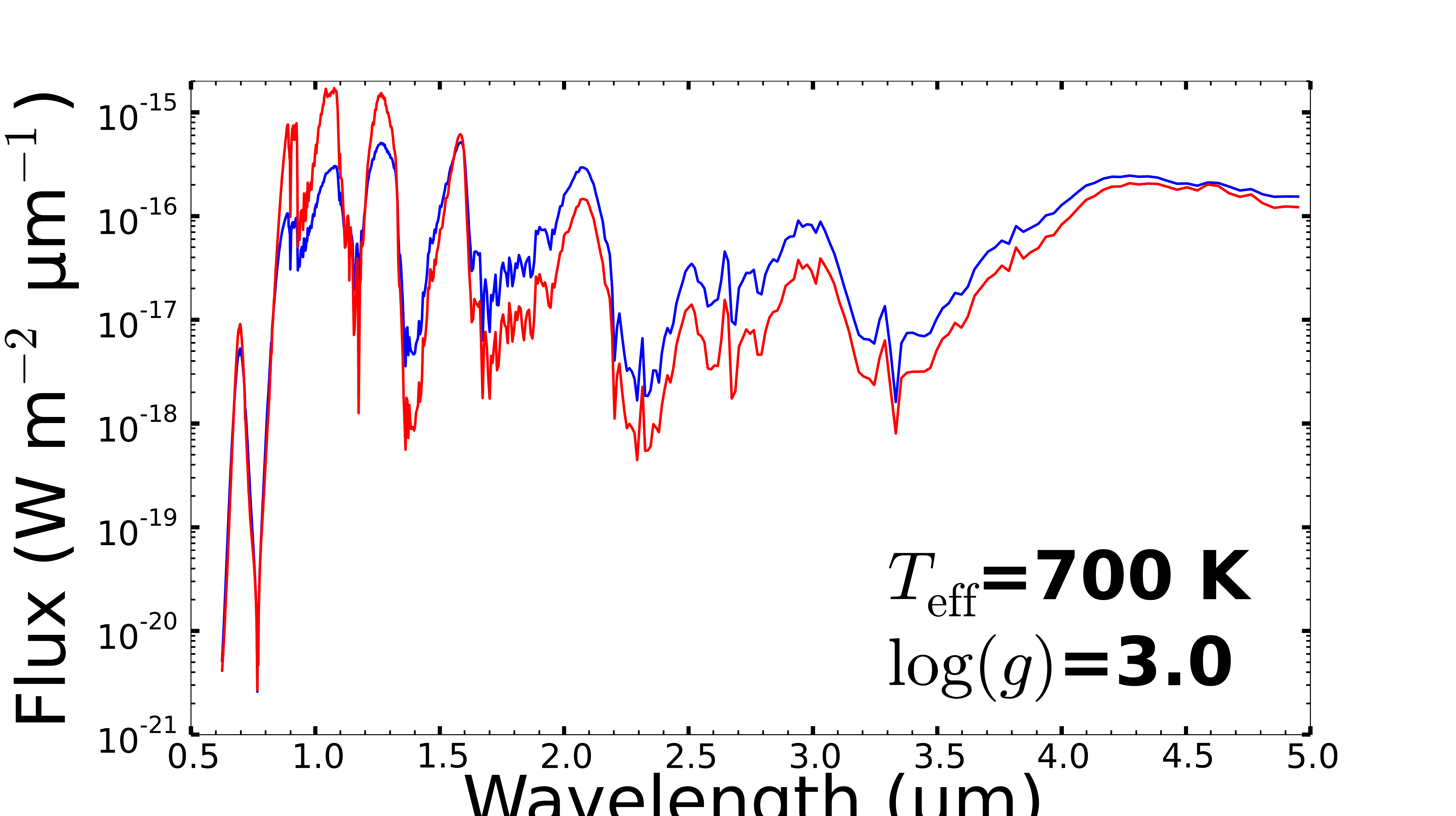} & \includegraphics[width=0.31\textwidth, trim = 3.9cm 1.9cm 4.9cm 3cm, clip]{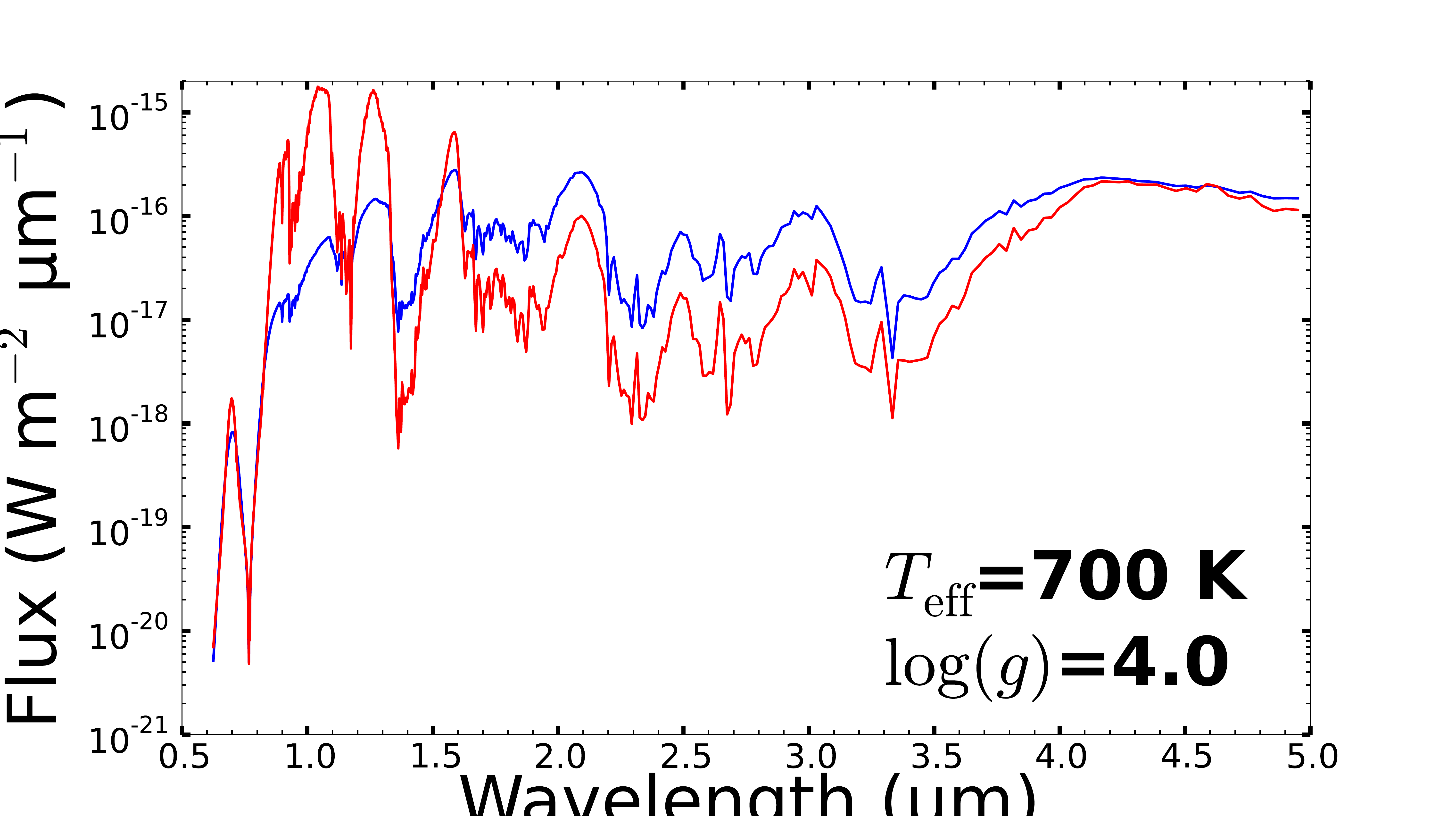} & \includegraphics[width=0.31\textwidth, trim = 3.9cm 1.9cm 5cm 3cm, clip]{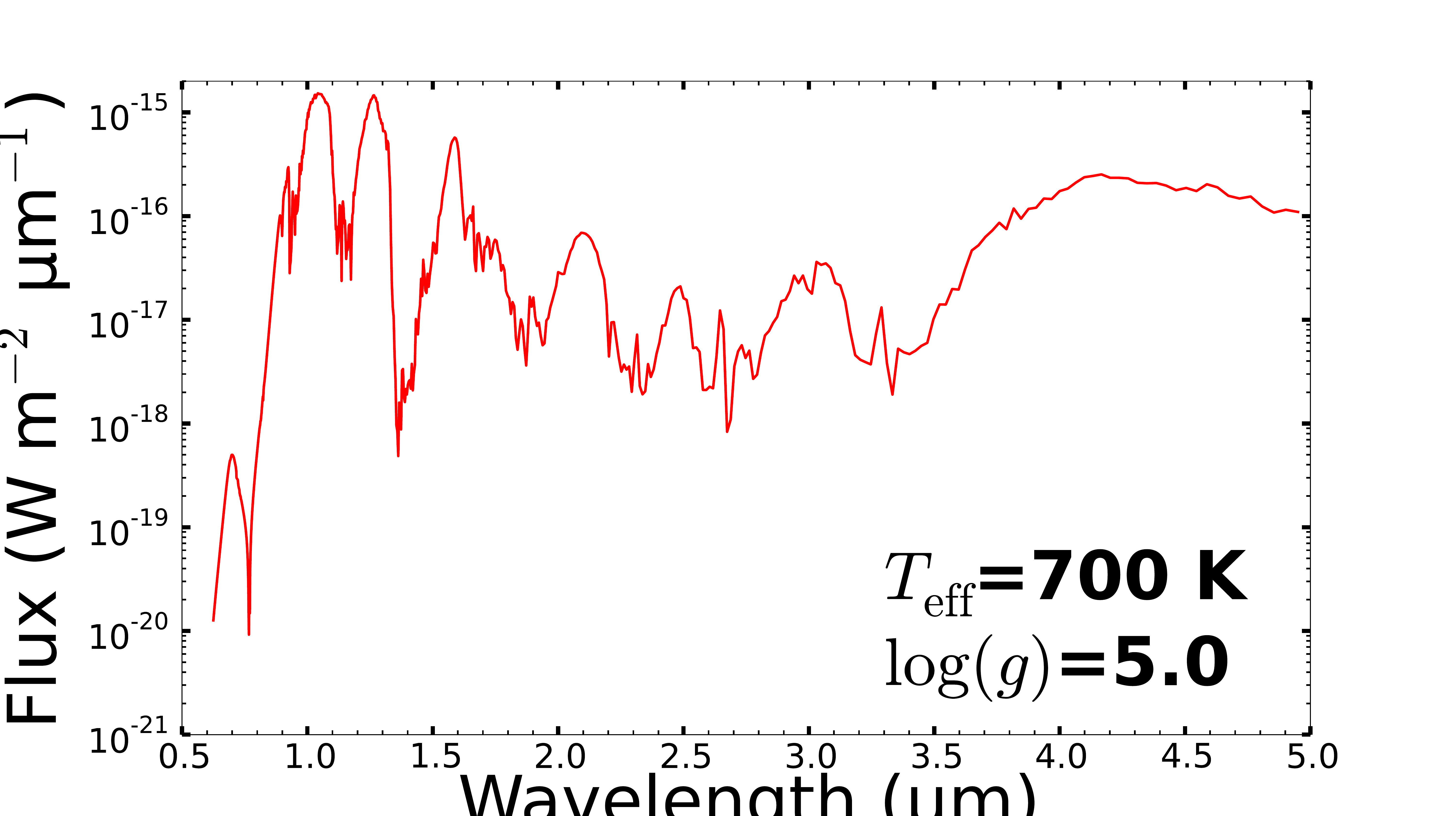}\\
                  \includegraphics[width=0.332\textwidth, trim = 0cm 1.9cm 5cm 3cm, clip]{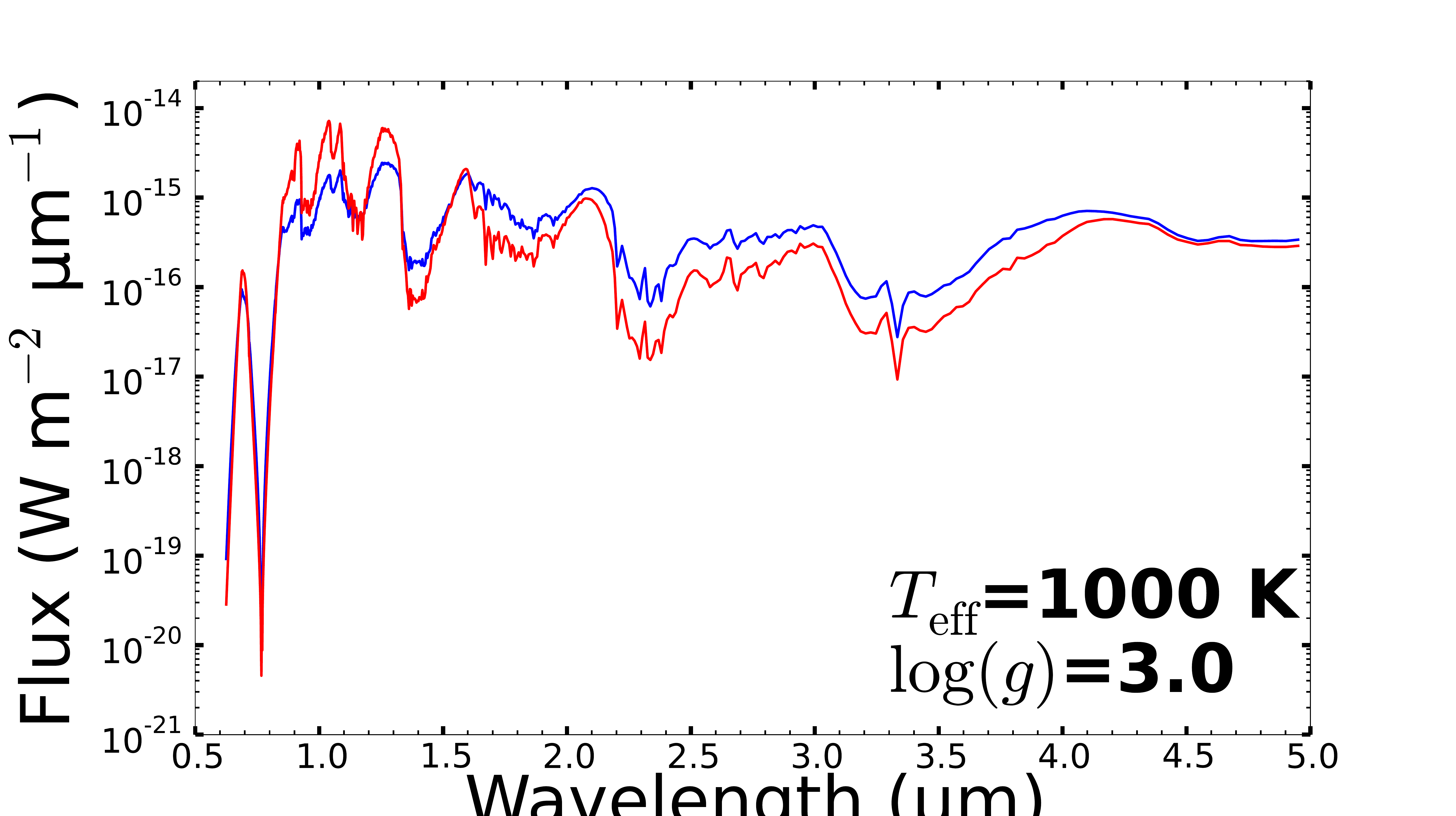} & \includegraphics[width=0.31\textwidth, trim = 3.9cm 1.9cm 4.9cm 3cm, clip]{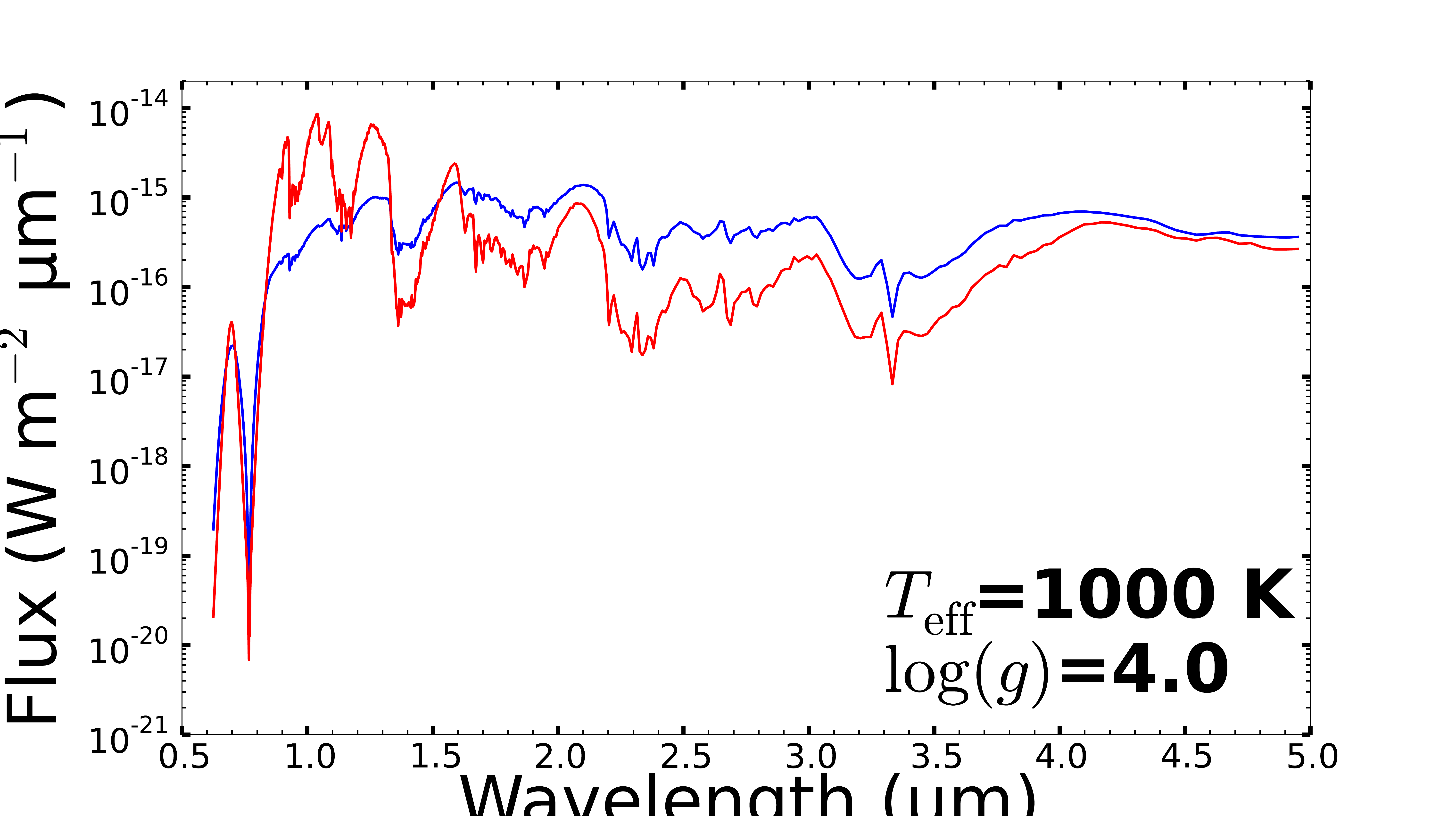} & \includegraphics[width=0.31\textwidth, trim = 3.9cm 1.9cm 5cm 3cm, clip]{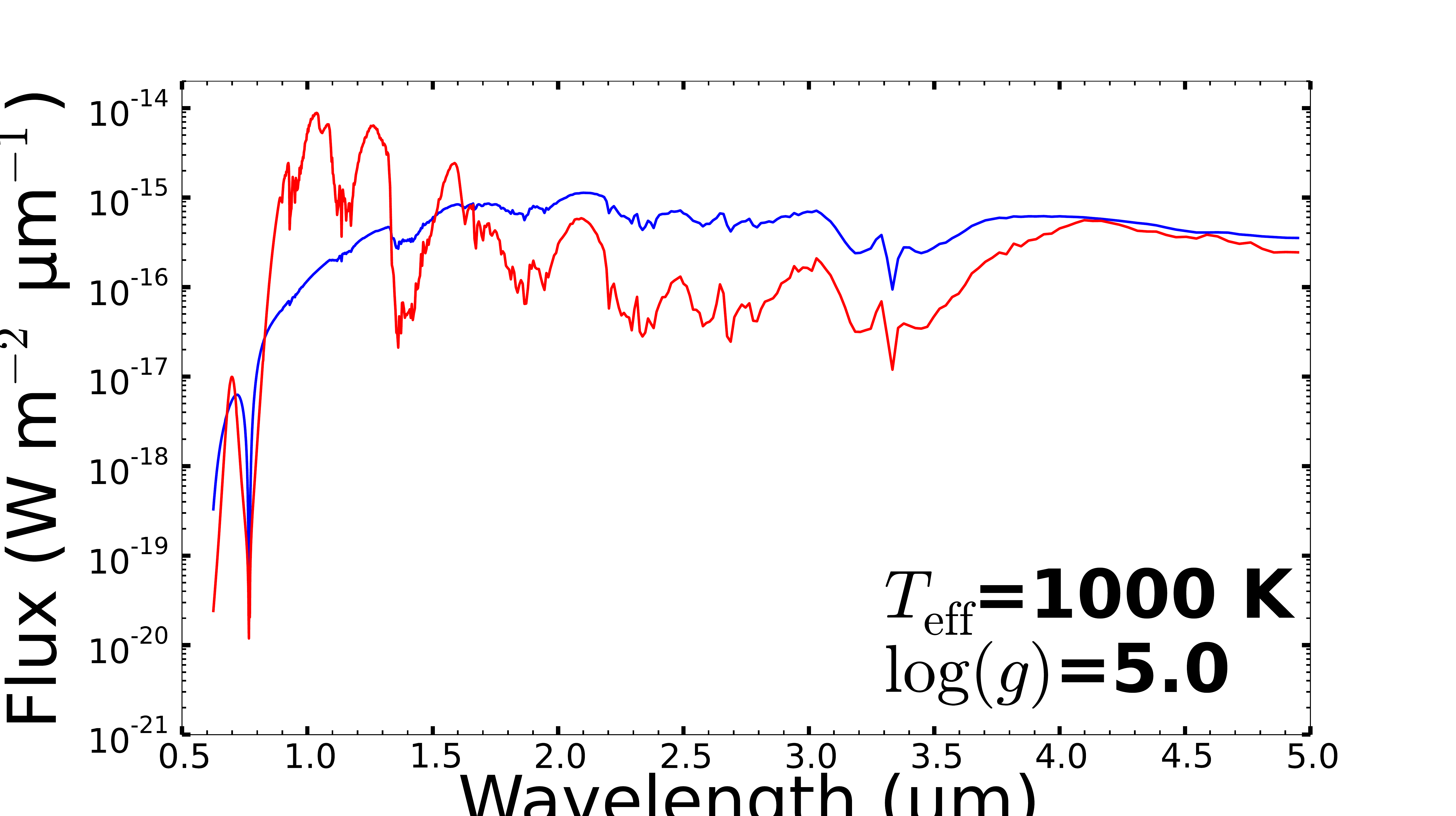}\\
                  \includegraphics[width=0.332\textwidth, trim = 0cm 1.9cm 5cm 3cm, clip]{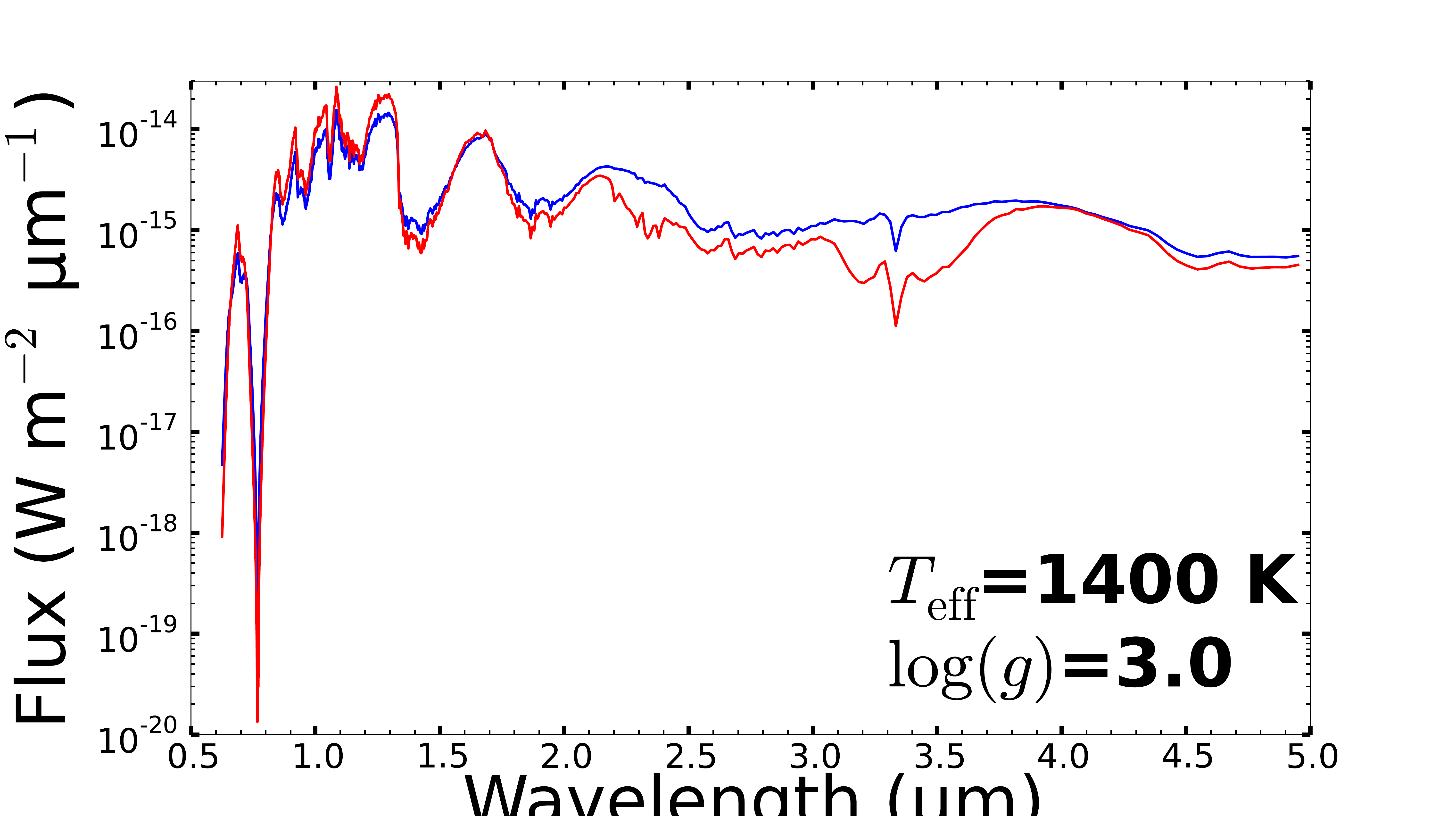} & \includegraphics[width=0.31\textwidth, trim = 3.9cm 1.9cm 4.9cm 3cm, clip]{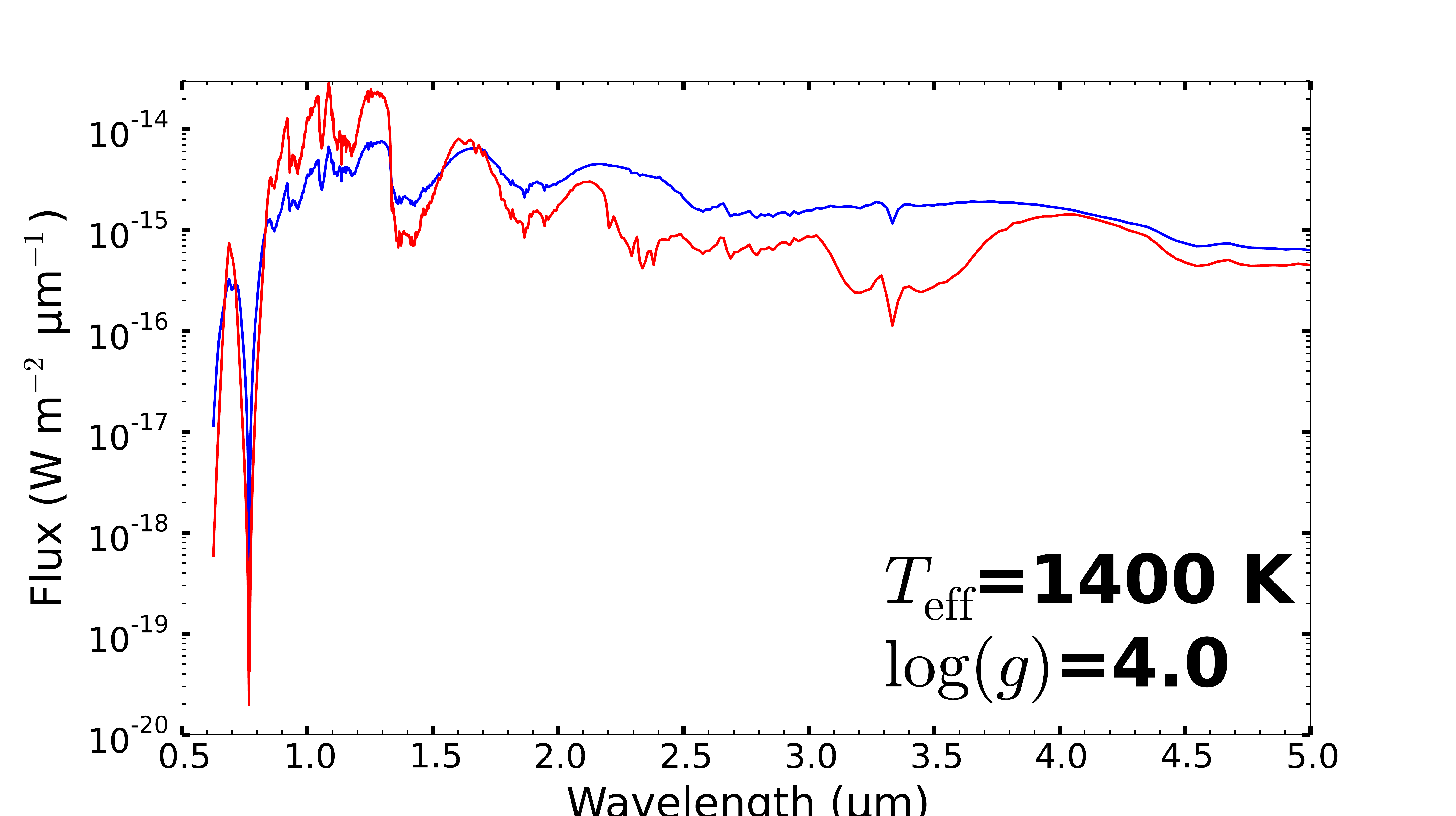} & \includegraphics[width=0.31\textwidth, trim = 3.9cm 1.9cm 5cm 3cm, clip]{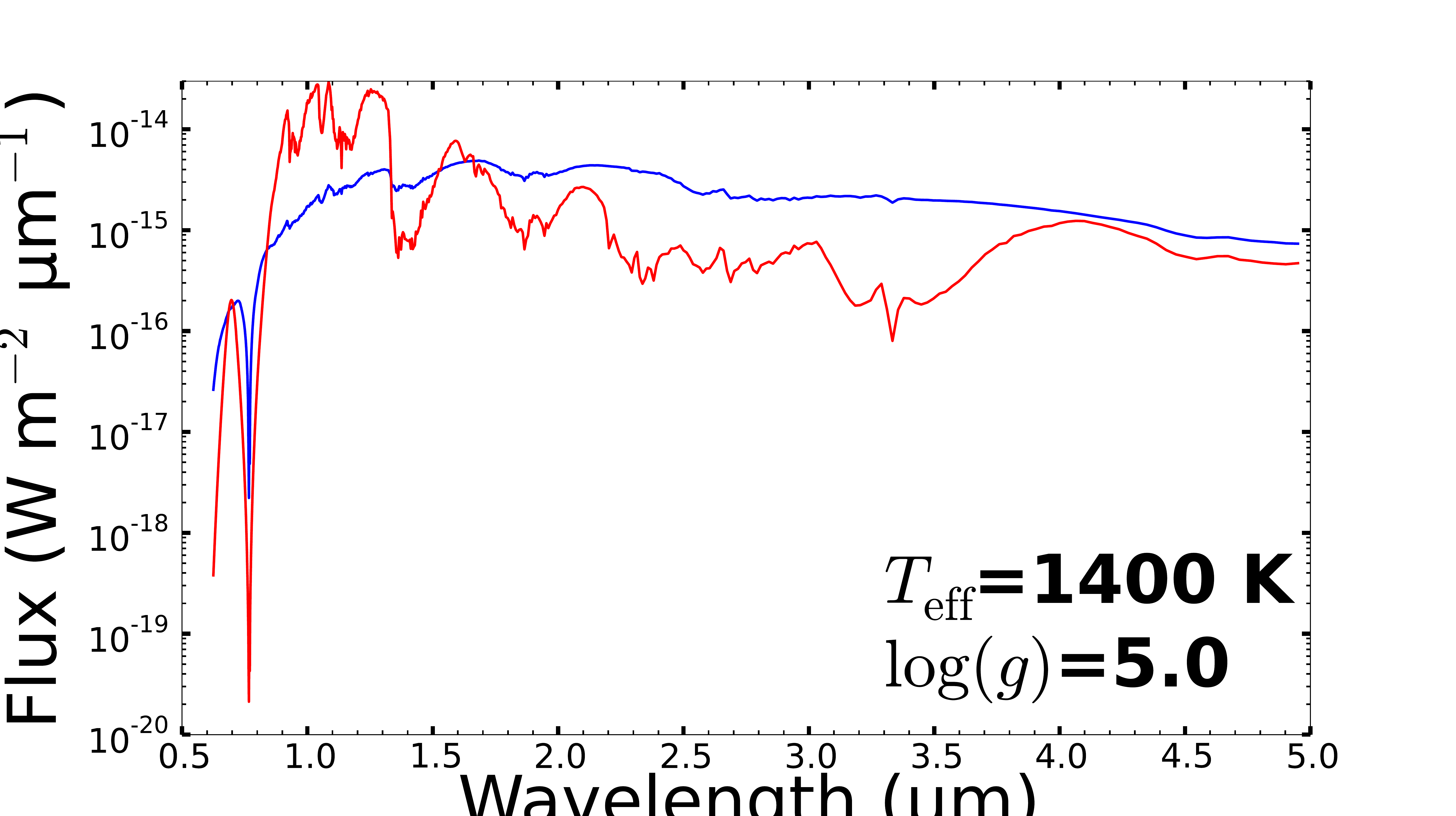}\\
                  \includegraphics[width=0.332\textwidth, trim = 0cm 0.5cm 5cm 3cm, clip]{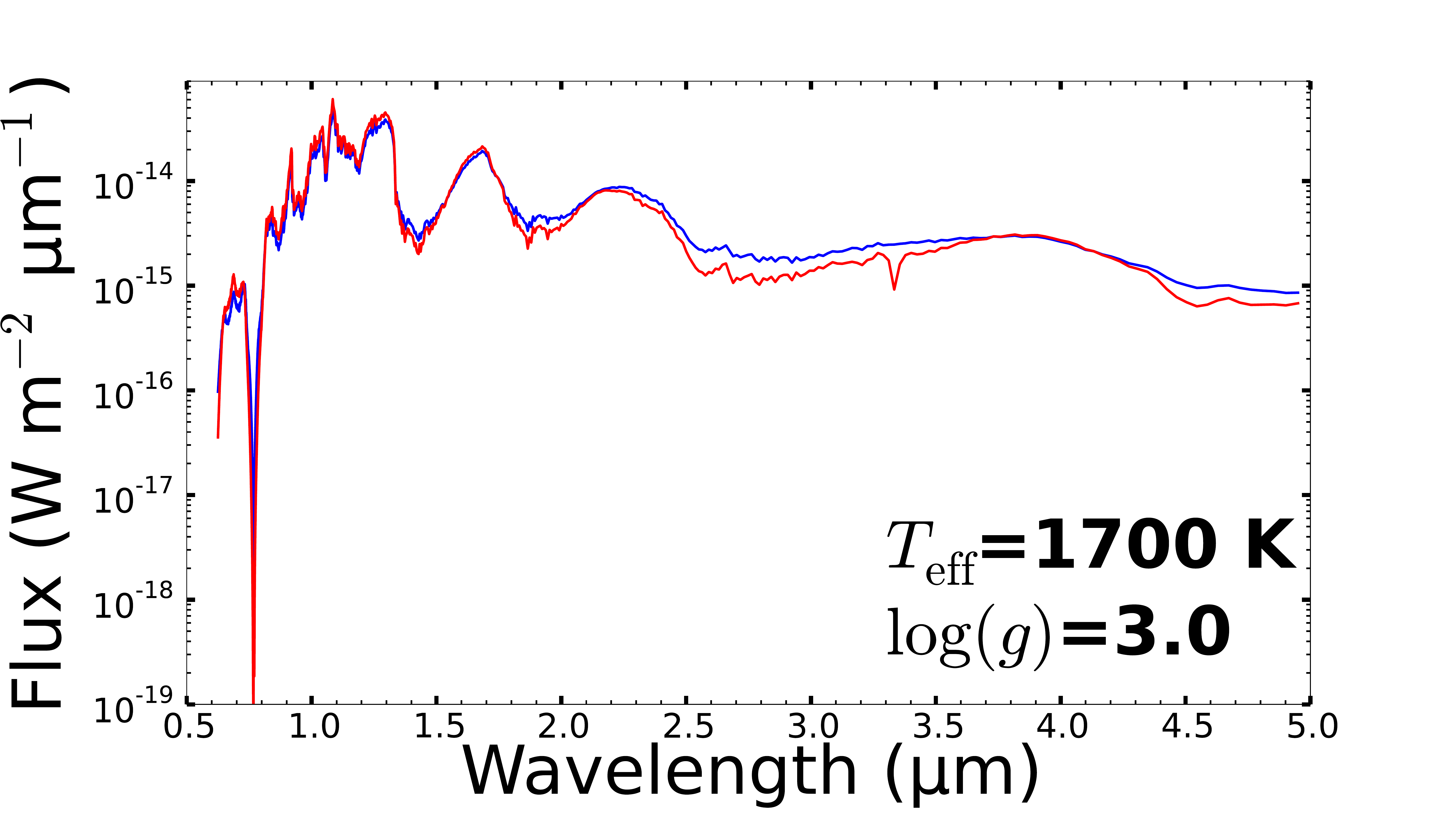} & \includegraphics[width=0.31\textwidth, trim = 3.9cm 0.5cm 4.9cm 3cm, clip]{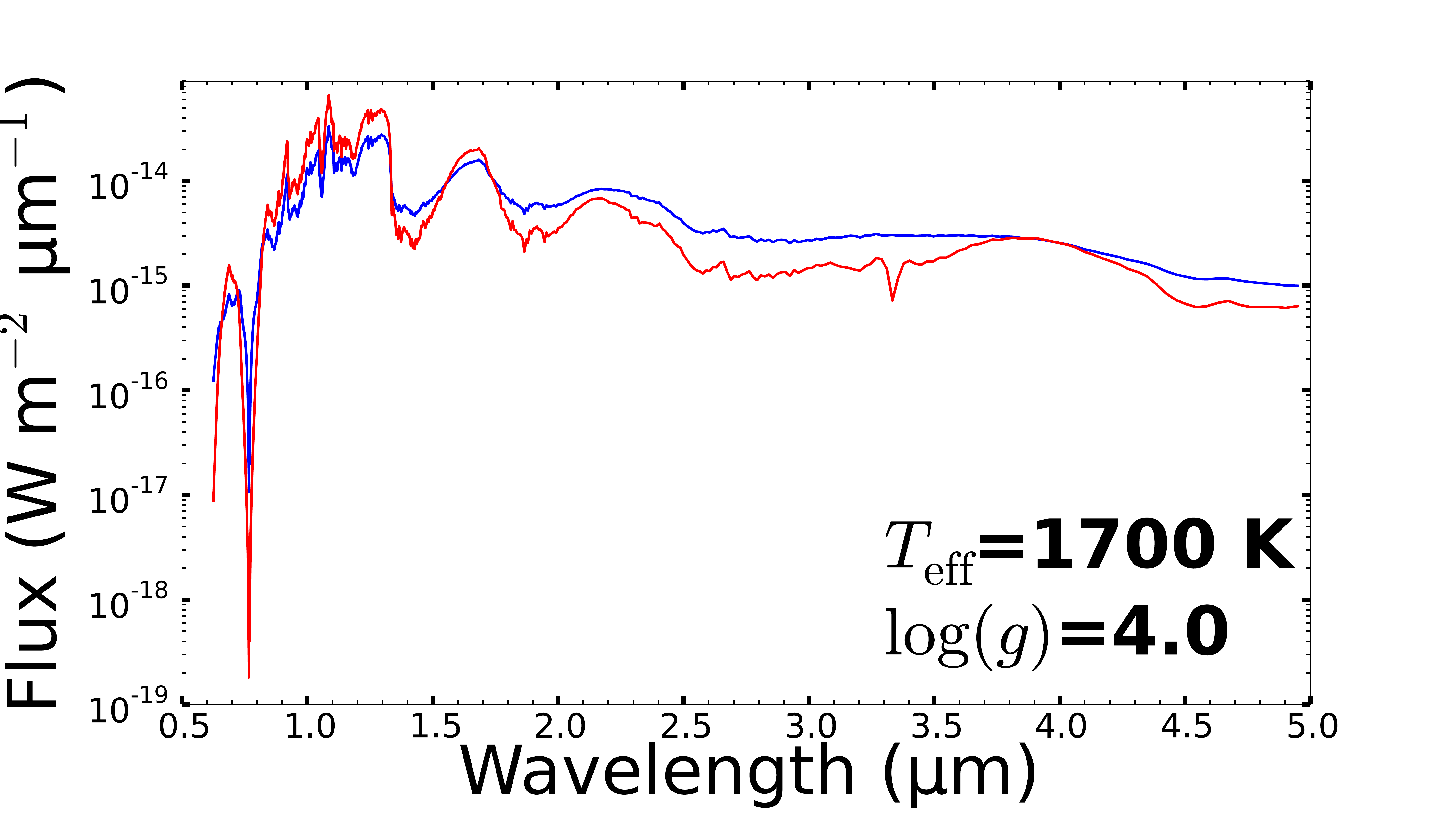} & \includegraphics[width=0.31\textwidth, trim = 3.9cm 0.5cm 5cm 3cm, clip]{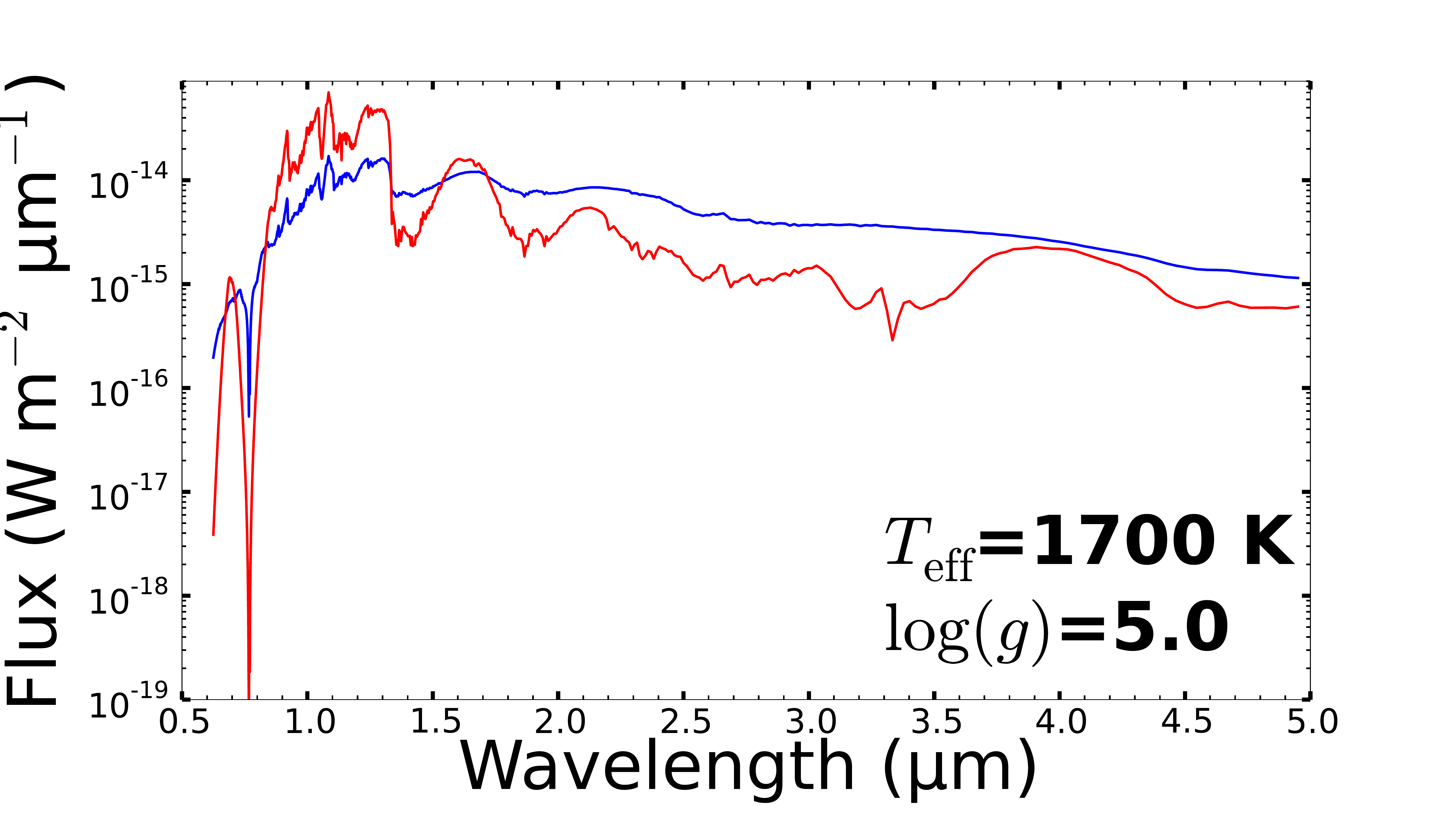}
               \end{array}$
            \end{center}
            \caption{Spectra for a range of models without (red) and with (blue) clouds at 
            the distance of 10 pc and with a radius of 1 $R_\mathrm{Jup}$ and with a radius 
            of 1 $R_\mathrm{Jup}$. For 700, 1000, 1400 and 1700 K the maximum of the 
            blackbody is respectively around 4.1, 2.9, 2.1, and 1.7 $\mu$m}
            \label{CloudEffect}
         \end{figure*}    
   
   All these models were used to constrain the main parameters, such as the  effective 
   temperature, surface gravity, atmospheric compounds, or radius of the detected planets
   \citep{Neuhauser2005, Marley2012a, Bonnefoy2013f, Galicher2014a, Bonnefoy2014}
   and to predict abilities of new instruments \citep{Boccaletti2005b, 
   Hanot2010,Vigan2010c}.
   Most of these models are developed for BD and applied to exoplanets. Although, the radius 
   of exoplanets can be similar to that of BD, they have lower masses. 
   Therefore, the range of surface gravity considered in BD atmospheric models 
   ($\log (g)$>3.5) does not necessarily cover the entire range expectable for YEGP 
   ($\log (g)$>2).\\

   Direct imaging of YEGP is characterized by low flux, low signal to 
   noise and low spectral resolution. In that respect, the models that are 
   used to interpret these images should be representative of the level of 
   data quality.
   For that purpose we specifically developed a model to analyze direct 
   imaging of YEGP for instruments like SPHERE. It is a 
   radiative-convective equilibrium model, assuming thermochemical equilibrium for self -luminous planets in which stellar heating is neglected. It allows us to explore low 
   surface gravity, 
   i.e. low-mass YEGP.
   
   The radiative-convective equilibrium model is described in Section 2.
   In Section 3, 
   we apply Exo-REM to the well-known planet \object{$\beta$ Pictoris $b$}, derive
   physical parameters from existing measurements, and compare our results to previously 
   published investigations. In Section 4, 
   we analyze the uncertainties in the derived physical parameters as a function of 
   photometric errors in the context of SPHERE observations.  
   The conclusion is drawn in Section 5.

\section{Model description}

   \subsection{Radiative-convective equilibrium model}
      \subsubsection{Numerical method}
         We solve for radiative-convective equilibrium, assuming that the net flux 
         (radiative + convective) is conservative, and neglecting the stellar flux impinging 
         on the planet. The net fluxes are calculated between 20 cm$^{-1}$ and 16000 
         cm$^{-1}$ over 20-cm$^{-1}$ intervals using the radiative transfer equation with no 
         scattering and a $k$-correlated distribution method to represent molecular opacity. 
         The atmospheric grid consists of 64 logarithmically equally spaced pressure levels. 
         The system of equations fixing the constancy of the net flux over the atmospheric 
         grid is solved iteratively through a constrained linear inversion method. Details 
         are given in Appendix A.

      \subsubsection{Spectroscopic data}
       
         As mentioned above, the spectral flux was calculated over 20 cm$^{-1}$ intervals
         using a $k$-correlated distribution method. For each molecule and each interval, we 
         calculated a set of $n_\mathrm{k}$ = 16 $k$-coefficients ($l$=1, $n_\mathrm{k}$), 8 
         for the interval [0:0.95] of the normalized frequency $g^\star$, and 8 for the 
         interval [0.95:1.00]. The values of $g_\mathrm{l}^\star$ and associated weights 
         $\varpi_\mathrm{l}$ are those of the 8-point Gaussian-Legendre quadrature for each 
         of the two $g_\mathrm{l}^\star$ intervals. The $k$-coefficients were calculated
         for a set of 15 pressures between 100 bar and 0.01 mbar (2 values per decade) and, 
         for each pressure, a set of 6 temperatures, increasing with pressure to encompass 
         model temperature profiles encountered in the literature for exoplanets with 500 K 
         < $T_\mathrm{eff}$ < 2000 K. Absorptivity spectra for a given pressure and 
         temperature were calculated using a line-by-line radiative transfer program with a 
         frequency step equal to the Doppler half-width of the lines.

         We considered the eight most important molecules and atoms in terms of opacity for
         relatively cool exoplanets (500 K < $T_\mathrm{eff}$ < 2000 K): H$_2$O, CO, CH$_4$, 
         NH$_3$, TiO, VO, Na, and K. The origin of the line lists and the intensity cutoff 
         used to calculate the absorptivity spectra are given in Table \ref{tabRef}. In the previous version of the model used by \cite{Baudino2013, Baudino2014, 
         Baudino2014c}, \cite{Galicher2014a}, and \cite{Bonnefoy2014}, the methane line list 
         originated from \cite{Albert2009c}, \cite{Boudon2006a}, \cite{Daumont2013}, and 
         \cite{Campargue2012} for CH$_4$, and from \cite{Nikitin2002k, Nikitin2006o, 
         Nikitin2013d} for CH$_3$D. Our new methane line list now comes from the Exomol 
         database \citep{Yurchenko2014a}.

       \begin{table}[htb!]           
            \caption{Compounds considered in thermochemical equilibrium calculations}
            \begin{center}
               \begin{tabular}{ll} 
               \hline\hline
                 \text{Compounds of interest$^1$} & \text{Species included in chemical} \\
                  & \text{equilibrium calculations$^1$} \\ \hline
                  \text{H$_2$O, CO, CH$_4$} &\text{H$_2$O, H$_2$O*, CO, CH$_4$}\\
                  \text{NH$_3$} & \text{NH$_3$, N$_2$, NH$_4$SH*, H$_2$S}\\
                  \text{Na, K} & \text{Na, Na$_2$S*, H$_2$S, HCl, NaCl,}\\
                    & \text{K, KCl, KCl*, NH$_3$, NH$_4$Cl*}\\
                  \text{TiO, VO} & \text{Ti, TiO, TiO$_2$, V, VO, VO$_2$,}\\
                    & \text{Ca, CaTiO$_3$*, VO*, H$_2$O}\\
                  \text{Mg$_2$SiO$_4$*, MgSiO$_3$*, SiO$_2$*}& \text{Mg, SiO, H$_2$O, Mg$_2$SiO$_4$*,}\\
                  & \text{MgSiO$_3$*, SiO$_2$*}\\
                  \text{Fe*} & \text{Fe, Fe*}\\\hline                   
               \end{tabular}\\
           \end{center}
           $^1$: Species marked with asterisks are condensates
            \label{tabCompounds}
         \end{table}

         For all species except alkali, we calculated line absorption up to 120 cm$^{-1}$ 
         from line center using a Voigt profile multiplied by a $\chi$ factor to account for 
         sub-Lorentzian far wings. For $\chi$, we used the profile derived by 
         \cite{Hartmann2002al} for H$_2$-broadened lines of methane. The far wing absorption 
         of Na and K has been shown to strongly affect the near-infrared spectra of brown 
         dwarfs and extra-solar giant planets \citep{Burrows2000y}. For Na and K, we used a
         Voigt profile V($\sigma$-$\sigma_0$) in the impact region, up to a detuning 
         frequency ($\delta \sigma$) of $30 (T/500)^{0.6}$ cm$^{-1}$ for Na and $50 
         (T/500)^{0.6}$ cm$^{-1}$ for K, following \cite{Burrows2000y}. The Lorentz 
         half-widths, calculated from the impact theory,
         are $0.27 (T/296)^{-0.70}$ cm$^{-1}$ atm$^{-1}$ for Na and $0.53 (T/296)^{-0.70}$ 
         cm$^{-1}$ atm$^{-1}$ for K. Beyond the detuning frequency, we used a profile in the 
         form
         
         \begin{multline}
          F(\sigma-\sigma_0)=V(\delta \sigma) [\delta \sigma/(\sigma-\sigma_0)]^{3/2}\\
          \exp[-(hc(\sigma-\sigma_0)/kT)(\sigma-\sigma_0)/\sigma_F],
         \end{multline}
         
         where $\sigma-\sigma_0$ is distance from line center, $V(\delta \sigma)$ is the 
         Voigt profile at the detuning frequency $\delta \sigma$, and $\sigma_F$ a parameter 
         that we adjusted to best reproduce the absorption cross sections calculated by 
         \cite{Burrows2003m} for the red wings of the Na/K + H$_2$ systems as shown in their 
         Fig. 6. We derived $\sigma_F = 5000$ cm$^{-1}$ for Na and $1600$ cm$^{-1}$ for K 
         from best fitting of the 0.6-0.9 $\mu$m and 0.8-1.0 $\mu$m regions for Na and K, 
         respectively. Profiles were calculated up to 9000 cm$^{-1}$ of line center.

         \begin{figure}[htb!]
            \centering
            \includegraphics[trim = 1cm 0cm 2cm 1cm, clip,width=0.5\textwidth]{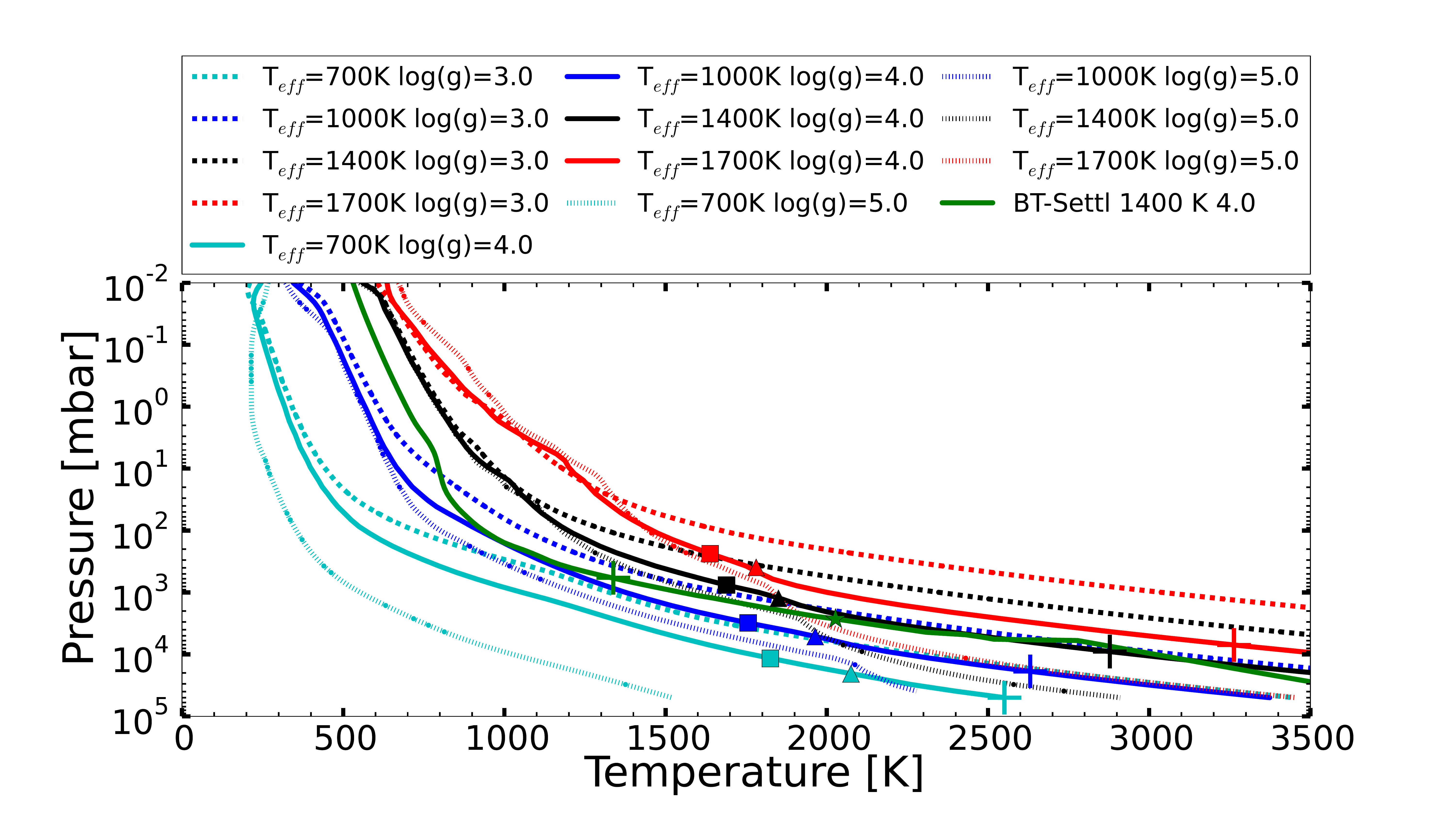} 
            \caption{Temperature profiles calculated for the set of $T_\mathrm{eff}$ and 
            $\log (g)$ parameters used to generate the spectra in Fig.~\ref{CloudEffect}. 
            Triangles and squares on the $\log (g)$ = 4 profiles indicate the bottom of the 
            iron and silicate clouds respectively. One BT-Settl profile (green) is 
            also shown for comparison, the star corresponding to the bottom of the dust location.
            The radiative-convective boundary is shown as a cross on the same profiles.}
            \label{profils}
         \end{figure}          
         
         Besides line opacity, we added the collision-induced absorption from H$_2$-H$_2$ 
         and H$_2$-He using data files and subroutines provided by A. 
         Borysow\footnote{\url{http://www.astro.ku.dk/~aborysow/programs/}}. These are based 
         on publications by \cite{BorysowU.G.Jorgensen2001} and \cite{Borysow2002} for 
         H$_2$-H$_2$, and \cite{Borysow1988, Borysow1989a}  and \cite{Borysow1989d} for 
         H$_2$-He.     
         
         We finally added absorption by cloud particles, discarding scattering. We 
         considered condensates from Si and Fe, the two most abundant condensing elements in 
         exoplanets with $T_\mathrm{eff}$ in the range 500-2000 K \citep{Lunine1989e}. For 
         silicates particles, we used the optical constants of crystalline forsterite 
         Mg$_2$SiO$_4$ published by \cite{Jager2003a} and for Fe liquid particles those
         from \cite{Ordal1988}.

   \subsection{Atmospheric model}
   
      \subsubsection{Gas composition}

         \begin{figure*}[htb!]
            \begin{center}
               $\begin{array}{cccc}
                  &\includegraphics[trim = 1.7cm 0cm 5.0cm 0cm, clip,width=0.327\textwidth]{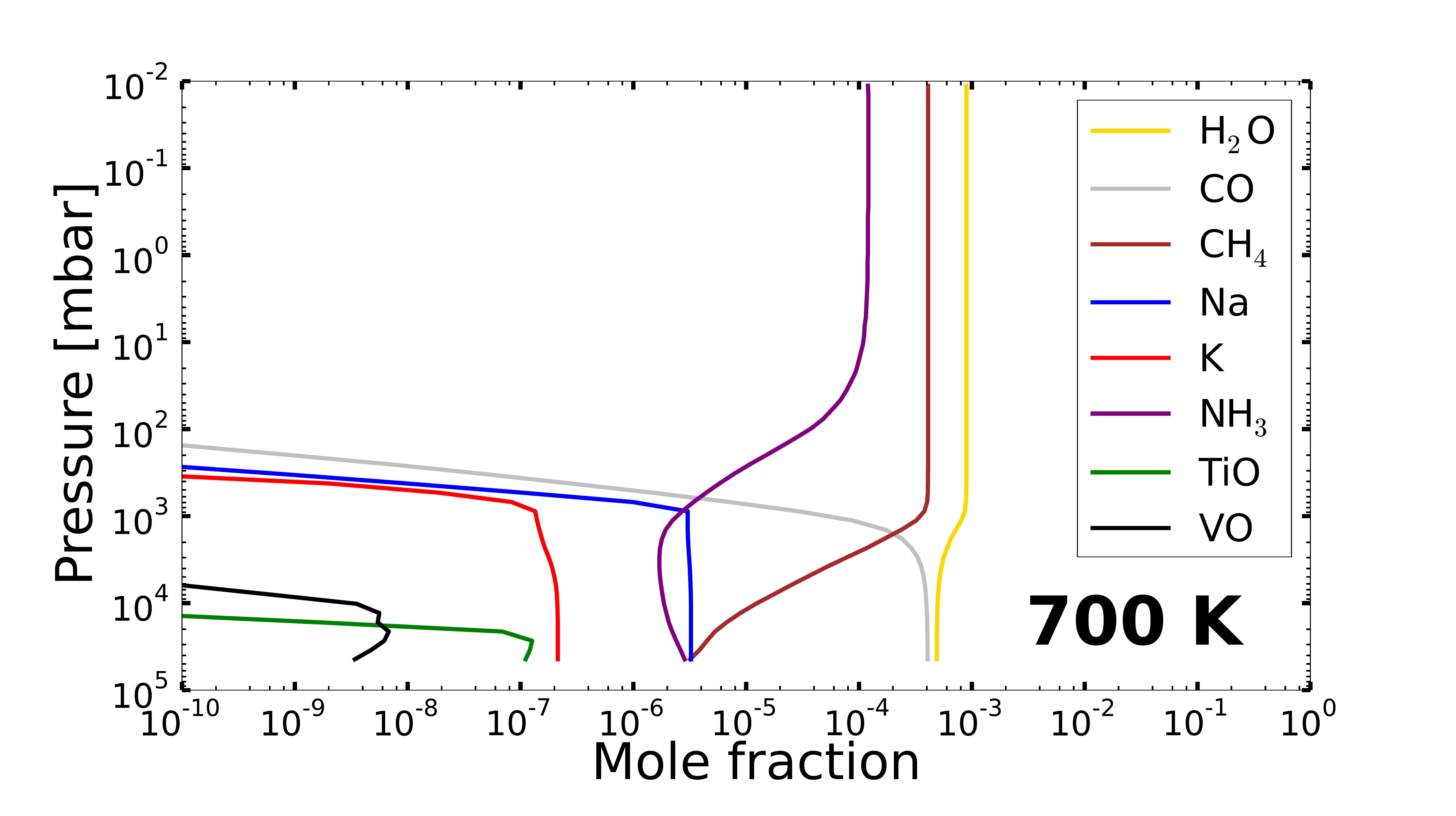}
                  &\includegraphics[trim = 4.5cm 0cm 5.0cm 0cm, clip,width=0.31\textwidth]{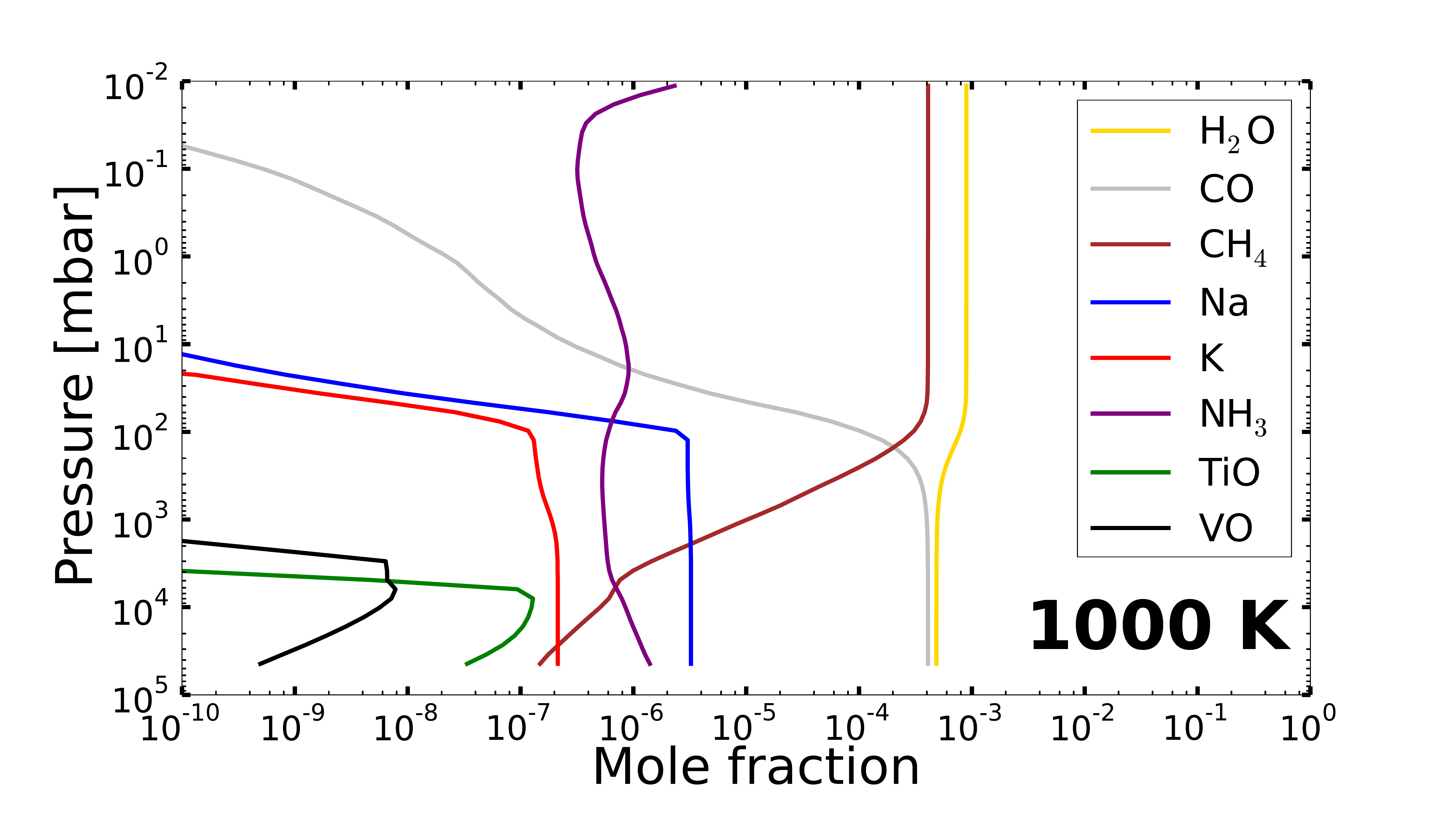} 
                  &\includegraphics[trim = 4.5cm 0cm 5.0cm 0cm, clip,width=0.31\textwidth]{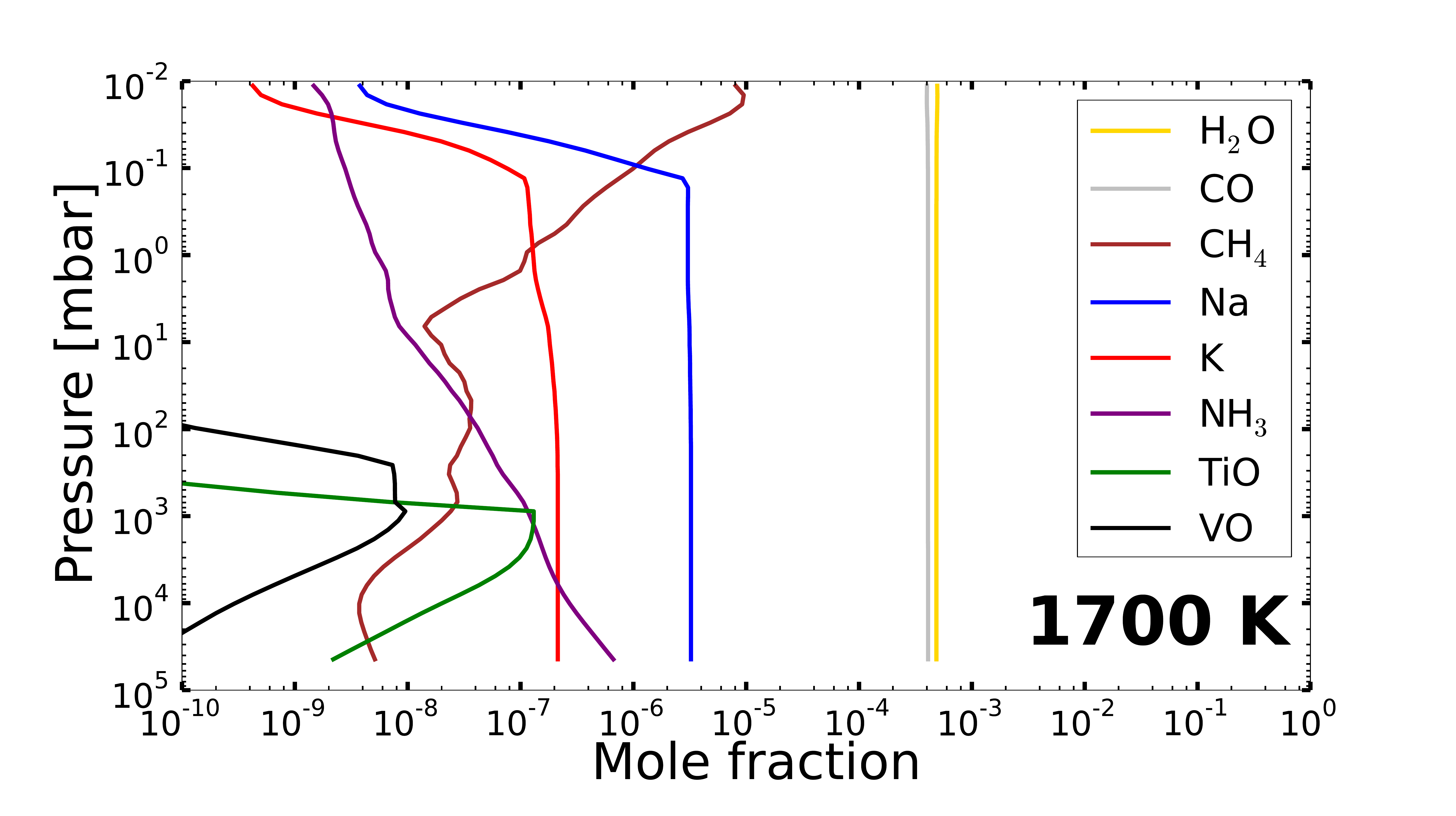}
               \end{array}$
            \end{center}
            \caption{Mixing ratio profiles of important molecules assuming a cloudy 
            atmosphere,  $\log (g)$ = 4, and three different effective temperatures 700, 
            1000, and 1700 K.}
            \label{MixProf}
         \end{figure*} 
         
         The vertical profiles of H$_2$O, CO, CH$_4$, NH$_3$, TiO, VO, Na, and K are 
         calculated at each iteration from thermochemical equilibrium, assuming a 0.83/0.17 
         H$_2$/He volume mixing ratio and solar system elemental abundances from Table 3 of 
         \cite{Lodders2010}. In the model, it is also possible to use enrichment 
         factors over the solar values, independently for C, O, N, and heavier elements.
         We considered only the species that significantly affect the profiles of the above 
         mentioned molecules in cool giant exoplanets according to \cite{Burrows1999i} and 
         \cite{Lodders2006}. These species are given in Table \ref{tabCompounds} (Column 2). 
         We also included species involved in the formation of silicate and iron clouds to 
         determine their condensation levels. Equilibrium abundances are derived from the 
         equations of conservation for each element and using the standard molar free 
         energies $\Delta G^0 (T)$ listed in \cite{chase1998monograph} to calculate 
         equilibrium constants involving the species in Table \ref{tabCompounds} 
         (Column 2). Calculation is done level by level, starting from the deepest level of 
         our grid, at highest pressure and temperature, and moving upward in the grid. When 
         a condensate appears in a given layer, its constituent elements are partly removed
         from the gas phase and the new elemental abundances in the gas phase are used to 
         calculate equilibrium abundances in the overlying layer. If no condensation 
         occurs, the same elemental abundances are used in the overlying layer. As in 
         \cite{Lodders2006}, we take  the dissolution of VO in perovskite 
         (CaTiO$_3$) into account, assuming an ideal solid solution and Henry's law.

      \subsubsection{Cloud model}
      
        Absorption by silicate and iron clouds is included above their respective 
        condensation level $p_\mathrm{c}$ up to one hundredth of this pressure level. A 
        particle-to-gas scale height ratio of 1 is assumed. We assumed spherical particles 
        and used the Mie theory to calculate the absorption Mie efficiency $Q_\mathrm{abs}$ 
        as a function of wavelength. The particle size distribution follows a gamma 
        distribution with a mean radius $r$ and an effective variance of 0.05.
         
        As discussed by \cite{Ackerman2001g} and \cite{Marley2012a}, the cloud opacity is 
        expected to be proportional to the pressure $p_\mathrm{c}$ at the condensation 
        level, proportional to the total concentration of the condensing element (Si or Fe) 
        embedded in various molecules at level $p_\mathrm{c}$, and inversely proportional to 
        the gravity $g$. This relates to the available column density of condensing material 
        at the condensation level. We write the optical depth of the cloud as
        
        \begin{equation}
           \tau_\mathrm{cloud} = \tau_\mathrm{ref}\frac{p_\mathrm{c}}{p_\mathrm{ref}}
           \label{taucloud}
        ,\end{equation}
        
        where $p_\mathrm{ref}$ = 1 bar. Because the solar elemental ratios Si/H and Fe/H are 
        about the same, we assumed the particle column densities of the silicate and iron 
        clouds are in the ratios of the pressure of their condensation levels, and thus that 
        their $\tau_\mathrm{ref}$ at any wavelength are in the ratios of their 
        $Q_\mathrm{abs}$ at this wavelength. We then just keep one free parameter in this 
        cloud model, which is $\tau_\mathrm{ref}$ for the Fe cloud at some reference 
        wavelength.
            
        In running Exo-REM, we found that in some cases, when $\tau_\mathrm{ref}$ is
        large and the condensation curve is close to the solution temperature profile, the 
        model is instable through the iteration process and does not converge toward a 
        radiative equilibrium solution. This is because adding particulate opacity increases 
         the temperature significantly just above
        the condensation level. The temperature in this region may then become larger than 
        the condensation temperature and no self-consistent solution can be found. This 
        instability was also seen by \cite{Morley2012} who advocated a patchy atmosphere to 
        solve this problem and reach a radiative equilibrium state.
        
   \subsection{Input and output parameters}
         The input parameters of the model are the effective temperature $T_\mathrm{eff}$, 
         the acceleration of gravity $g$ at 1 bar, which affects the atmospheric scale 
         height and thus the optical depth profiles, and the oversolar enrichment factors 
         $\alpha$ for C/H, N/H, O/H and heavier elements X/H ($\alpha = 1$ for solar system 
         values). The other set of free parameters are the optical depth of the iron cloud
         $\tau_\mathrm{ref}$ at 1.2 $\mu$m and a reference condensation level of 1.0 bar, 
         and the mean radius $r$ of the cloud particles.
        
         For output, the model provides the radiative-convective equilibrium temperature 
         profile $T(p)$, the corresponding vertical profiles of the absorbers at chemical 
         equilibrium, and the spectrum at the resolution of the $k$-correlated coefficient 
         distribution, i.e., 20 cm$^{-1}$.
         
   \subsection{Examples of model outputs}
         This section shows examples of model outputs (spectra, temperature and abundance 
         profiles) for various input parameters, allowing us to investigate the effect of 
         surface gravity, effective temperature, and clouds. All models here, unless 
         specified, assume a solar metallicity. For models with clouds (silicates and iron), 
         we used $\tau_\mathrm{ref}$ =1 and a mean particle size of 30 $\mu$m 
         \citep{Ackerman2001g}. We do not consider water vapor (H$_2$O) condensation and 
         thus formation of ice clouds, which would occur in the upper atmospheres of planets 
         with $T_\mathrm{eff}$ less than $\sim$~600 K.
         
         Figure~\ref{CompareModel} allows us to compare a typical case of atmospheric models,
         with and without clouds, with two models in the literature: BT-Settl and Drift-PHOENIX. 
         The BT-Settl spectrum is relatively close to our case with clouds, 
         except for some difference in the wings of the alkali lines and for 
         the presence of an absorption band near 4.3 micron (outside of the wavelength 
         range accessible to SPHERE), presumably due to CO$_2$, which was not included
         in our model. Other minor differences may originate from differences in the 
         line lists or missing trace compounds in our model.
         The corresponding temperature profile of this BT-Settl model is shown
         in Fig. 4. On the other hand,  the Drift-PHOENIX spectrum is very different from 
         both the BT-Settl one and ours. The molecular absorption features are much less visible 
         because the clouds are thicker than in the other models and the spectrum 
         ressembles that of a blackbody.

         Figure~\ref{Identification} presents spectra calculated for a cloud-free 
         atmosphere, $\log (g)$ = 4 and two values of $T_\mathrm{eff}$, 700, and 2000 K. 
         Water vapor absorbs all over the spectral range for both effective temperatures, 
         with strongest bands centered at 0.94, 1.14, 1.38, 1.87, and 2.7 $\mu$m. Other 
         absorption features are due to other compounds as indicated in the figure. Besides 
         H$_2$O, K and Na absorption are visible both in low- and high-temperature spectra 
         through their resonant lines at 767 and 589 nm, respectively. On the other hand, 
         TiO, VO, and CO absorptions are only important  for large $T_\mathrm{eff}$, while 
         CH$_4$ bands play a significant role at low temperatures.  NH$_3$ has a weaker 
         effect,  is only visible in planets with a low $T_\mathrm{eff}$, and provides 
         additional absorption around 3.0, 2.0, and 1.5 $\mu$m.

         Figure~\ref{CloudEffect} shows spectra calculated for $T_\mathrm{eff}$ varying from 
         700 to 1700 K and $\log (g)$ varying from 3 to 5. The fluxes correspond to a planet 
         having a Jupiter radius and located at 10 pc. For each set of parameters, a 
         cloud-free
          model and a model with clouds are shown. For the case with $\log (g)$ = 5 and 
         $T_\mathrm{eff}$ = 700 K, only the cloud-free case is shown because cloud 
         condensation occurs below our pressure grid and cannot be taken into account.
         The temperature profiles corresponding to these sets of parameters and to a cloudy 
         atmosphere are shown in Fig.~\ref{profils}. The locations of the iron and silicate 
         clouds and of the radiative-convective boundary are also indicated. 
         The BT-Settl temperature profile for $T_\mathrm{eff}$ = 1400 K and $\log(g)$ = 4 
         (solid green line) is shown for comparison with the Exo-REM profile 
         (solid black line). Above the radiative-convective boundary, the two 
         profiles are different, with a given temperature  reached one or 
         two pressure scale heights deeper in the BT-Settl model. The reason 
         for this discrepancy is unknown but probably lies in different 
         vertical distributions of the opacity.

         For a given $T_\mathrm{eff}$, adding cloud absorption yields  a smoother spectrum,  
         decreasing the contrast between absorption bands and spectral windows. This is 
         because cloud opacity, concentrated near the cloud base in the 1600-2100 K range, 
         (depending on $\log (g)$ and $T_\mathrm{eff}$, as shown in Fig.~\ref{profils}) more strongly
         reduces  the flux in the windows, which originate from deeper levels 
         than the flux in the absorption bands. Cloud opacity also affects the relative 
         fluxes in the various photometric bands. Essentially, for a given $T_\mathrm{eff}$, 
         the flux is reduced below $\sim$1.7 $\mu$m and increased longward. Therefore, the 
         flux is significantly lower in the Y and J bands, at which the atmosphere is the 
         most transparent, and higher in the K, L, and M bands where atmospheric opacity is 
         larger. In the set of examples we show in Fig.~\ref{CloudEffect}, the strongest 
         cloud effects are seen for $\log (g)$ = 5 and $T_\mathrm{eff}$ = 1000, 1400, or 
         1700 K.  In these cases, the emission resembles that of a blackbody at temperature 
         $T_\mathrm{eff}$.
         The pressure levels where $T$ = $T_\mathrm{eff}$, representative 
         of the mean emission level, are the deepest of the set in Fig.~\ref{profils}, 
         $\sim$ 0.4 bar, and thus the cloud opacity at this level is the largest, 
         according to Eq.~\eqref{taucloud}.    

         \begin{figure}[htb!]
            \includegraphics[width=0.5\textwidth]{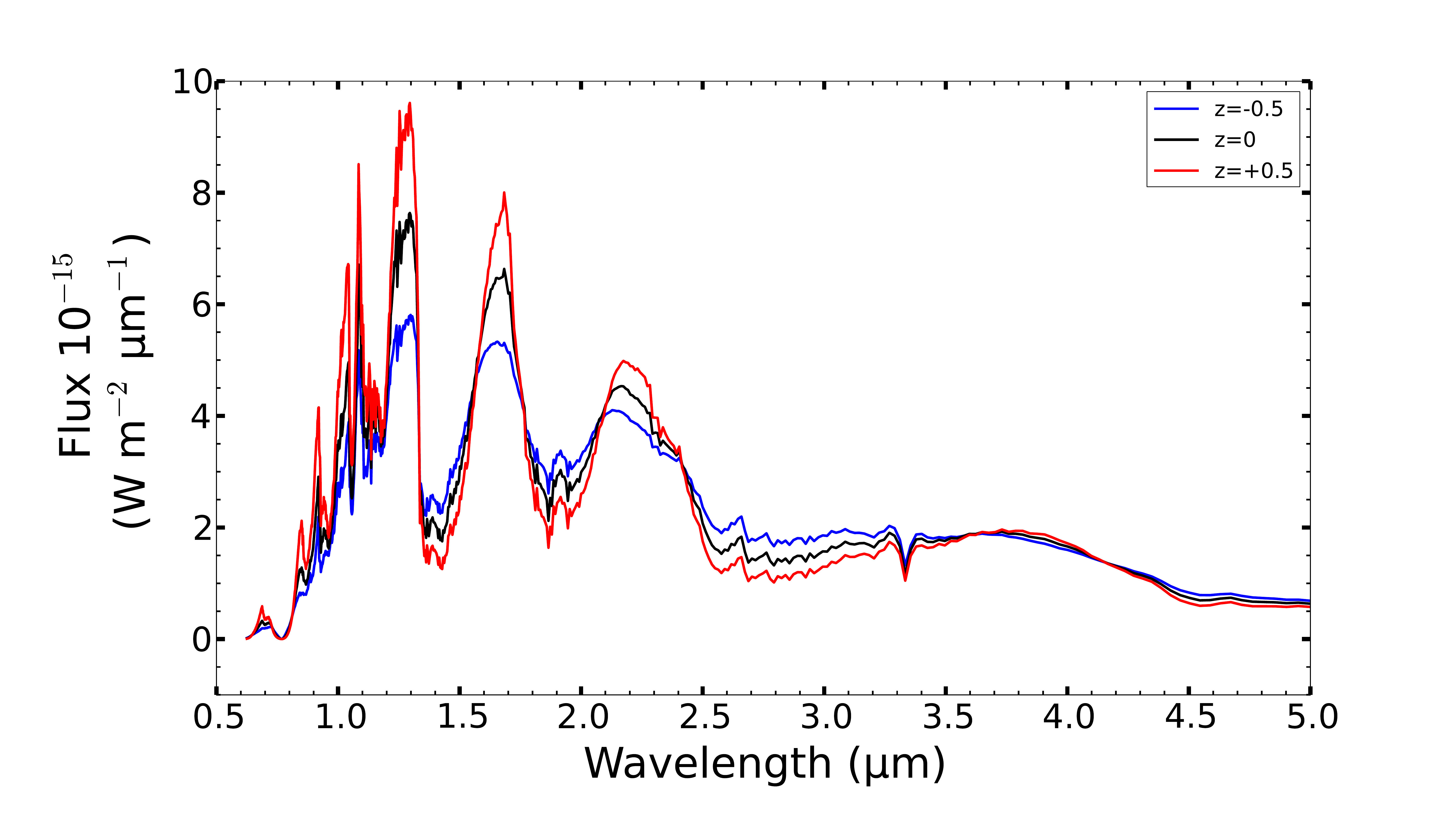}
            \caption{Effect of varying the metallicity $Z$ in a cloudy model with 
            $T_\mathrm{eff}$ = 1400 K and $\log (g)$ = 4.0}
            \label{MetalEffect} 
         \end{figure}         
         
         We only show here the effect of clouds for a single set of parameters: 
         $\tau_\mathrm{ref}$ = 1 and a particle scale height equal to the pressure scale 
         height. For this parametrization, the cloud optical depths at a given pressure 
         level are the same for any location of the condensation level (see 
         Eq.~\ref{taucloud}) and any value of $T_\mathrm{eff}$  or $\log (g)$. Therefore, we 
         cannot draw any conclusion on the relative effect of cloud opacity as a function of 
         $T_\mathrm{eff}$ or  $\log (g)$ from the calculations shown in 
         Fig.~\ref{CloudEffect}. For example, it could be reasonable to assume that 
         $\tau_\mathrm{ref}$ varies as 1/$g$, as does the pressure scale height, following 
         the parametrization of \citeauthor{Ackerman2001g} (\citeyear{Ackerman2001g}, e.g., their Eq.~18). This would reduce 
         the increasing effect of clouds with increasing gravity. Also, if the cloud is more 
         confined near the condensation level than assumed here, i.e., a particle-to-gas 
         scale height ratio lower than 1, the effect of clouds would be much reduced for 
         cases with $T_\mathrm{eff}$ = 700 or even 1000 K since particles would be confined 
         to levels well below the mean emission level.         

         \begin{figure}[htb!]
            \includegraphics[clip,width=0.5\textwidth]{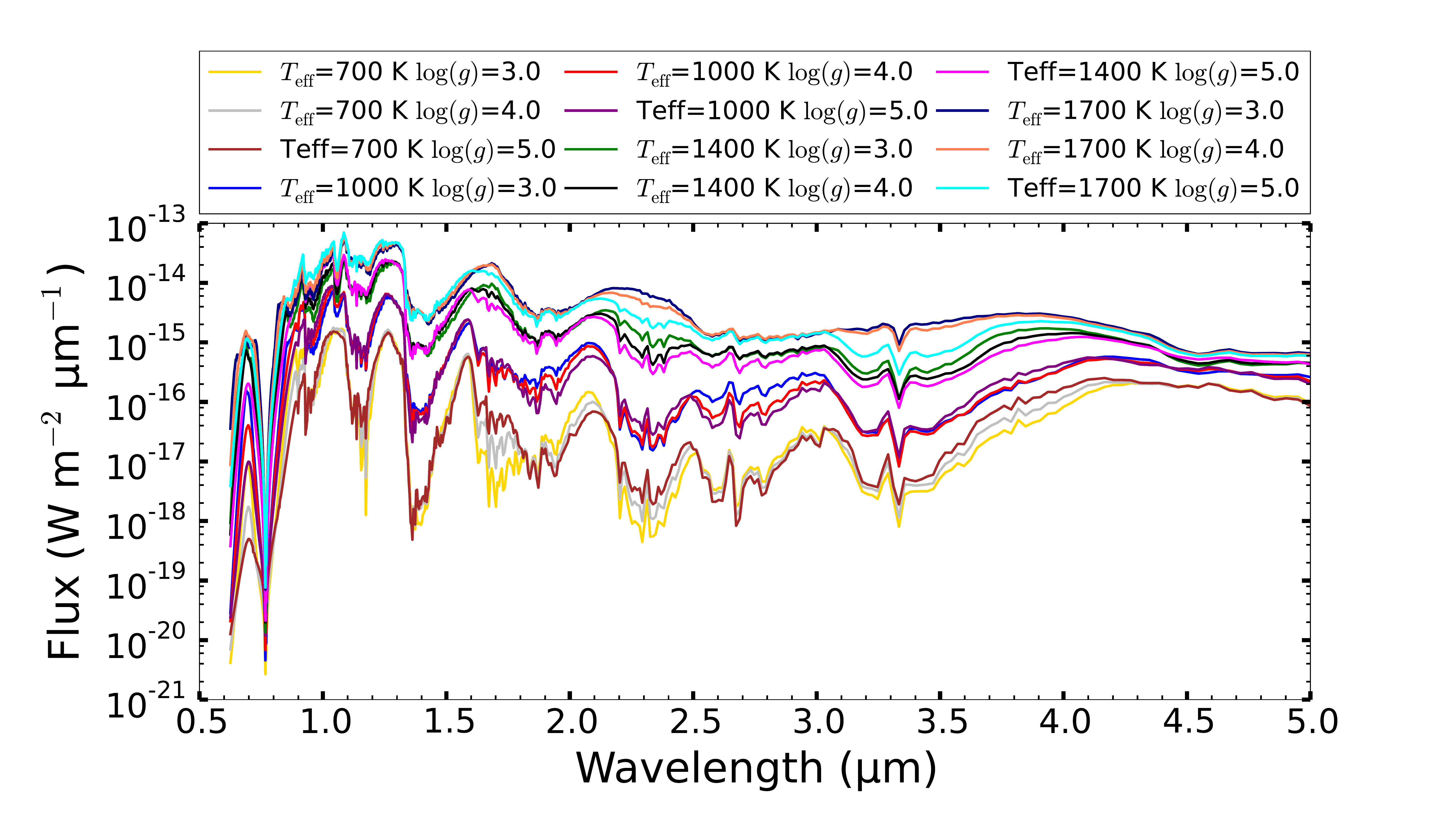}
            \caption{Series of spectra calculated for a cloud-free atmosphere, 
            $T_\mathrm{eff}$ varying from 700 to 1700 K, and $\log (g)$ from 3 to 5.}
            \label{CloudEffectNOcloud} 
         \end{figure}

         \begin{figure*}[htb!]
            \begin{center}
               $\begin{array}{lll}
                  \includegraphics[trim = 2.5cm 2cm 15.0cm 2cm, clip,scale=0.12]{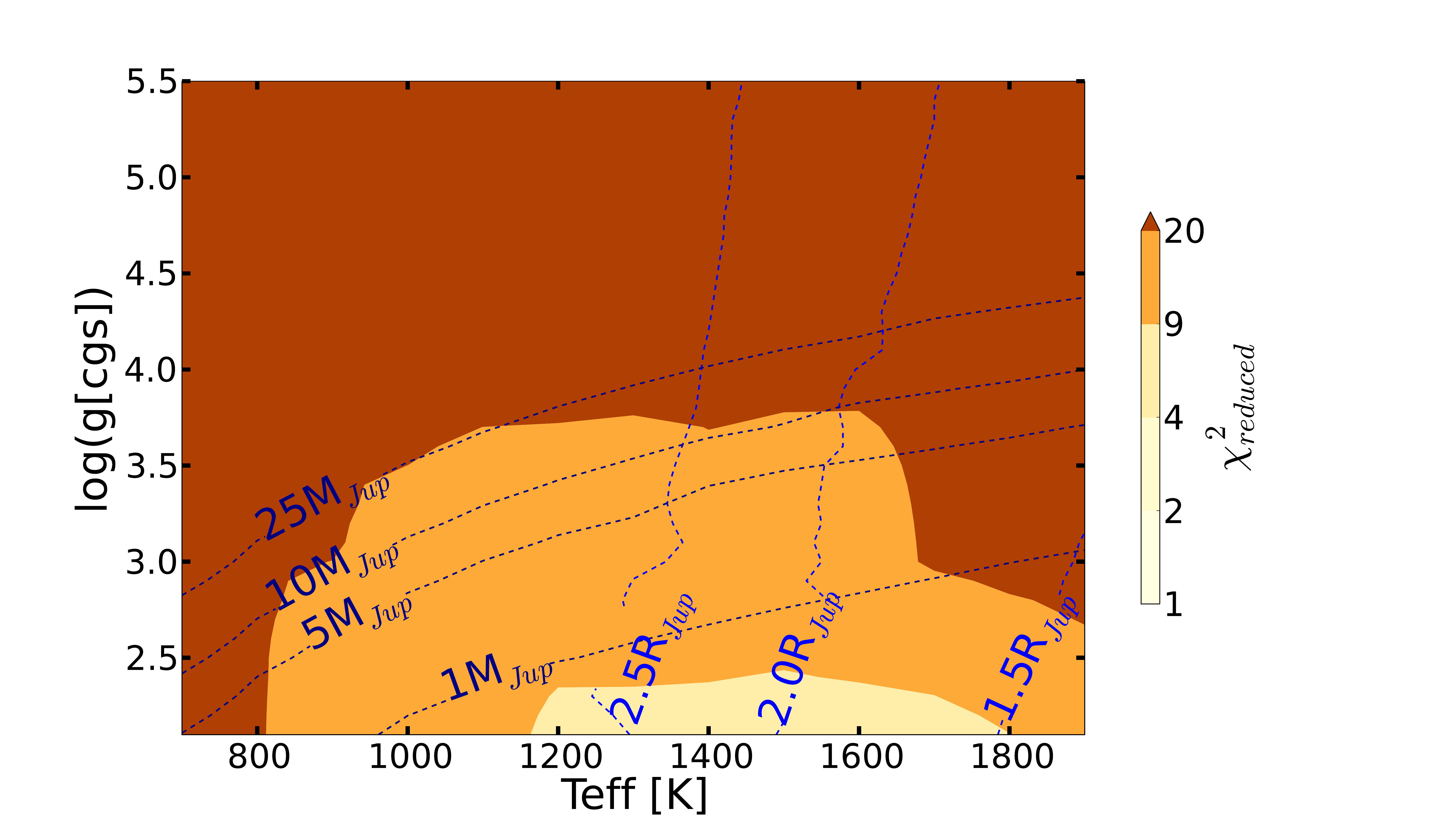}
                  &\includegraphics[trim = 5.2cm 2cm 5.0cm 2cm, clip,scale=0.12]{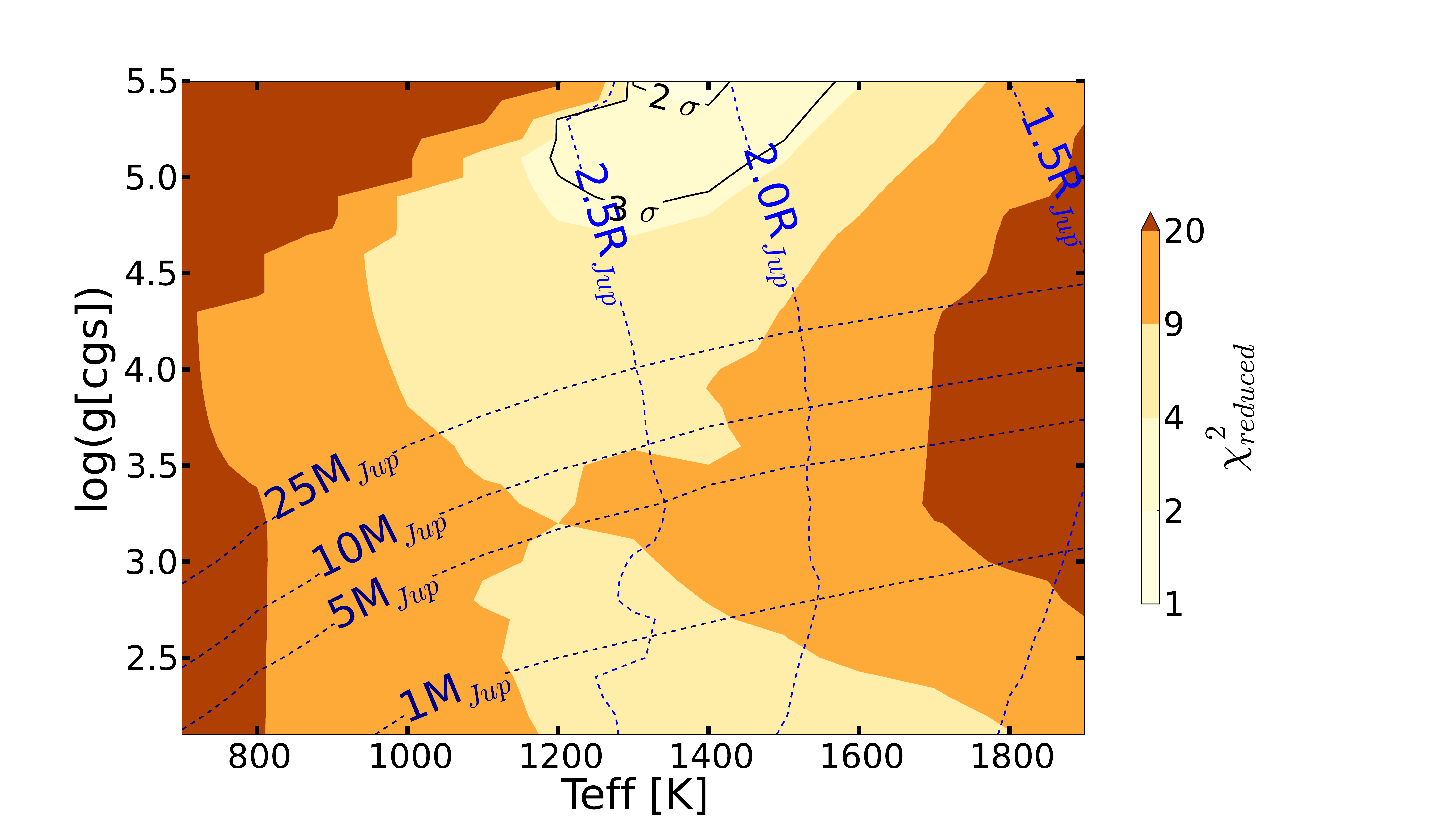} 
                  &{ }
               \end{array}$
               $\begin{array}{lll}
                   \includegraphics[trim = 2.5cm 0cm 15.0cm 2cm, clip,scale=0.12]{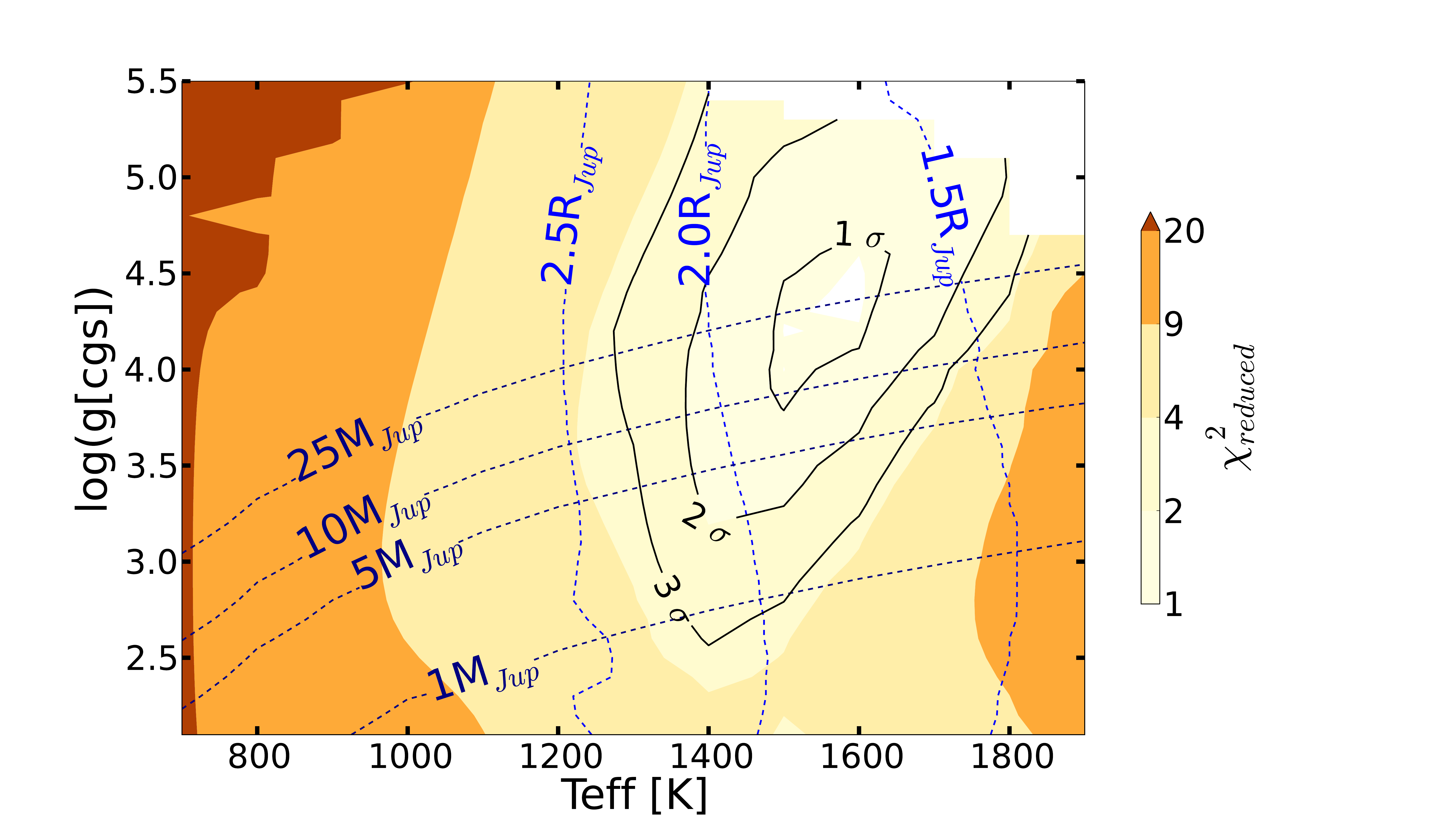} 
                  &\includegraphics[trim = 5.2cm 0cm 15.0cm 2cm, clip,scale=0.12]{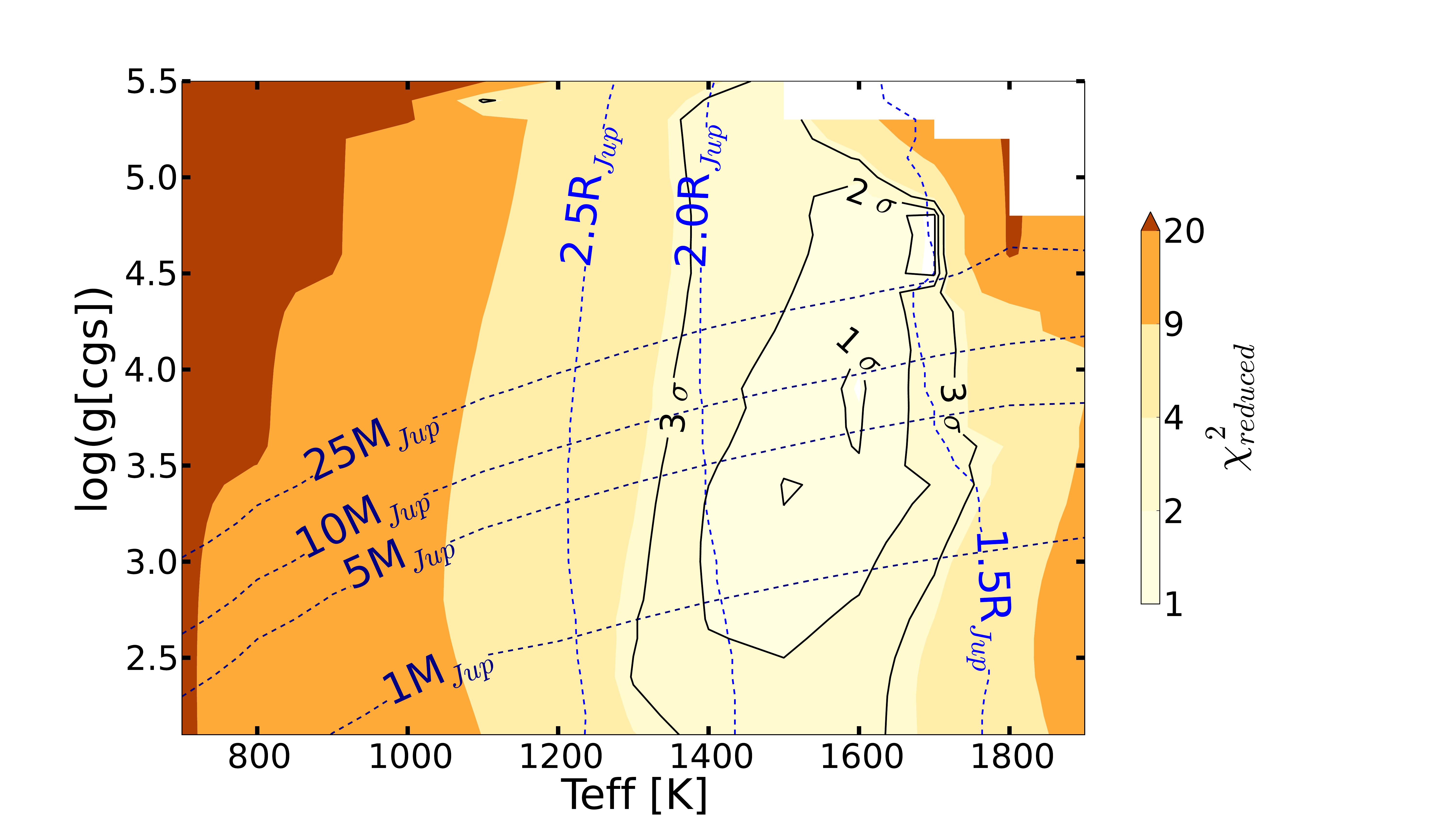} 
                  &\includegraphics[trim = 5.2cm 0cm 5.0cm 2cm, clip,scale=0.12]{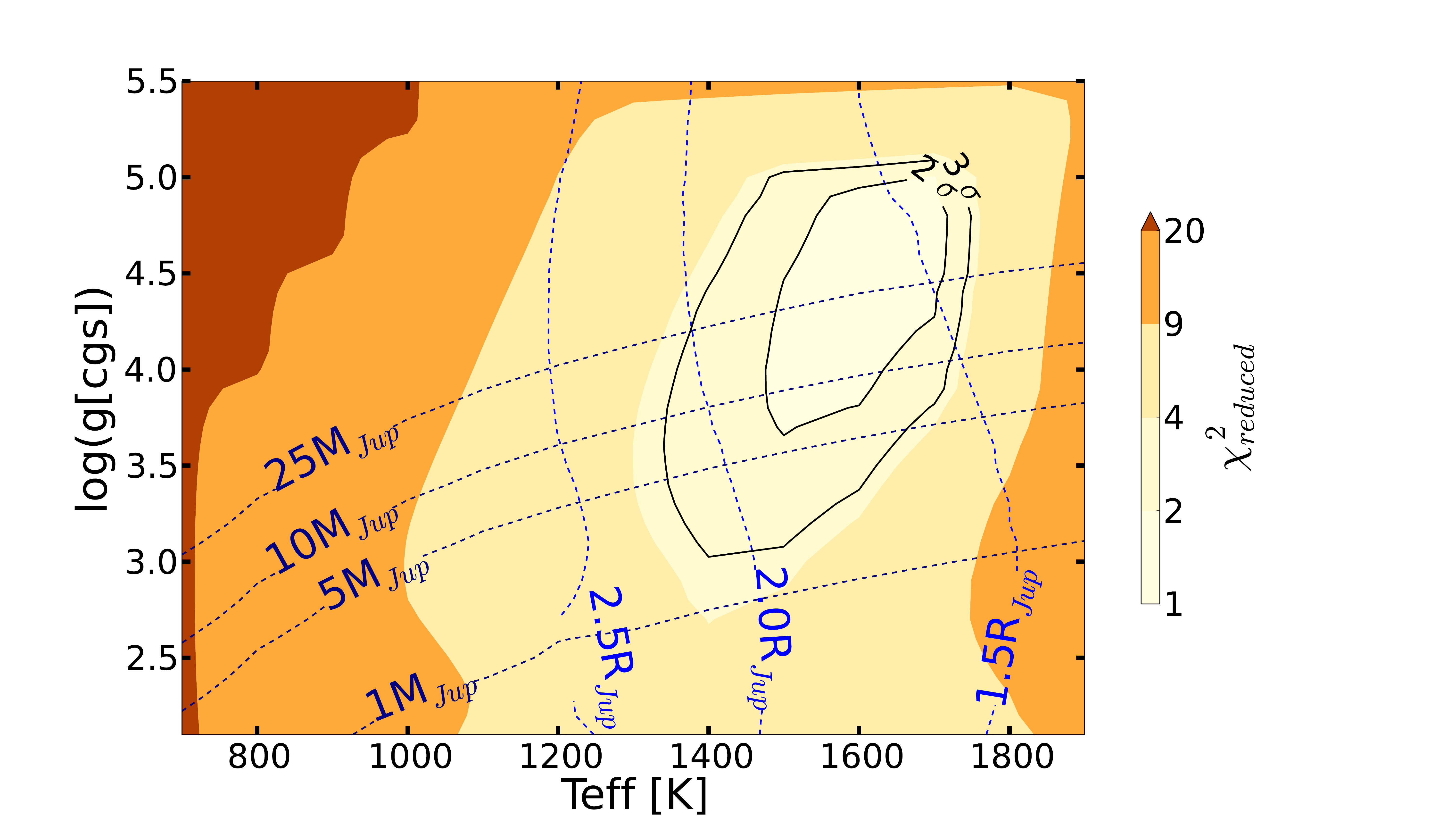} 
               \end{array}$
            \end{center}
            \caption{Maps of reduced $\chi^2$ for \object{$\beta$ Pictoris $b$} SED. 
            Vertical blue lines indicate radii of 2.5, 2.0, and 1.5  $R_\mathrm{Jup}$. 
            Horizontal blue lines indicate masses of 25, 10, 5, and 1 $M_\mathrm{Jup}$.  From 
            top left to bottom right, models correspond to: no cloud, $\tau$ = 0.1 and $<r>$ 
            = 30 $\mu$m, $\tau$ = 1 and $<r>$ = 30 $\mu$m, $\tau$ = 3 and $<r>$ = 30 $\mu$m, 
            $\tau$ = 1 and $<r>$ = 3 $\mu$m, respectively.}
            \label{betpicSEDxi2maps}
         \end{figure*}          
         
         Obviously, the main effect of increasing the effective temperature is to increase 
         the emitted flux but, in addition, changes in the spectral shape  between 1 and 4 
         $\mu$m can be noted. As $T_\mathrm{eff}$ increases, CH$_4$ bands, mostly visible at 
         1.7, 2.3, and 3.3 $\mu$m, become less and less intense, while the CO band at 4.7 
         $\mu$m as well as TiO and VO absorption below 1.3 $\mu$m become visible. These large 
         variations may easily be detected from narrowband photometry or low-resolution 
         spectroscopy: TiO and VO signatures occur in the J-band, CO affects the M-band at 
         high $T_\mathrm{eff}$ while, at low $T_\mathrm{eff}$, CH$_4$ has a strong effect in 
         the H, K, and L-bands, and NH$_3$ has a marginal effect in the K-band.   

         \begin{table}[htb!]
          \centering
          \caption{Clouds parameters used in the five test grids}
          \begin{tabular}{cccc} 
          \hline \hline
             $<r>$  & $\tau_\mathrm{ref}$ & $\tau_\mathrm{Mg_2 SiO_4}$ & $\tau_\mathrm{Fe}$\\
             {($\mu$m)}&{}& {\tiny ($\lambda$=1.2 $\mu$m; p = 1 bar)} & {\tiny ($\lambda$=1.2 $\mu$m; p = 1 bar)}\\ \hline             
              & 0 & 0 & 0\\
             30 & 0.1 & 0.015 & 0.1\\
             30 & 1 & 0.15 & 1\\
             30 & 3 & 0.45 & 3\\
             3 & 1 & 0.018 & 1\\
             \hline
          \end{tabular}
          \label{5grids}
       \end{table} 
        
         The spectral variations with effective temperature are, of course, due to changes 
         in composition as illustrated in Fig.~\ref{MixProf}. Carbon is partitioned between 
         CO and CH$_4$, with a CO/CH$_4$ ratio depending on temperature and, to a lesser 
         extent, on pressure. In the 700-K planet, methane dominates over carbon monoxide 
         above the 1-bar level whereas in the 1700-K planet, CO dominates over the whole 
         pressure grid. Similarly, nitrogen is partitioned between N$_2$ and NH$_3$, the 
         latter being abundant in the observable atmosphere only for relatively low 
         effective temperatures ($\leq$ 800 K).  Also, as the effective temperature 
         decreases, TiO and VO get confined to deeper levels and have thus less influence on 
         the outgoing flux. The depletion of TiO and VO in the upper (colder) atmosphere is 
         due to perovskite (CaTiO$_3$) formation and VO condensation, respectively. Alkali Na 
         and K affect all spectra in our grid but are confined at deeper levels in the 
         case of low $T_\mathrm{eff}$ atmospheres. They are removed from the upper 
         atmosphere through the formation of Na$_2$S condensate and KCl condensation, respectively.

         Fig.~\ref{MetalEffect} shows the effect of varying the metallicity for given 
         $T_\mathrm{eff}$ and $\log (g)$ assuming no clouds. As expected, increasing the 
         metallicity increases the depth of all absorption bands. For example, considering 
         the water vapor band at 2.7 $\mu$m, a metallicity of $Z$ = +0.5 produces a band 
         depth (2.2 / 2.7 $\mu$m) twice as large as in the case with $Z$ = -0.5. In 
         principle, the metallicity of an observed exoplanet could thus be deduced from low-resolution spectroscopy provided that the temperature profile modeled from 
         radiative-convective equilibrium is reliable, which also requires that 
         $T_\mathrm{eff}$ and $\log (g)$ can be accurately derived from the spectra. The
         unknown effect of clouds may be a stronger limitation in some cases since cloud 
         absorption reduces the band depths and may mimic some decrease in metallicity.

        The gas scale height is inversely proportional to the acceleration of gravity $g$. 
        As a result, a given optical depth at a given wavelength is found at deeper pressure 
        levels when $g$ increases. This explains the general behavior of the temperature 
        profiles as a function of $\log (g)$ for a given effective temperature as seen in 
        Fig.~\ref{profils}. As $\log (g)$ increases, the temperature profile generally moves 
        downward along with the cloud condensation levels. The situation is however more 
        complicated because of the presence of clouds and  the dependence of molecular 
        absorptivity with pressure. The effect of gravity on the calculated spectral shape 
        is more subtle than that of effective temperature. It is best seen in spectra of 
        Fig.~\ref{CloudEffectNOcloud} having no cloud opacity. Because thermochemical 
        equilibrium at a given temperature depends on pressure, the gas abundances at a 
        given temperature level depend on the pressure at this level and thus indirectly on 
        the gravity. For example, the CH$_4$/CO ratio at a given temperature varies as the 
        square of pressure so that the methane mixing ratio at and above 
        the atmospheric level where $T$ = $T_\mathrm{eff}$, representative of the mean 
        emission level, is larger for larger $g$. This explains the large 
        increase in the depth of the methane bands for the $T_\mathrm{eff}$ = 1700 K (and to 
        a lesser extent 1400 K) profiles when $\log (g)$ increases from 3 to 5. In this 
        case, the CH$_4$ mixing ratio is two orders of magnitude larger at the $T$ = 1700 K 
        level for $\log (g)$ = 5 than for $\log (g)$ = 3. These calculations suggest that, 
        among objects with $T_\mathrm{eff} \sim$ 1600-1800 K, methane absorption would be 
        detectable at 2.3 or 3.3 $\mu$m in brown dwarfs, but would probably not be 
        detectable at 2.3 or 3.3 $\mu$m in Jupiter-mass 
        planets. On the other hand, for the $T_\mathrm{eff}$ = 700 K profiles, the 
        CH$_4$/H$_2$ mixing ratio is similar for all $\log (g)$ at and above the $T$ = 
        $T_\mathrm{eff}$ level,  at its maximum value, which is twice the C/H elemental 
        ratio. In conclusion, the effect of gravity on the spectra is significant but may be 
        difficult to disentangle from compositional variations.

         \begin{figure*}[htb!]
            \begin{center}
               $\begin{array}{lll}
                  \includegraphics[trim = 2.5cm 2cm 15.0cm 2cm, clip,scale=0.12]{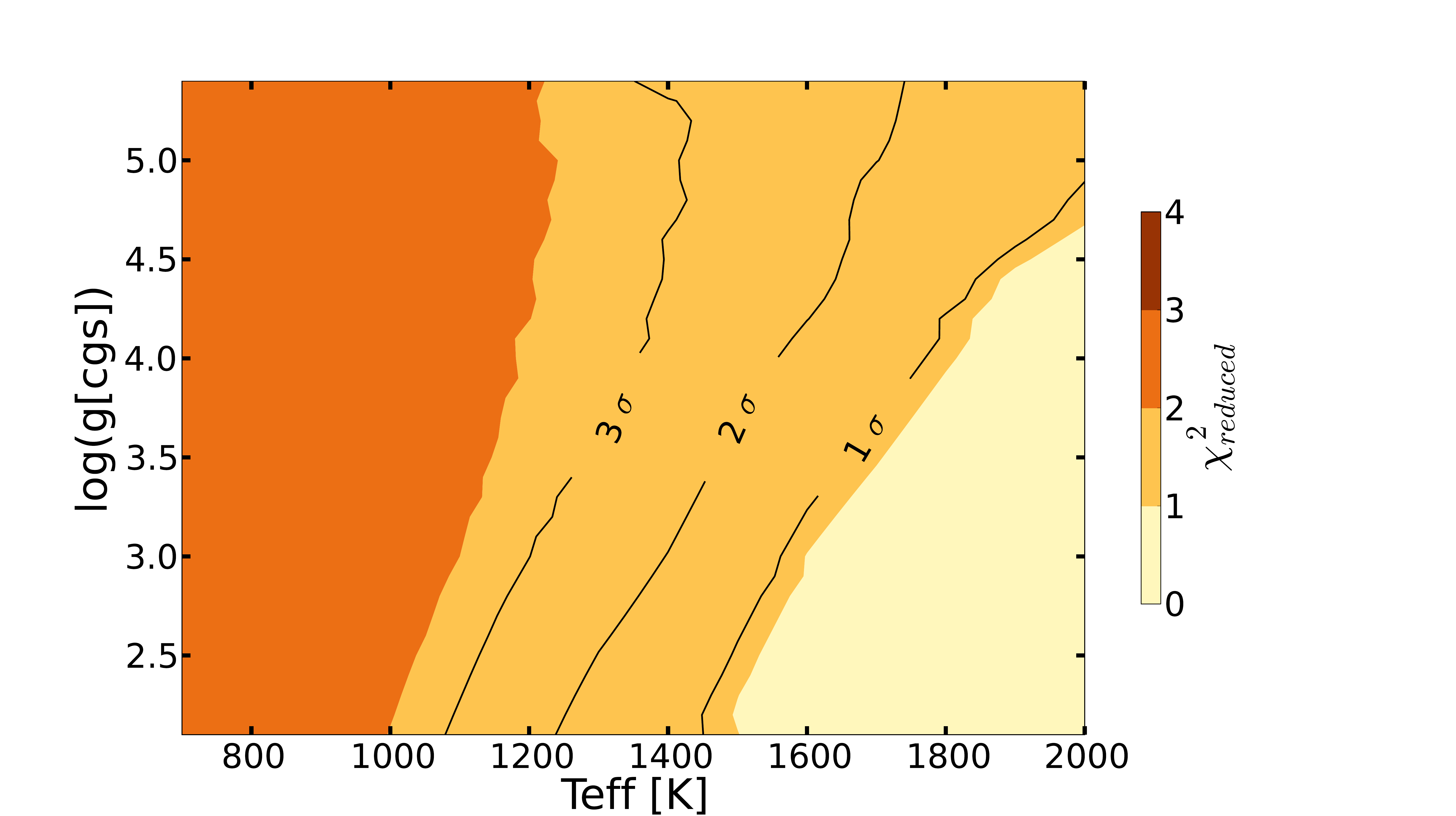}
                  & \includegraphics[trim = 5.2cm 2cm 5.0cm 2cm, clip,scale=0.12]{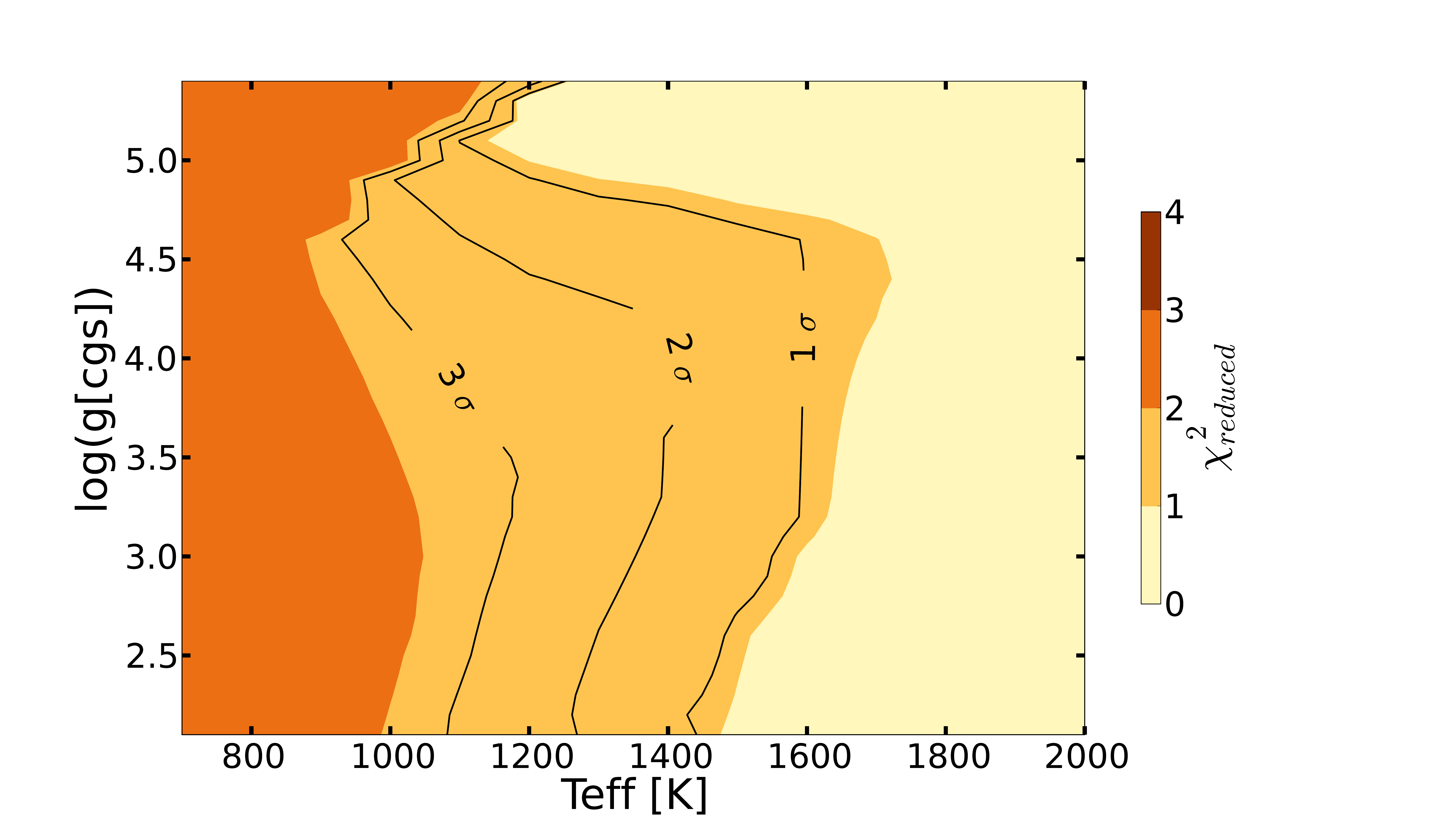} 
                  & { }
               \end{array}$
               $\begin{array}{lll}
                   \includegraphics[trim = 2.5cm 0cm 15.0cm 2cm, clip,scale=0.12]{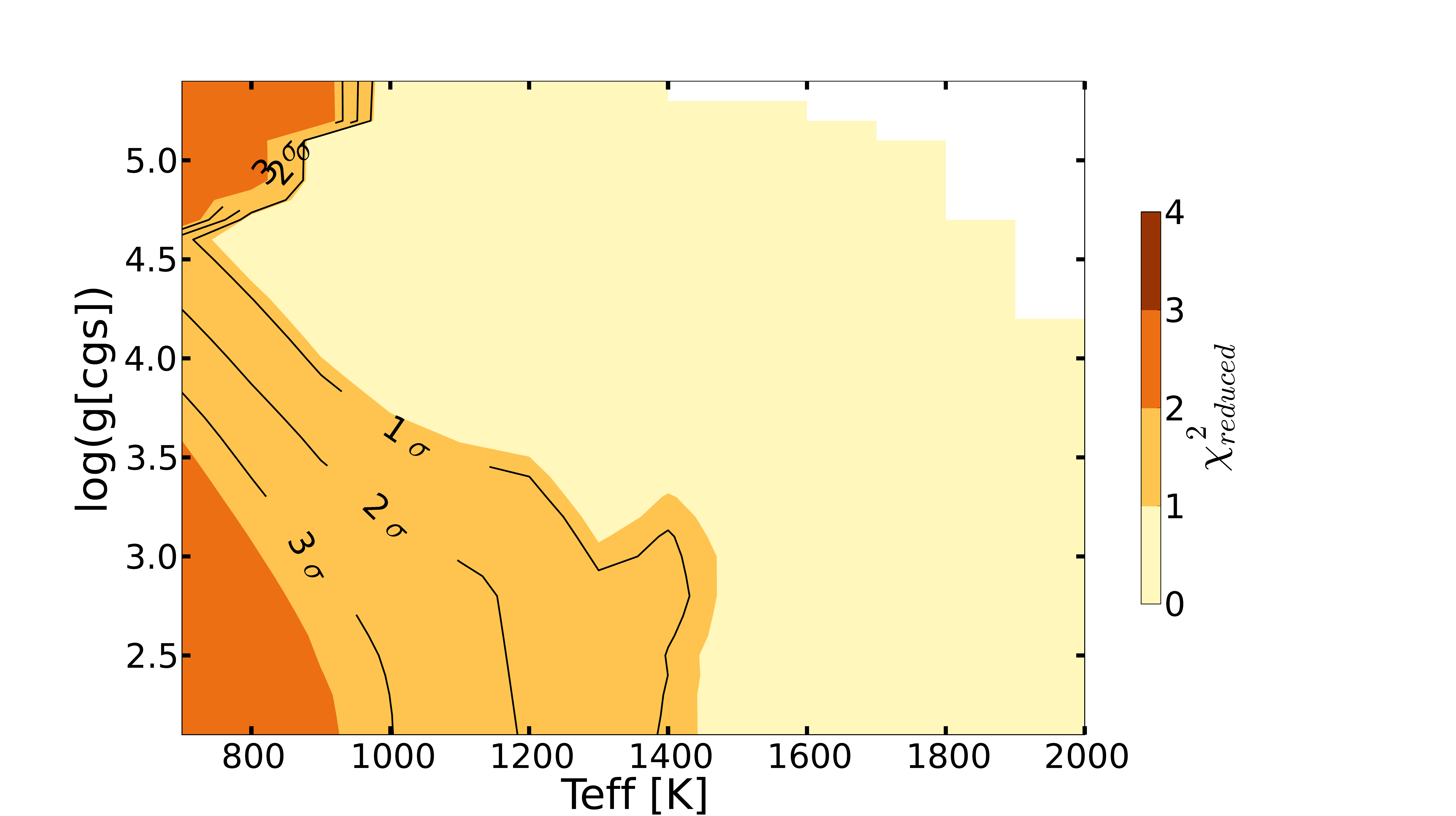} 
                  & \includegraphics[trim = 5.2cm 0cm 15.0cm 2cm, clip,scale=0.12]{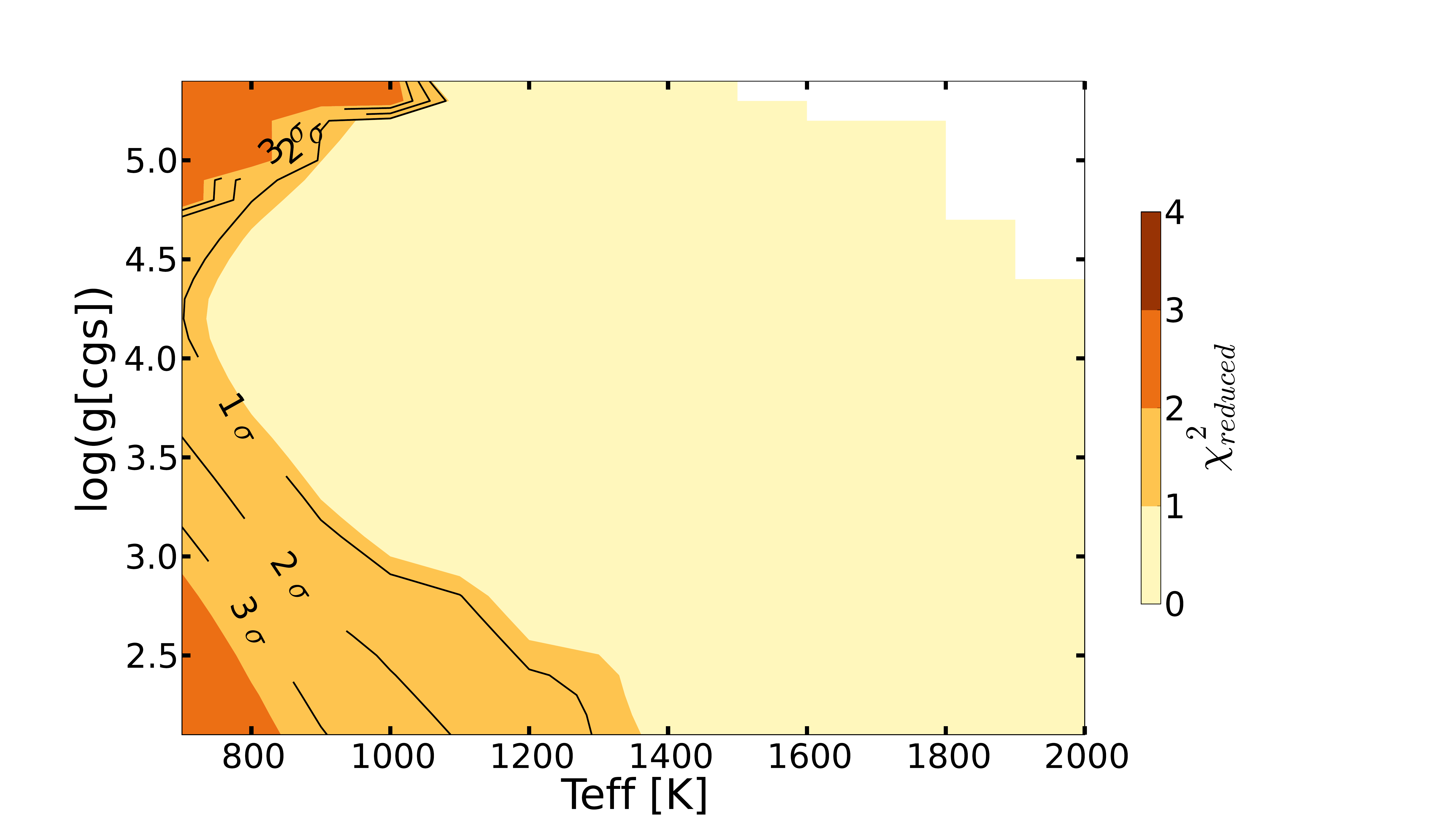} 
                  & \includegraphics[trim = 5.2cm 0cm 5.0cm 2cm, clip,scale=0.12]{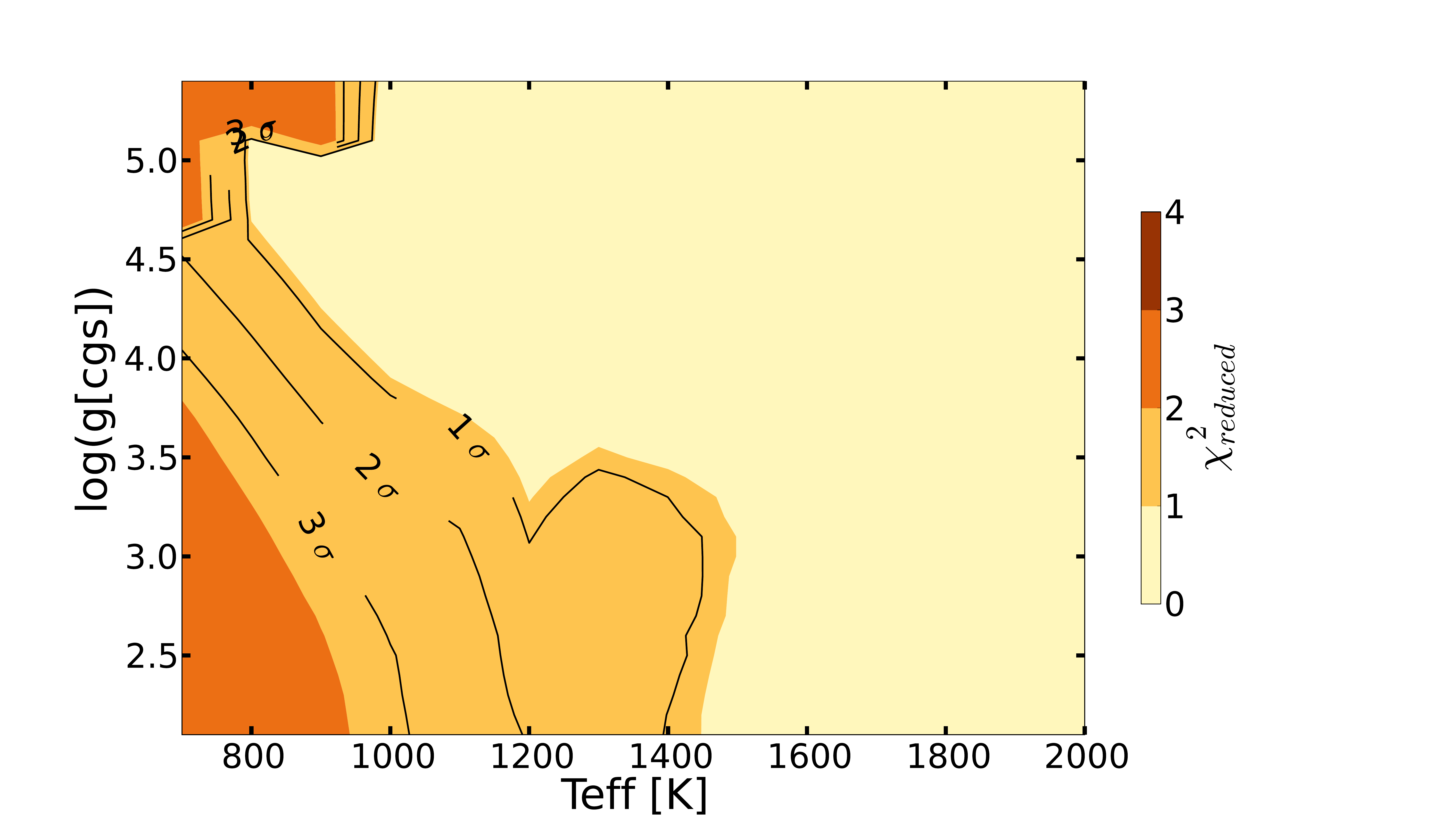}                
               \end{array}$
            \end{center}
            \caption{Maps of reduced $\chi^2$ for \object{$\beta$ Pictoris $b$} J-band spectrum. 
            From top left to bottom right, models correspond to: 
            no cloud, $\tau$ = 0.1 and $<r>$ = 30 $\mu$m, $\tau$ = 1 and $<r>$ = 30 $\mu$m,
            $\tau$ = 3 and $<r>$ = 30 $\mu$m, $\tau$ = 1 and $<r>$ = 3 $\mu$m, respectively.}%
            \label{betpicspecxi2maps}
         \end{figure*} 
         
\section{Application of the model to actual observations}
   \subsection{Method}
         
       We now describe how we exploit existing data to derive characteristics of planets.       
       As a first step, the model generates a grid of spectra for a range of physical 
       parameters, $\log(g)$ between 2.1 and 5.5 with a step of 0.1 and $T_\mathrm{eff}$ 
       between 700 and 2000 K with a step of 100 K. Importantly, the explored parameter 
       space must be large enough to encompass all acceptable solutions. For simplicity, we 
       fixed the planet radius to one Jupiter radius ($R_\mathrm{Jup}$), and leave the 
       determination of the planet radius to the minimization part (see below). 
          
       Besides the direct geometrical effect on the observed flux, the radius also affects 
       the variation of the acceleration of gravity with altitude, and thus the scale height 
       at a given pressure level. We tested this effect in a few test cases in our grid by 
       solving for radiative equilibrium for two different radii and comparing the 
       corresponding spectra. We only observed  very small modifications of the shape of the 
       spectrum, negligible compared with other error bars. Hence the radius may be 
       considered as an independent scaling parameter, only affecting the observed flux 
       through the area $\pi R^2$ seen from Earth. Physical parameters $T_\mathrm{eff}$ and 
       $g$ are derived with associated 1- or 2-$\sigma$ error bars (68$\%$ and 95$\%$ 
       confidence level, respectively) from a $\chi^2$ analysis with $n$-1 degrees of freedom 
       \citep{Bevington2003}, where $n$ is the number of independent observation points (one 
       degree of freedom is removed by the determination of $R$, see below).      
       
       Five types of clouds were considered with characteristics  given in Table 
       \ref{5grids}: one without cloud, three with a mean particle radius of 30 $\mu$m and 
       $\tau_\mathrm{ref}$ = 0.1, 1, and 3, and one with a mean particle radius of 3 $\mu$m 
       and $\tau_\mathrm{ref}$ = 1. The differences in the absorption efficiency $Q_{abs}$  
       between iron and forsterite explain the differences in the optical depth calculated 
       for each compound. As already mentioned in section 2.2.2, the model may be 
       unstable for $\tau_\mathrm{ref} > 3$.

       The data consist of a series of either photometric points (broadbands and/or narrowbands) expressed in magnitudes, a normalized spectrum, or both. We compute the 
       $\chi^2$ between the data $X_\mathrm{Observed}$ and each synthetic spectrum in our 
       grid, once integrated over the photometric filters or convolved to the spectrograph 
       resolution ($X_\mathrm{Model}$), with the following relation:
       
       \begin{equation}
          \chi^2 = \sum (\frac{X_\mathrm{Observed}-X_\mathrm{Model}}{ \Delta 
          X_\mathrm{Observed}}) ^2 ,
       \end{equation}
       
       where $\Delta X_\mathrm{Observed}$ are the uncertainties in the planet photometry.  
       Then, in the case of photometric measurements, we derive the radius that minimizes 
       the $\chi^2$ metric
       
       \begin{equation}
          5 \log_\mathrm{10} (R)=-\frac{\sum (\frac{X_\mathrm{Observed}-X_\mathrm{Model}}
          { \Delta X_\mathrm{Observed}^2}) }
                 {\sum (\frac{1}{ \Delta X_\mathrm{Observed}^2}) }.\end{equation}

       Therefore, the radius $R$ (given in $R_\mathrm{Jup}$ unit) is considered  a global 
       scaling parameter that does not influence the shape of the synthetic spectra. 
       Finally, additional constraints based on models and measurements can be introduced 
       in the analysis. Considering the core-accretion model \citep{Mordasini2012d} and the 
       hot-start model \citep{Spiegel2012a}, assuming a given age of the star, the radius 
       range can be restrained within lower and upper boundaries. In addition, radial 
       velocity measurements, when available, can be used to put constraints on mass and thus on 
       $g$ thanks to the relation,
       
       \begin{equation}
          g = \frac{GM}{R^2} .
       \end{equation}

      \begin{figure}[htb!]
         \includegraphics[trim = 0.8cm 2.5cm 2cm 1cm, clip,width=0.5\textwidth]{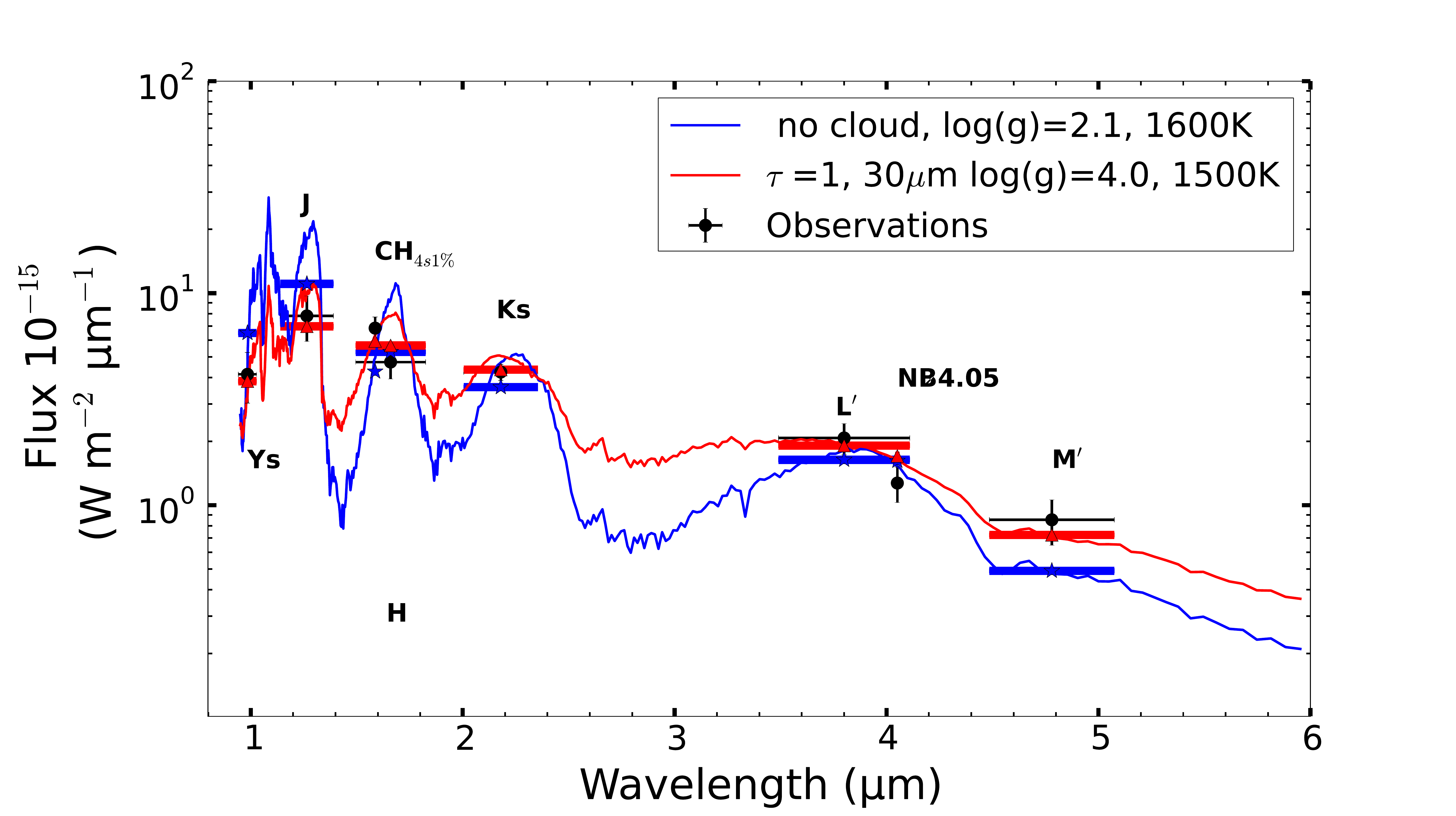}
         \includegraphics[trim = 1.3cm 0cm 2cm 1cm, clip,width=0.5\textwidth]{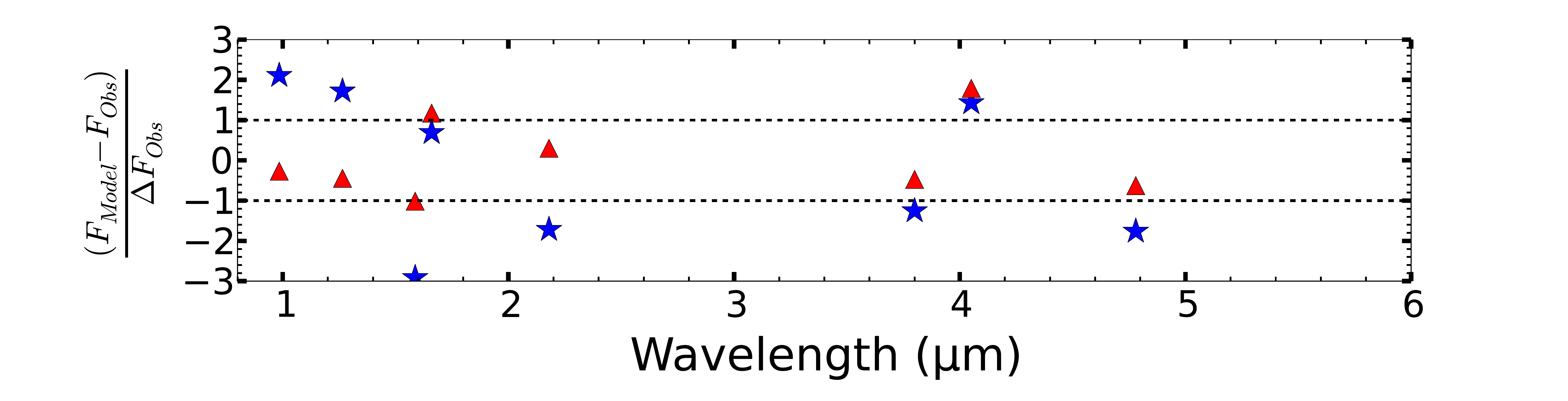}
         \caption{Top: best $\chi^2$ for \object{$\beta$ Pictoris $b$} SED (black dots) with 
         (red triangles) and without (blue stars) clouds after radius and mass selection.
         Thick lines indicate the widths of the filters. CH$_4$ $_{S,1\%}$ and NB 4.05 are 
         very narrow filters, less than the width of plotted observation points. Bottom: 
         difference between synthetic and observed fluxes divided by the uncertainty on the 
         observed flux. The two horizontal lines indicate the $\pm$1 standard deviation 
         level.}
         \label{betpicBestPlot} 
      \end{figure}
      
      \begin{figure}[htb!]
         \includegraphics[trim = 1cm 0cm 2cm 1cm, clip,width=0.5\textwidth]{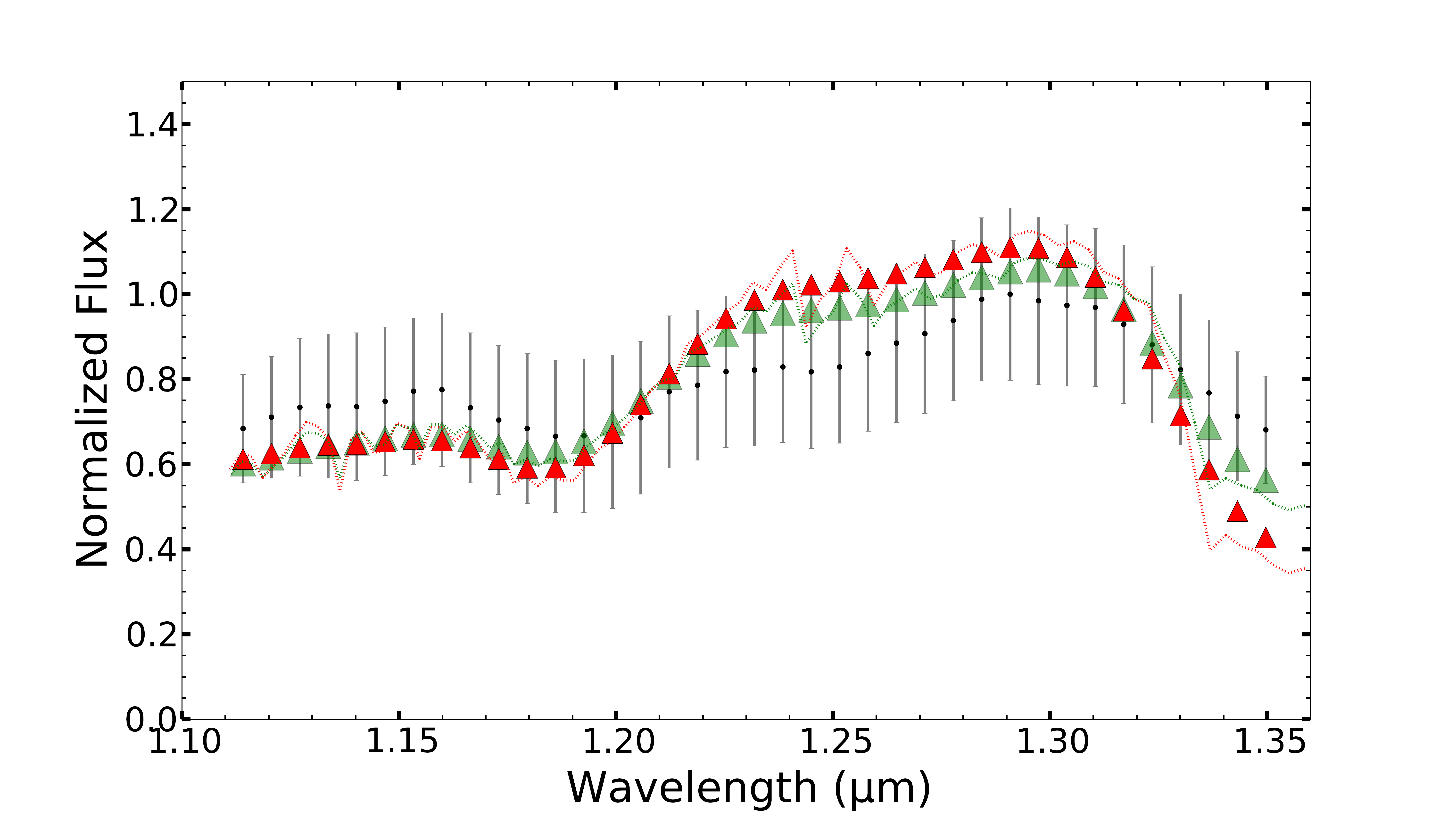}
         \caption{The normalized GPI J-band spectrum of \object{$\beta$ Pictoris $b$} 
         (black) is plotted against the best fit of the SED also plotted in Fig.~\ref{betpicBestPlot} with clouds 
         (red triangles). Also plotted in green is a model with $T_\mathrm{eff}$ = 1500 K, 
         $\log(g)$ = 4.1 and $\tau_\mathrm{ref}$ = 3, which correctly reproduces the GPI H-band spectrum
         (Fig. \ref{betpicHband}).
         The dotted lines represent the calculated spectra prior to convolution.}
         \label{betpicBestSpectrumPlot} 
      \end{figure} 
      
   \subsection{\object{$\beta$ Pictoris $b$}}

       \begin{table}[htb!]
          \small
          \caption{Photometric measurements of Planet \object{$\beta$ 
          Pictoris $b$}.}
          \begin{center}
          \begin{tabular}{c|c|c}
          \hline\hline
             {Filter} & {Apparent Magnitude} & {References}\\ \hline
             Ys & 15.53 $\pm$ 0.34 & \cite{Males2014c}\\
             J & 14.0 $\pm$ 0.3 & \cite{Bonnefoy2013f}\\
             CH$_4$ $_{S,1\%}$ & 13.18 $\pm$ 0.15 & \cite{Males2014c}\\
             H & 13.5 $\pm$ 0.2 & \cite{Bonnefoy2013f}\\
             Ks & 12.6 $\pm$ 0.1 & \cite{Bonnefoy2011}\\
             L\' & 11.02 $\pm$ 0.2 & \cite{Bonnefoy2011, Bonnefoy2013f}\\
             NB 4.05 & 11.20 $\pm$ 0.23 & \cite{Quanz2010b}\\
             M\' & 11.0 $\pm$ 0.3 & \cite{Bonnefoy2013f}\\\hline
          \end{tabular}
          \end{center}
          \label{betaPicObs}
       \end{table}

      \begin{figure}[htb!]
         \includegraphics[trim = 0cm 0cm 8cm 6cm, clip,width=0.5\textwidth]{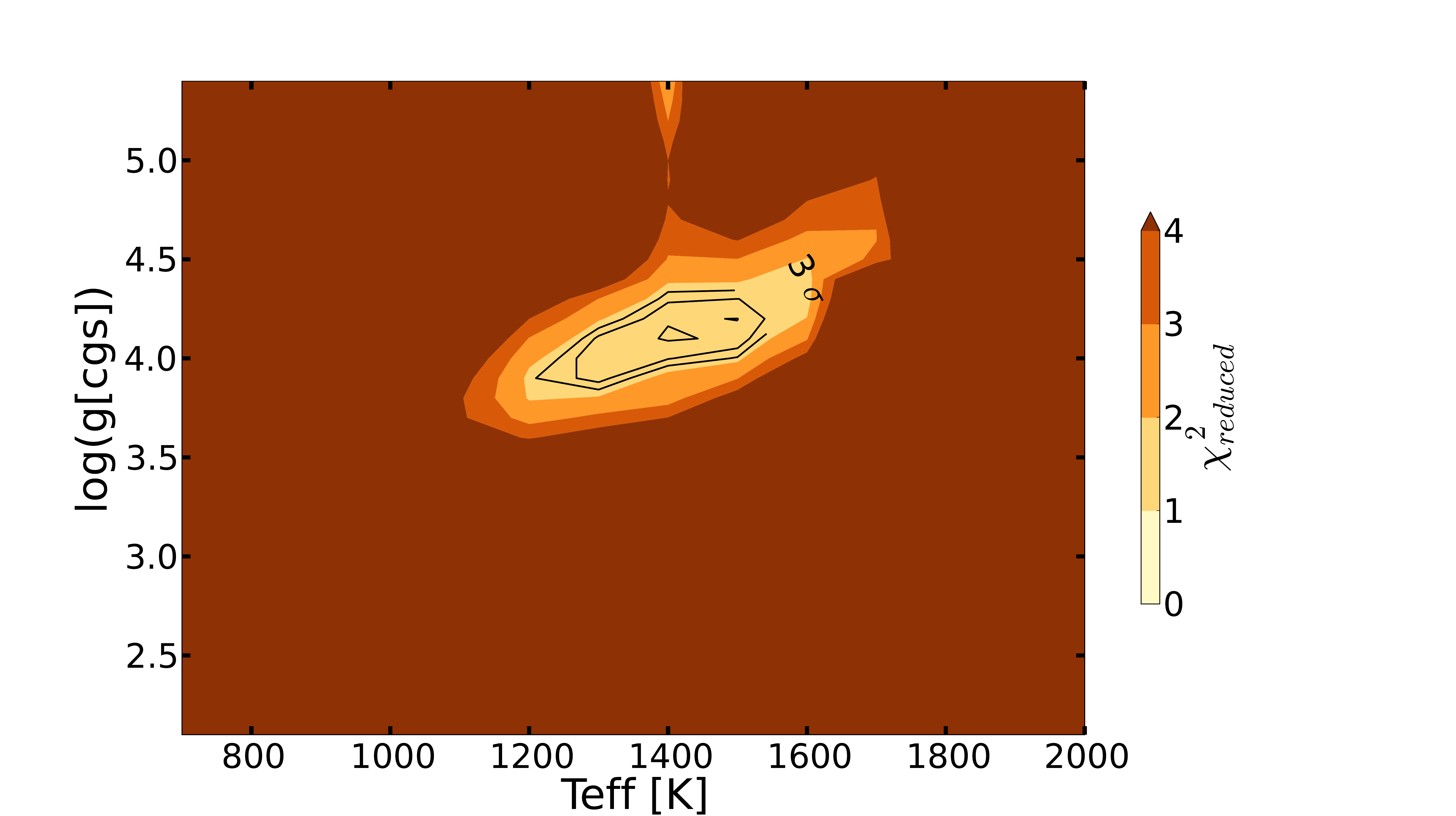}
         \includegraphics[trim = 1cm 0cm 2cm 1cm, clip,width=0.5\textwidth]{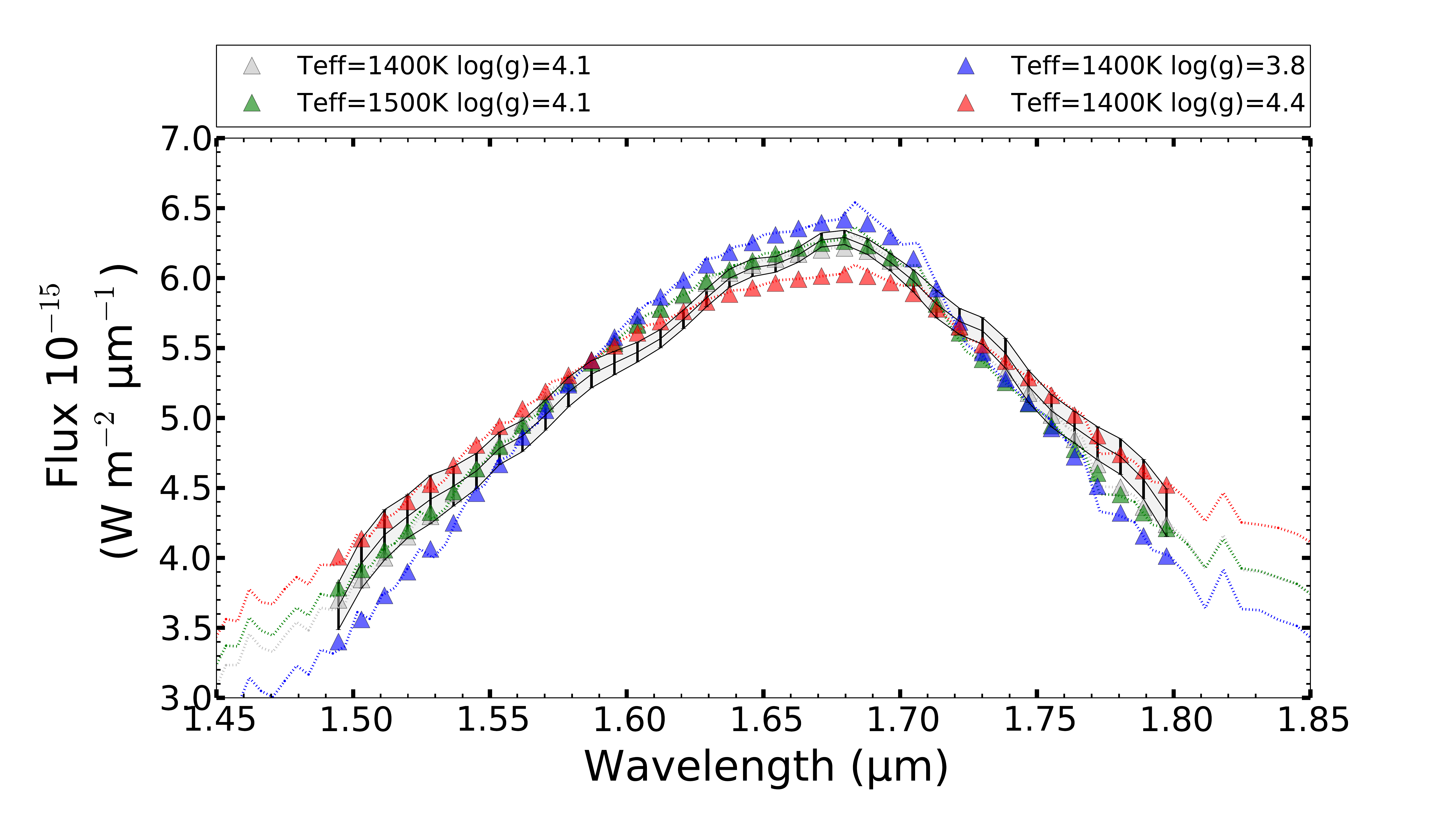}
         \caption{Top: map of reduced $\chi^2$ for \object{$\beta$ Pictoris $b$} H-band spectrum with 
         $\tau$ = 3 and $<r>$ = 30 $\mu$m.
         Bottom: \object{$\beta$ Pictoris $b$} GPI H-band spectrum (black dots) is compared 
         to models with clouds with $\tau_\mathrm{ref}$ = 3. Best fit, in a least-squares 
         sense, is shown as gray triangles ($T_\mathrm{eff}$=1400 K, $\log(g)$ = 4.1). 
         Models in which the gravity is changed by $\pm$ 0.3 dex are plotted for comparison.
         A model with a larger effective temperature (1500 K) and $\log(g)$ = 4.1, which fits
         the spectrum at the 1.2-$\sigma$ confidence level, is also shown. The dotted lines
          represent the calculated spectra prior to 
         convolution.}
         \label{betpicHband} 
      \end{figure}       
       
       For illustration, we now apply Exo-REM to the case of the planet 
       \object{$\beta$ Pictoris $b$} located at 19.44$\pm$0.05 pc \citep{2007h}. Discovered 
       back in 2008 \citep{Lagrange2009}, this object is a case study. As it orbits 
       relatively close to its young \citep[21$\pm$4 Myr][]{Binks2014a} parent star, a 
       precise followup enables a careful determination of the semimajor axis, which is 
       8.9$^{+0.4}_{-0.6}$ AU \citep{Bonnefoy2014, Lagrange2013d, Lagrange2014a}. The 
       planet resides inside the circumstellar disk detected in 1987 \citep{Smith1987ed}. 
       Importantly, \cite{Lagrange2012a} demonstrated that its orbital plane is in fact 
       aligned with the warp observed by \cite{Mouillet1997b} instead of the main disk 
       plane, providing an unambiguous evidence for the disk/planet interaction. A 
       photometric event reported in 1981 could have been produced by the transit of this 
       planet in front of the star \citep{LecavelierDesEtangs1997b}.

       We considered the whole set of available photometric measurements covering 
       the near-IR wavelengths, all the way to the mid-IR. Observations (Table 
       \ref{betaPicObs}) were collected with NaCo \citep{Lenzen2003, Rousset2003} at the VLT 
       in the J, H, K$_S$, L', NB\_4.05, M'  bands \citep{Bonnefoy2013f, Bonnefoy2011, 
       Quanz2010b} and with MagAO \citep{Close2012g} in the Ys and CH$_{4s}$ bands as well 
       \citep{Males2014c}. Recently, J-band (between 1.12 and 1.35 $\mu$m for a 
       resolution of 35-39) and H-band (between 1.51 and 1.79  $\mu$m for a 
       resolution of 44-49) spectra  were obtained during the GPI \citep{Macintosh2014c} 
       commissioning \citep{Bonnefoy2014, Chilcote2015}.
     
       First, the grids of models were generated as explained hereabove without any 
       constraint on radius and mass. We started with the analysis of the photometric data 
       points alone. The models with no cloud and thin clouds (Fig.~\ref{betpicSEDxi2maps}, 
       top left and right panels) do not allow us to achieve a decent minimization, the 
       regions of minima being located at the boundaries of the parameter space. If thicker 
       clouds are introduced (Fig.~\ref{betpicSEDxi2maps} bottom), the model is able to 
       reproduce the data points (reduced $\chi^2$ is lower) and the region limited by the 
       1-$\sigma$ contour falls within the grid boundaries. We can constrain the effective 
       temperature to 1500-1700K, while only a lower limit is derived for gravity 
       ($\log(g)$ > 4). Calculations with $<r>$ = 3 $\mu$m do not yield a minimum $\chi^2$ 
       value as low as in the case of $<r>$ = 30 $\mu$m and do not provide acceptable 
       solutions at the 1-$\sigma$ uncertainty level (Fig.~\ref{betpicSEDxi2maps} bottom 
       right).

      \begin{table*}[htb!]
      \caption{Derived $T_\mathrm{eff}$ and $\log (g)$ of \object{$\beta$ Pictoris $b$} in 
      each step of this analysis and shown by other studies. }
      \tiny
      \begin{center}
         \begin{tabular}{c|c|c|c|c|c|c|c}
         \hline\hline
            {Reference} & {Data type} & {Constraints} & {Error size} & {$T_\mathrm{eff}$ [K]} & {$\log (g)$} & Radius [$R_\mathrm{Jup}$] & {Best-fit model}\\\hline
            {This work} &{SED} & {no} & {1 $\sigma$} & {1600 $\pm$ 100} & {4.0 $\pm$ 0.8} & {1.68 $\pm$ 0.22} & {clouds, 30 $\mu$m}\\
            {This work} &{SED} & {no} & {2 $\sigma$} & {1550 $\pm$ 250} & {>3.2} & {1.82 $\pm$ 0.44} & {clouds}\\\hline
            {This work} &{H-band} & {no} & {2 $\sigma$} & {1400 $\pm$ 100} & {4.1 $\pm$ 0.3} & {2.2 $\pm$ 0.4} & {clouds, $\tau_\mathrm{ref}$ = 3}\\\hline
            {This work} &{SED} & {radius and mass} & {1 $\sigma$} & {1550 $\pm$ 50} & {3.8 $\pm$ 0.6} &{1.73 $\pm$ 0.12}& {clouds, 30 $\mu$m}\\
            {This work} &{SED} & {radius and mass} & {2 $\sigma$} & {1550 $\pm$ 150} & {3.5 $\pm$ 1.0} &{1.76 $\pm$ 0.24}& {clouds}\\\hline
            {\cite{Bonnefoy2013f}} &{SED} & {no} & {best $\chi^2$} & {1700 $\pm$ 100} & {4.0 $\pm$ 0.5 } & {1.22-1.76} & {clouds, Drift Phoenix}\\\hline
            {\cite{Currie2013}} &{SED} & {no} & {1 $\sigma$} & {1575-1650} & {3.8 $\pm$ 0.2} & {1.65 $\pm$ 0.06} & {clouds, Modified AMES-Dusty}\\\hline
         \end{tabular}
       \end{center}
         \label{ParamDeriv}
      \end{table*}
           
       The same work was carried out using the J-band spectrum. In that case, the reduced 
       $\chi^2$ values are much lower across the grid as a result of larger flux uncertainties 
       (more models can fit the data)
       (Fig.~\ref{betpicspecxi2maps}). For models with clouds, the best $\chi^2$ correspond
       to an effective temperature of 1500 $\pm$ 100 K  but large gravities ($\log (g)$ > 3.5). In the
       $\tau$ = 3 grid of models, the minimum reduced $\chi^2$ is as low as 0.14, which suggests that
       the error bars are overestimated or, more likely, strongly correlated in wavelength. 
       With these large error bars, cloud-free models are also capable of fitting the spectrum
       at the 1-$\sigma$ level with $T_\mathrm{eff}$ = 2000 K, $\log (g)$ = 2.3, corresponding to a minimum reduced
       $\chi^2$ of 0.63. \\

       Assuming an age for the $\beta$ Pictoris system of 15-25 Myr, evolutionary models 
       predict a radius between 0.6 and 2 $R_\mathrm{Jup}$. Including this constraint in the 
       $\chi^2$ minimization implies a lower limit for the effective temperature of about 
       1400 K.           

       Radial velocity measurements presented in \cite{Lagrange2012d} yield constraints 
       on the planet mass. For separations of 8, 9, 10, 11, and 12\,AU, the detection limit
       corresponds to a mass of 10, 12, 15.5, 20, 25 $M_\mathrm{Jup}$, while the 
       model-dependent mass derived from photometry compared to evolutionary models
       is $\geq$ 6 $M_\mathrm{Jup}$ \citep{Bonnefoy2013f}. Recently, \cite{Bonnefoy2014} used  
       an up-to-date compilation of radial velocity measurements and constrained the mass 
       limit to 20 $M_\mathrm{Jup}$ (at 96\% confidence level). For the purpose of being 
       conservative, we retain a maximum mass of 25 $M_\mathrm{Jup}$. These new constraints 
       remove the models with highest gravities in the $\chi^2$ maps.

       The mass and radius constraints allow us to  more accurately determine the physical 
       parameters $T_\mathrm{eff}$ and $\log(g),$ as presented in Table~\ref{ParamDeriv}. 
       Therefore, we propose a new determination of  $T_\mathrm{eff} = 1550\pm150$ 
       K and $\log(g)=3.5\pm1.0$ at the 2-$\sigma$ confidence level.

       These values are in good agreement with the former analyses by \cite{Bonnefoy2013f} 
       and \cite{Currie2013}, who used the PHOENIX models like BT-Settl, Drift-PHOENIX or
       Ames-Dusty. 

       The AMES-Cond and AMES-Dusty models represent extreme cases. Both solve for 
       thermochemical equilibrium assuming level-by-level element conservation. In the 
       AMES-Cond model, all the dust is removed for opacity calculation while in the AMES-Dusty 
       model the amount of dust is that derived from thermochemical equilibrium with no 
       depletion process.
       In BT-Settl, the dust particle properties (number density and mean radius) are 
       derived from a comparison of timescales of various microphysical and transport 
       processes. Finally, Drift-PHOENIX really solves for the formation and evolution of 
       cloud particles, taking dust microphysics and atmospheric convection into account. \cite{Bonnefoy2014} provided a detailed comparison with these and other models
       as concerns the J-band spectrum.

        Therefore, Exo-REM yields similar results as
       other models, either using photometric data or low-resolution spectra, although it is far
       less complex. A more recent study by  \cite{Males2014c}, using bolometric luminosity 
       compared to evolutionary models, found an effective temperature of  $T_\mathrm{eff}$ 
       = 1643 $\pm$ 32 K. In our work, error bars are larger probably because we consider 
       2-$\sigma$ confidence level and possibly because  we probe a larger parameter space. 
       The models with the best $\chi^2$ are shown in Figs.~\ref{betpicBestPlot} and 
       \ref{betpicBestSpectrumPlot}. We confirm that we need clouds to reproduce the 
       contrast between all photometric points. In the case with clouds we observe two bad 
       fitting locations, in H band and NB4.05. As concerns the spectrum we have  difficulty 
       fitting the observations after 1.33 $\mu$m.

       Recently, \cite{Chilcote2015} published a spectrum of \object{$\beta$ Pictoris $b$}
       in the H band (Fig.~\ref{betpicHband}). We find that only models with thick clouds 
       ($\tau_\mathrm{ref}$ = 3) can fit the data at a 3-$\sigma$ confidence level.
       Note that published error bars of the spectrum
       only account for random errors (at the 1-2\% level) and do not incorporate 
       systematic uncertainties, such as an estimated 10\% uncertainty on the overall 
       spectral slope. Our best fit, in a least-squares sense, is obtained for 
       $T_\mathrm{eff}$ = 1400 K, $\log(g)$ = 4.1 and $\tau_\mathrm{ref}$ = 3 
       (Fig.~\ref{betpicHband}) with a reduced $\chi^2_\mathrm{red}$ of 1.12. However, 
       this case would imply a too large radius of 2.1 $R_\mathrm{Jup}$ to reproduce 
       the absolute flux, which is based on the photometric measurement of \cite{Males2014c}.
       We also show a case with $T_\mathrm{eff}$= 1500 K, $\log(g)$ = 4.1 and 
       $\tau_\mathrm{ref}$ = 3 yielding $\chi^2_\mathrm{red}$ = 1.18, which agrees 
       within error bars with the parameters derived from photometric measurements 
       (Table~\ref{ParamDeriv}) with mass and radius constraints and from the GPI J-spectrum 
       (Fig. \ref{betpicBestSpectrumPlot}).  
       The agreement of both synthetic spectra with the GPI H-spectrum is very good. 
       In Table~\ref{ParamDeriv}, we give the model parameters that fit the spectrum 
       at the 2-$\sigma$ level and not for the 1-$\sigma$ level because, as mentioned above, 
       the error bars of Chilcote et al. may miss some systematic uncertainty. Adding a 10\% 
       uncertainty on the overall spectral shape would significantly increase the error bars
       on the derived parameters. The fits we obtain are closer to the observations than models
       presented in \cite{Chilcote2015}. The differences between the models possibly result from
       different modeling of the cloud opacity. To fit the shape of the observed spectrum, 
       models with thick clouds are needed. Models with no or thin clouds produce too much 
       contrast between the peak and both ends of the observed spectrum. 
       In Fig.~\ref{betpicHband}, we also show cases where $\log(g)$ is varied by 
       0.3 dex around the previous case to illustrate the strong sensitivity of the
       shape of the spectrum to this parameter (as gravity increases the spectrum broadens). 
       Spectral observations in this band thus 
       provide a way to constrain the gravity of the planet, although one would need to 
       investigate to what extent it can be disentangled from cloud opacity and metallicity.

\section{SPHERE expected observations}

       In this section we discuss the ability of SPHERE to put useful constraints on gravity 
       and effective temperature according to the quality of the data. The purpose is to 
       link the photometric errors to the uncertainties in the physical parameters of 
       planetary atmospheres. A related analysis to derive $\log(g)$ and  $T_\mathrm{eff}$ 
       was performed by \cite{Vigan2010c} with the narrowband differential filters of  
       SPHERE combined with AMES-Cond/Dusty \citep{Allard2001a,Allard2003j}, BT-Settl 
       \citep{Allard2007p} and Burrows models \citep{Burrows2006w}  .
       
       In the following, we consider 12 test cases for which the model has a robust 
       convergence, and which cover a representative range of $\log(g)$ = 2.5, 3.5, 4.5,  
       $T_\mathrm{eff}$ = 800, 1100, 1400, 1700 K, and cloud properties ($\tau_\mathrm{ref}$ 
       = 1 for 30 $\mu$m particles). For this preliminary analysis, we focus on the near- IR 
       broadband filters Y, J, H, Ks, offered in IRDIS, the SPHERE camera, as well as the 
       Y-H mode of IFS (39 wavelengths, with 0.014-0.020 $\mu$m between adjacent pixels), 
       the near-IR spectrograph (Table~\ref{sphereFilters}).

\begin{figure}[htb!]
       \centering
       \includegraphics[width=0.36\textwidth]{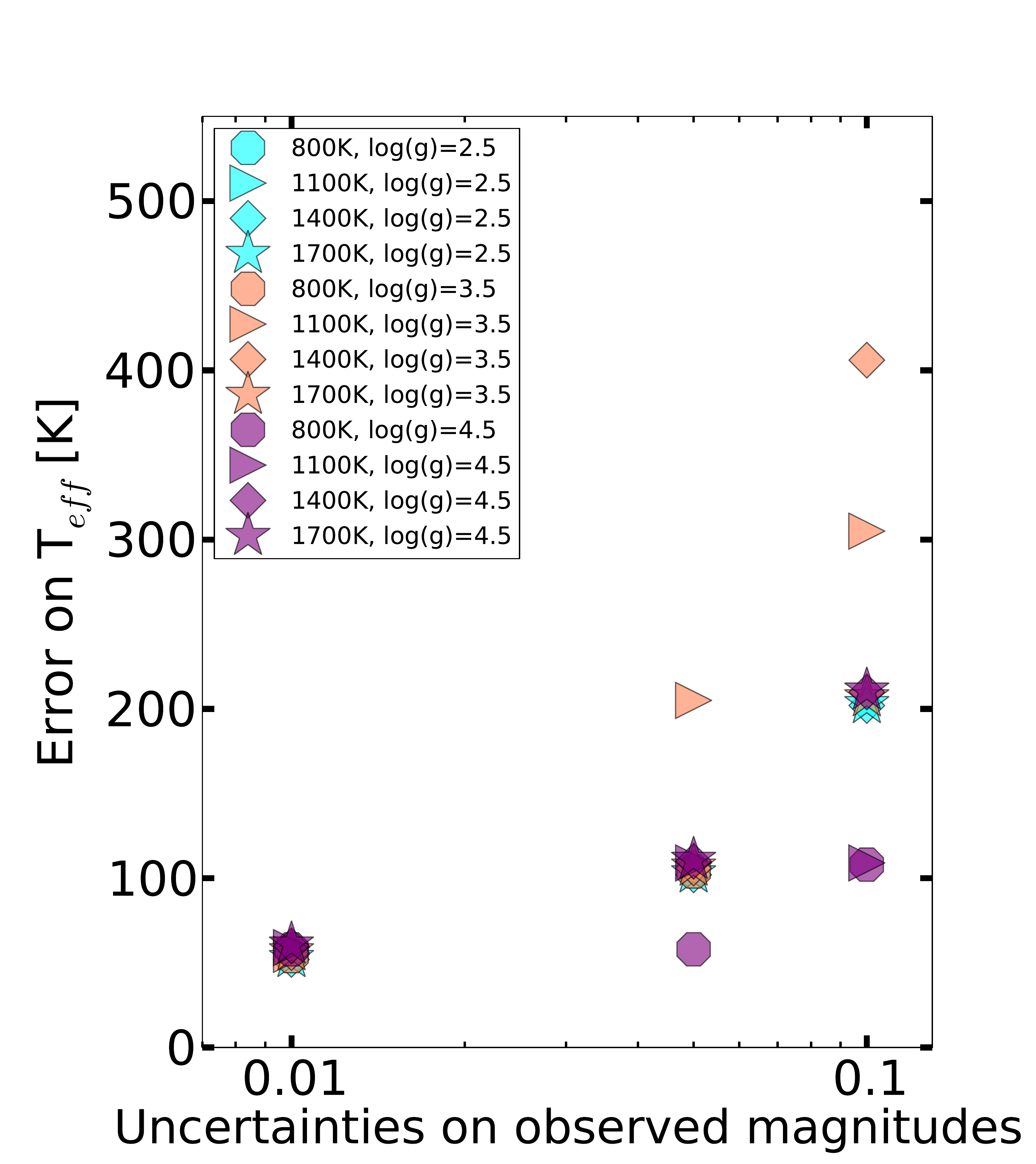}
       \includegraphics[width=0.36\textwidth]{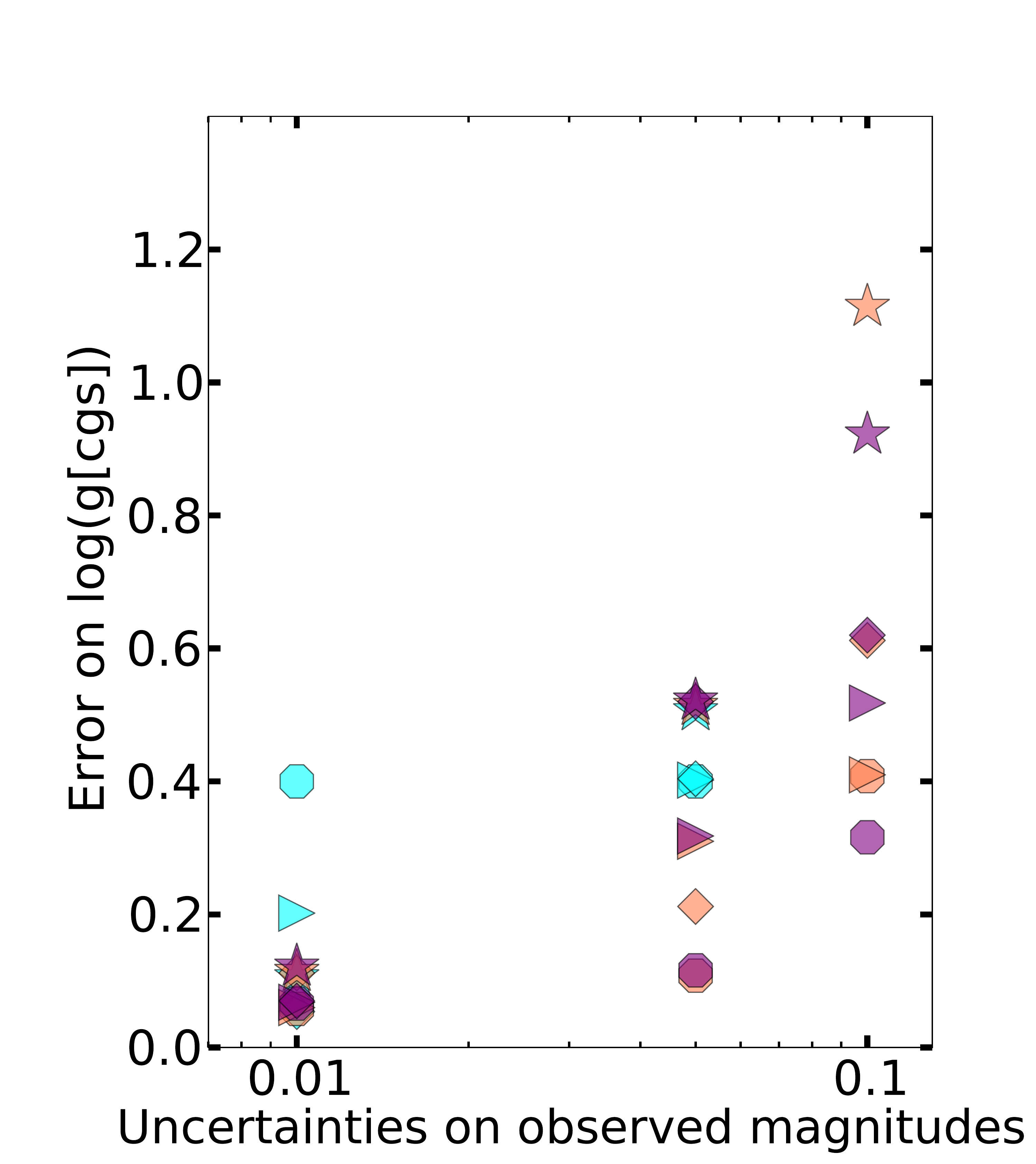}
       \caption{Effect of uncertainties in the magnitude of photometric data points 
       upon uncertainties in derived $T_\mathrm{eff}$ (top) and $\log(g)$ (bottom). Cases 
       with 2-$\sigma$ error bars exceeding our test grid are not plotted.}
       \label{sedmag}
      \end{figure}

       The spectra of test cases were integrated over IRDIS filters and a photometric error 
       was added to the integrated flux to mimic an actual photometric measurement. We 
       considered several error amplitudes (in magnitude), $\Delta$m = 0.01, 0.05, 0.1, 0.5, 
       and 1.0, corresponding to very good to very poor data. The same error amplitude was 
       applied to all filters, and we did not consider data with various qualities. 
       In fact, the flux error in the SPHERE measurements is expected 
       to show some correlation between wavelengths that affect the spectral 
       shape to some extent. We did not take these systematic errors  into 
       account in the present analysis. As in the previous section, the photometry 
       of test cases was compared to the grid of 
       models using the $\chi^2$ minimization (no mass or radius constraint).
       The uncertainties in physical parameters, $\Delta T_\mathrm{eff}$ and $\Delta 
       \log(g)$, were derived from the 2-$\sigma$ contours. Results are shown in 
       Fig.~\ref{sedmag}. We observe that $\Delta T_\mathrm{eff}$ decreases as $\log(g)$ 
       increases and conversely, $\Delta \log(g)$ increases as  $T_\mathrm{eff}$ increases.

       When photometric errors are small, on the order of 0.01 mag, the errors are often smaller or 
       similar to the step of the grid. At the other extreme, when the photometric error is 
       as large as 1 mag, all models contained in the grid match the observation, hence the 
       errors on the physical parameters exceed the range of the grid. We conclude that the 
       effective temperature and  gravity can be constrained to 200 K and 0.5 dex, respectively, if an accuracy of 0.2 mag is achieved.

\begin{figure}[htb!]
       \centering
       \includegraphics[width=0.36\textwidth]{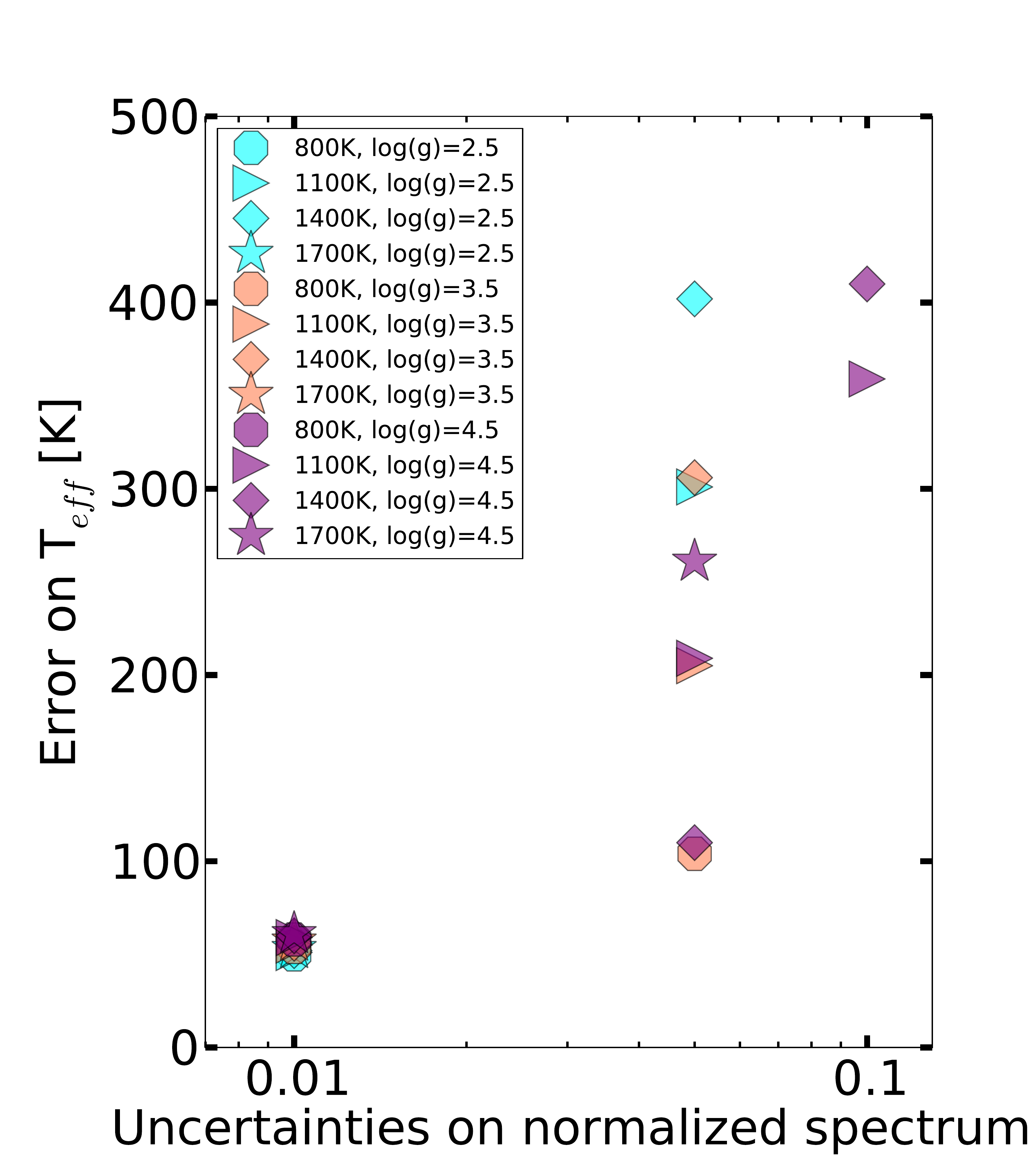}
       \includegraphics[width=0.36\textwidth]{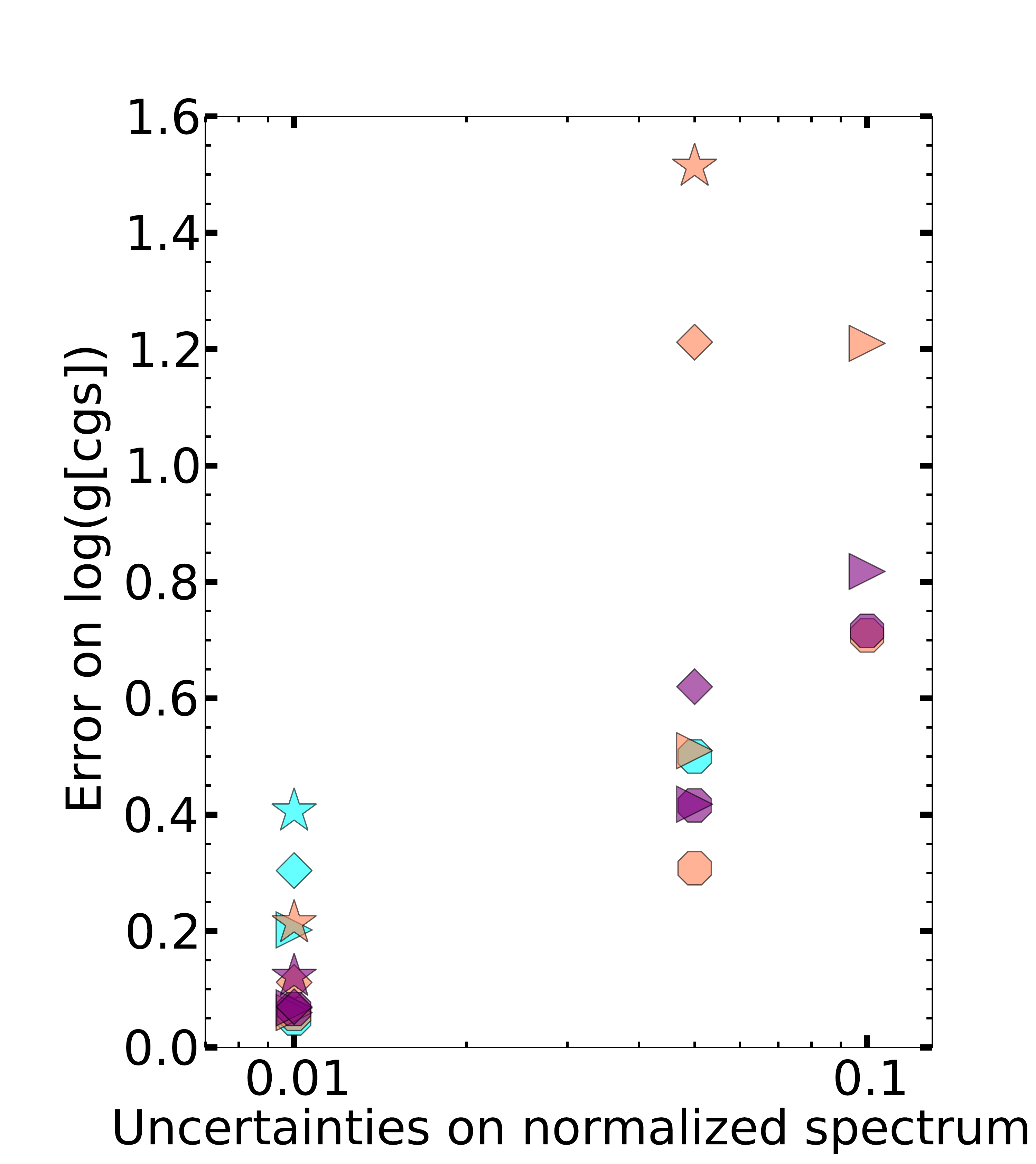}
       \caption{Effect of uncertainties in the normalized spectra upon uncertainties in 
       derived $T_\mathrm{eff}$ (top) and $\log(g)$ (bottom). Cases with 2-$\sigma$ error 
       bars exceeding our test grid are not plotted.}
       \label{specmag}
      \end{figure}

       We also considered flux spectra, normalized to unity at the peak, with an error of 
       0.01, 0.05, or 0.1, constant for all wavelengths. The same exercise performed with 
       our set of synthetic spectra (Fig.~\ref{specmag}) indicates accuracies of 200 K 
       for $T_\mathrm{eff}$ and 0.5 dex for $g$, assuming a precision of 0.1.
        
\begin{figure}[htb!]
       \centering
       \includegraphics[width=0.36\textwidth]{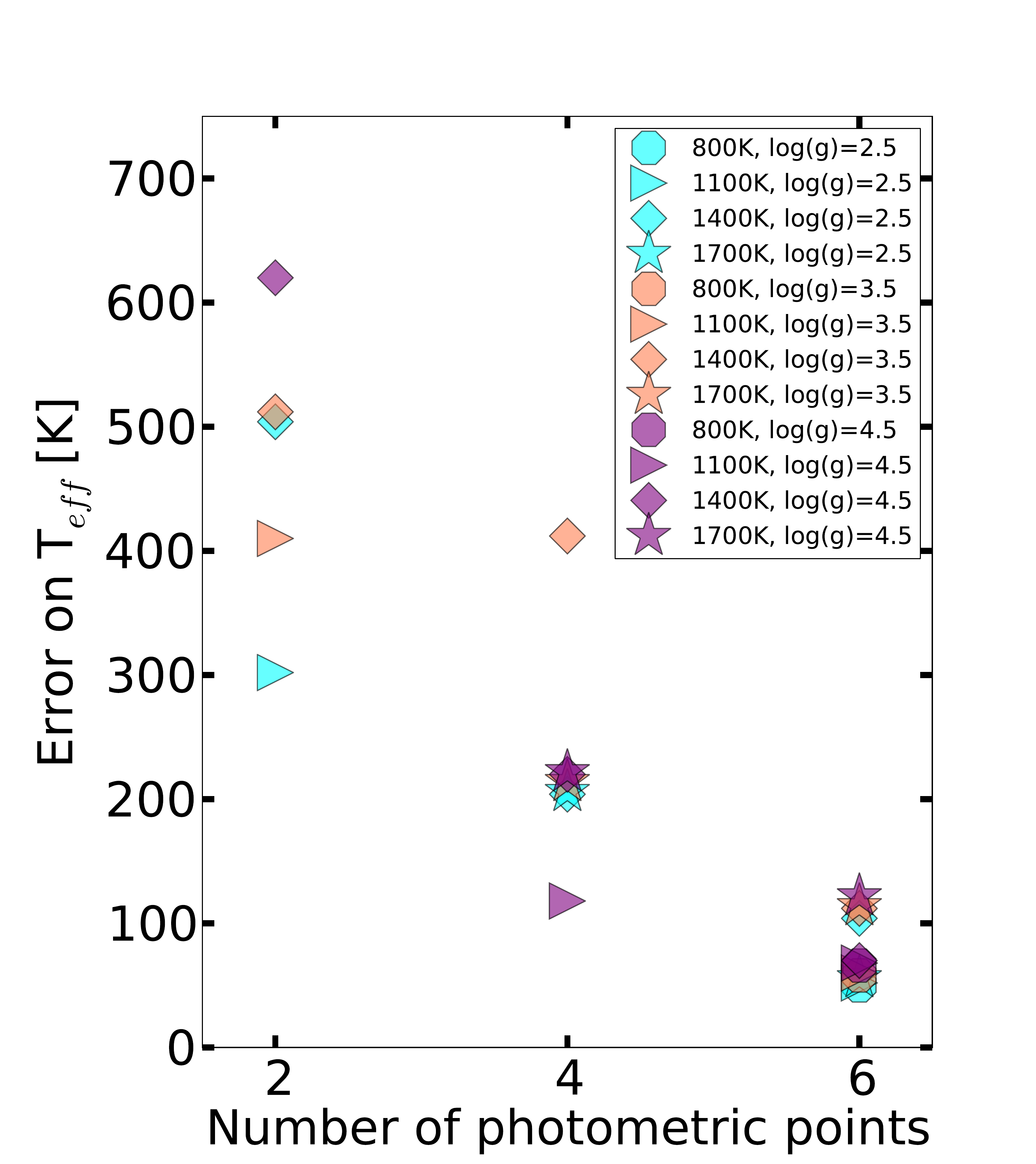}
       \includegraphics[width=0.36\textwidth]{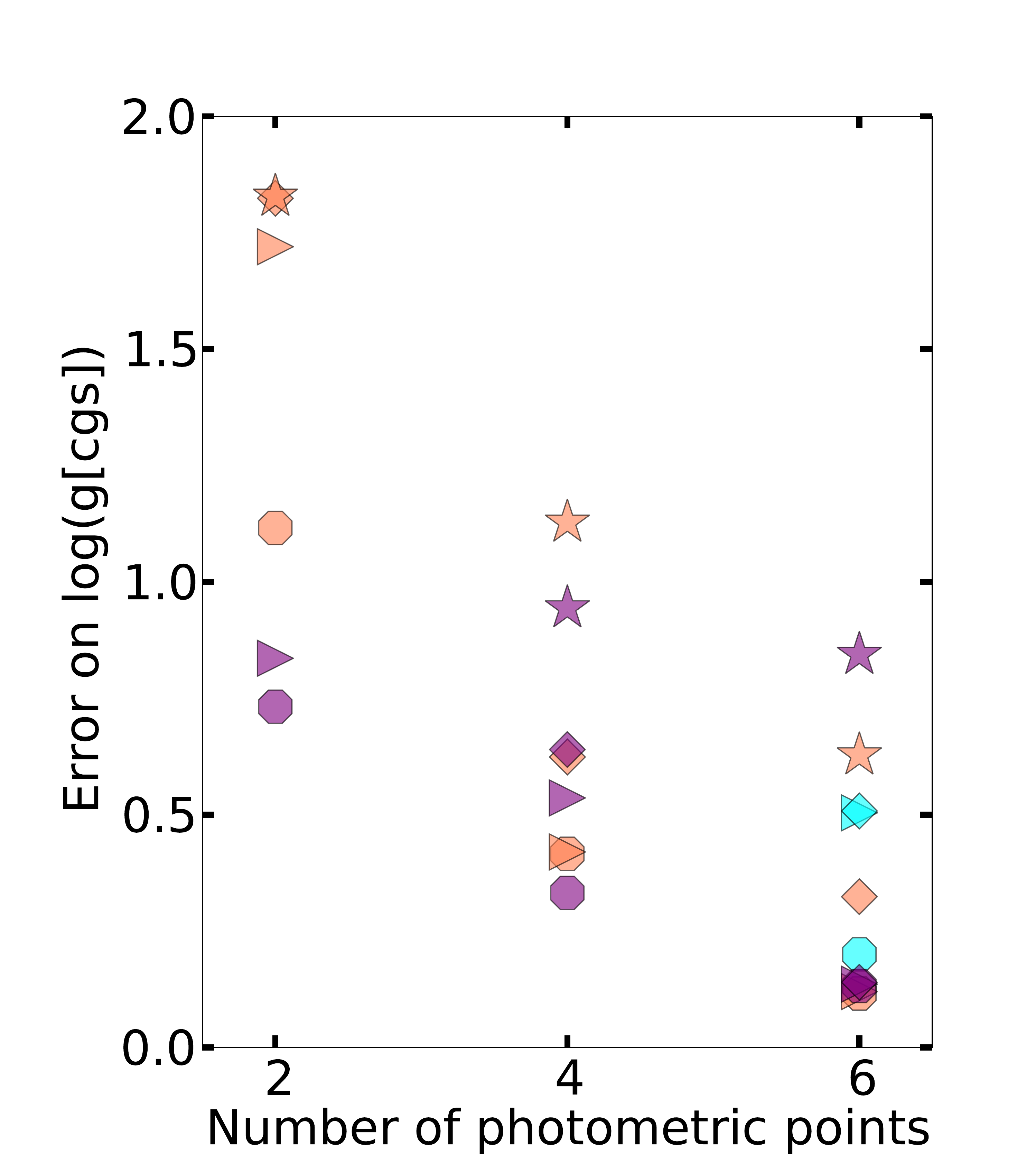}
       \caption{Effect of number of points in the SED upon uncertainties in derived 
       $T_\mathrm{eff}$ (top) and $\log(g)$ (bottom). The two-point case corresponds to the J 
       and Ks filters. Cases with 2-$\sigma$ error bars exceeding our test grid are not 
       plotted.}
       \label{nb}
      \end{figure}

       The number of available photometric data points, as well as the covered spectral 
       range, also have an impact on the accuracy of the retrieved physical parameters. For 
       instance, considering the combination of two broadband filters, the set H+Ks is more 
       appropriate to constrain the effective temperature (Fig.~\ref{nb} a) while for 
       gravity a large spectral range is preferable (such as J+Ks; Fig.~\ref{nb} b). We now 
       consider three possible sets of photometric data points: two SPHERE filters (H and 
       Ks, or J and Ks), four SPHERE filters (Y, J, H, Ks), and finally the same four SPHERE 
       filters with  two NaCo filters (L', M'). We assume an accuracy of 
       0.1 mag on SPHERE data. To achieve the same accuracy in 
       the physical parameters as previously achieved  (200 K for $T_\mathrm{eff}$ and 0.5 dex for $g$), 
       we conclude that at least three data points are required. In addition, a significant 
       improvement is achieved if the SPHERE photometry is complemented with the NaCo MIR 
       filters. With L' and M' filters we can expect uncertainties in $T_\mathrm{eff}$ lower 
       than 100 K and also a smaller error on $\log(g)$.
      
      \begin{table}[htb!]
            \caption{Characteristics of SPHERE IRDIS filters and IFS spectroscopic mode}
            \begin{center}
               \begin{tabular}{c|c|c}
               \hline\hline
                  {Name} & Central wavelength [$\mu$m] & FWHM$^\star$ [$\mu$m]\\\hline
                  {BB Y} & 1.0425 & 0.139 \\
                  {BB J} & 1.2575 & 0.197 \\
                  {BB H} & 1.6255 & 0.291 \\
                  {BB Ks} & 2.1813 & 0.3135 \\\hline
                  {Y-H} & 0.957-1.636 & R $\sim$ 30\\\hline
            \end{tabular}\\
            \end{center}  
            $\star$: Full width at half maximum
            \label{sphereFilters} 
      \end{table}      

\section{Conclusions}
   To analyze photometric and spectroscopic data from new instruments like SPHERE at 
   the VLT we have developed Exo-REM, a radiative-convective equilibrium model to simulate 
   the atmosphere of young Jupiters, which are privileged targets for direct imaging of 
   exoplanets. The model incorporates opacity from the molecules and atoms that are relevant 
   to observable levels for giant exoplanets having $T_\mathrm{eff}$ < 2000 K. It assumes 
   that vertical profiles of these species are governed by thermochemical equilibrium. Cloud 
   absorption by iron and silicate clouds is included through a simplified formalism and a 
   limited number of free parameters.\\
   
   We used Exo-REM to analyze data available for \object{$\beta$ Pictoris $b$} and to derive 
   physical parameters of the planet. We inferred an effective temperature $T_\mathrm{eff}$ 
   = 1550 $\pm$ 150 K, $\log(g)$ = 3.5 $\pm$ 1, and a radius $R$ = 1.76 $\pm$ 0.24 
   $R_\mathrm{Jup}$ (2-$\sigma$ error bars) from photometric measurements and considering 
   independent constraints on mass and radius. These results are similar to those previously 
   derived by other authors using different atmospheric models.
   The difference is that we considered 2-$\sigma$ error bars (rather than 1-$\sigma$) and 
   explored a wider range of parameters, in particular, with lower values of $g$, than in 
   previous studies. Our 2-$\sigma$ uncertainties include measurement error as well as model 
   dependence on our limited set of cloud parameters (optical depth, particle radius). 
   
   We were also able to reproduce the H-spectrum of Chilcote et al. (2015)  within their (small)
   error bars in contrast to other models displayed in that paper. Using this spectrum alone,
   the derived parameters are: $T_\mathrm{eff}$ = 1400 $\pm$ 100 K, $\log(g)$ = 4.1 $\pm$ 0.3, and 
   R = 1.9 $\pm$ 0.2 $R_\mathrm{Jup}$ (2-$\sigma$ error bars). \\
   
   We investigated the ability of SPHERE to characterize 
   exoplanets with the IRDIS broadband filters and the Y-H spectroscopic mode of IFS. A 
   couple of filters (H, Ks) appear best suited to constrain $T_\mathrm{eff}$, while the 
   couple (J, Ks) is more appropriate to constrain $\log(g)$. Combining MIR NaCo L' and M' 
   observations with SPHERE photometry enables us to obtain good constraints on both 
   $T_\mathrm{eff}$ and $\log(g)$.\\

   We plan to explore the set of free parameters of Exo-REM more systematically than shown 
   in this paper. In particular, we will  more extensively study the effect of metallicity 
   and of cloud parameters (scale height, reference optical depth, particle size). In future 
   works, we may consider constraints coming from ab-initio models like BT-Settl or 
   Drift-Phoenix, which provide guidelines for a range of realistic and physical cloud 
   parameters. We will also add absorption by water ice particles that are expected to form 
   in giant exoplanets having lower effective temperatures than studied here.
   
   SPHERE was commissioned successfully and the instrument is open to the community. Known 
   planets are prime targets for a thorough characterization (Lagrange et al. in prep.; Zurlo 
   et al in prep.).
   
\begin{acknowledgements}
      JLB PhD is funded by the LabEx "Exploration Spatiale des Environnements Plan\'etaires" 
      (ESEP) \# 2011-LABX-030.
      We thank G.-D. Marleau for useful discussion on exoplanet radii. We thank B. 
      Plez for his help with TiO and VO absorption. We thank J. Chilcote for providing a 
      file of the H-band spectrum of \object{$\beta$ Pictoris $b$}. 
      We thank the anonymous referee for useful remarks and suggestions.
\end{acknowledgements}
\bibliographystyle{aa}
\bibliography{biblio}{}

\begin{thebibliography}{93}
\expandafter\ifx\csname natexlab\endcsname\relax\def\natexlab#1{#1}\fi

\bibitem[{{Ackerman} \& {Marley}(2001)}]{Ackerman2001g}
{Ackerman}, A.~S. \& {Marley}, M.~S. 2001, \apj, 556, 872

\bibitem[{{Albert} {et~al.}(2009){Albert}, {Bauerecker}, {Boudon}, {Brown},
  {Champion}, {Lo{\"e}te}, {Nikitin}, \& {Quack}}]{Albert2009c}
{Albert}, S., {Bauerecker}, S., {Boudon}, V., {et~al.} 2009, Chemical Physics,
  356, 131

\bibitem[{{Allard} {et~al.}(2007){Allard}, {Allard}, {Homeier}, {Kielkopf},
  {McCaughrean}, \& {Spiegelman}}]{Allard2007p}
{Allard}, F., {Allard}, N.~F., {Homeier}, D., {et~al.} 2007, \aap, 474, L21

\bibitem[{{Allard} {et~al.}(2003){Allard}, {Guillot}, {Ludwig}, {Hauschildt},
  {Schweitzer}, {Alexander}, \& {Ferguson}}]{Allard2003j}
{Allard}, F., {Guillot}, T., {Ludwig}, H.-G., {et~al.} 2003, in IAU Symposium,
  Vol. 211, Brown Dwarfs, ed. E.~{Mart{\'{\i}}n}, 325

\bibitem[{{Allard} {et~al.}(2001){Allard}, {Hauschildt}, {Alexander},
  {Tamanai}, \& {Schweitzer}}]{Allard2001a}
{Allard}, F., {Hauschildt}, P.~H., {Alexander}, D.~R., {Tamanai}, A., \&
  {Schweitzer}, A. 2001, \apj, 556, 357

\bibitem[{{Barman} {et~al.}(2015){Barman}, {Konopacky}, {Macintosh}, \&
  {Marois}}]{Barman2015e}
{Barman}, T.~S., {Konopacky}, Q.~M., {Macintosh}, B., \& {Marois}, C. 2015,
  \apj, 804, 61

\bibitem[{{Barman} {et~al.}(2011){Barman}, {Macintosh}, {Konopacky}, \&
  {Marois}}]{Barman2011c}
{Barman}, T.~S., {Macintosh}, B., {Konopacky}, Q.~M., \& {Marois}, C. 2011,
  \apj, 733, 65

\bibitem[{{Baudino} {et~al.}(2014{\natexlab{a}}){Baudino}, {B{\'e}zard},
  {Boccaletti}, {Bonnefoy}, \& {Lagrange}}]{Baudino2014}
{Baudino}, J.-L., {B{\'e}zard}, B., {Boccaletti}, A., {Bonnefoy}, M., \&
  {Lagrange}, A.-M. 2014{\natexlab{a}}, in Proc. of the IAU, IAUS, Vol. 299,
  Exploring the Formation and Evolution of Planetary Systems, ed. M.~{Booth},
  B.~C. {Matthews}, \& J.~R. {Graham}, 277

\bibitem[{{Baudino} {et~al.}(2013){Baudino}, {B{\'e}zard}, {Boccaletti},
  {Lagrange}, \& {Bonnefoy}}]{Baudino2013}
{Baudino}, J.-L., {B{\'e}zard}, B., {Boccaletti}, A., {Lagrange}, A., \&
  {Bonnefoy}, M. 2013, in \baas, Vol.~45, 209.09

\bibitem[{{Baudino} {et~al.}(2014{\natexlab{b}}){Baudino}, {B{\'e}zard},
  {Boccaletti}, {Lagrange}, {Bonnefoy}, \& {Galicher}}]{Baudino2014c}
{Baudino}, J.-L., {B{\'e}zard}, B., {Boccaletti}, A., {et~al.}
  2014{\natexlab{b}}, in SF2A Proceedings, ed. J.~{Ballet}, F.~{Martins},
  F.~{Bournaud}, R.~{Monier}, \& C.~{Reyl{\'e}}, 53

\bibitem[{{Beuzit} {et~al.}(2008){Beuzit}, {Feldt}, {Dohlen}, {Mouillet},
  {Puget}, {Wildi}, {Abe}, {Antichi}, {Baruffolo}, {Baudoz}, {Boccaletti},
  {Carbillet}, {Charton}, {Claudi}, {Downing}, {Fabron}, {Feautrier},
  {Fedrigo}, {Fusco}, {Gach}, {Gratton}, {Henning}, {Hubin}, {Joos}, {Kasper},
  {Langlois}, {Lenzen}, {Moutou}, {Pavlov}, {Petit}, {Pragt}, {Rabou}, {Rigal},
  {Roelfsema}, {Rousset}, {Saisse}, {Schmid}, {Stadler}, {Thalmann}, {Turatto},
  {Udry}, {Vakili}, \& {Waters}}]{Beuzit2008}
{Beuzit}, J.-L., {Feldt}, M., {Dohlen}, K., {et~al.} 2008, in \procspie, Vol.
  7014, 18

\bibitem[{{Bevington} \& {Robinson}(2003)}]{Bevington2003}
{Bevington}, P.~R. \& {Robinson}, D.~K. 2003, {Data reduction and error
  analysis for the physical sciences} ({McGraw-Hill})

\bibitem[{{Binks} \& {Jeffries}(2014)}]{Binks2014a}
{Binks}, A.~S. \& {Jeffries}, R.~D. 2014, \mnras, 438, L11

\bibitem[{{Boccaletti} {et~al.}(2005){Boccaletti}, {Baudoz}, {Baudrand},
  {Reess}, \& {Rouan}}]{Boccaletti2005b}
{Boccaletti}, A., {Baudoz}, P., {Baudrand}, J., {Reess}, J.~M., \& {Rouan}, D.
  2005, Advances in Space Research, 36, 1099

\bibitem[{{Bonnefoy} {et~al.}(2013){Bonnefoy}, {Boccaletti}, {Lagrange},
  {Allard}, {Mordasini}, {Beust}, {Chauvin}, {Girard}, {Homeier}, {Apai},
  {Lacour}, \& {Rouan}}]{Bonnefoy2013f}
{Bonnefoy}, M., {Boccaletti}, A., {Lagrange}, A.-M., {et~al.} 2013, \aap, 555,
  A107

\bibitem[{{Bonnefoy} {et~al.}(2011){Bonnefoy}, {Lagrange}, {Boccaletti},
  {Chauvin}, {Apai}, {Allard}, {Ehrenreich}, {Girard}, {Mouillet}, {Rouan},
  {Gratadour}, \& {Kasper}}]{Bonnefoy2011}
{Bonnefoy}, M., {Lagrange}, A.-M., {Boccaletti}, A., {et~al.} 2011, \aap, 528,
  L15

\bibitem[{{Bonnefoy} {et~al.}(2014){Bonnefoy}, {Marleau}, {Galicher}, {Beust},
  {Lagrange}, {Baudino}, {Chauvin}, {Borgniet}, {Meunier}, {Rameau},
  {Boccaletti}, {Cumming}, {Helling}, {Homeier}, {Allard}, \&
  {Delorme}}]{Bonnefoy2014}
{Bonnefoy}, M., {Marleau}, G.-D., {Galicher}, R., {et~al.} 2014, \aap, 567, L9

\bibitem[{{Borysow } {et~al.}(2001){Borysow }, {Jorgensen}, \&
  {Fu}}]{BorysowU.G.Jorgensen2001}
{Borysow }, U., {Jorgensen}, A., \& {Fu}, Y. 2001, \jqsrt, 68, 235

\bibitem[{{Borysow}(2002)}]{Borysow2002}
{Borysow}, A. 2002, \aap, 390, 779

\bibitem[{{Borysow} \& {Frommhold}(1989)}]{Borysow1989d}
{Borysow}, A. \& {Frommhold}, L. 1989, \apj, 341, 549

\bibitem[{{Borysow} {et~al.}(1989){Borysow}, {Frommhold}, \&
  {Moraldi}}]{Borysow1989a}
{Borysow}, A., {Frommhold}, L., \& {Moraldi}, M. 1989, \apj, 336, 495

\bibitem[{{Borysow} {et~al.}(1988){Borysow}, {Frommhold}, \&
  {Birnbaum}}]{Borysow1988}
{Borysow}, J., {Frommhold}, L., \& {Birnbaum}, G. 1988, \apj, 326, 509

\bibitem[{{Boss}(2001)}]{Boss2001b}
{Boss}, A.~P. 2001, \apjl, 551, L167

\bibitem[{{Boudon} {et~al.}(2006){Boudon}, {Rey}, \& {Lo{\"e}te}}]{Boudon2006a}
{Boudon}, V., {Rey}, M., \& {Lo{\"e}te}, M. 2006, \jqsrt, 98, 394

\bibitem[{{Bowler} {et~al.}(2010){Bowler}, {Liu}, {Dupuy}, \&
  {Cushing}}]{Bowler2010}
{Bowler}, B.~P., {Liu}, M.~C., {Dupuy}, T.~J., \& {Cushing}, M.~C. 2010, \apj,
  723, 850

\bibitem[{{Burrows} {et~al.}(2000){Burrows}, {Marley}, \&
  {Sharp}}]{Burrows2000y}
{Burrows}, A., {Marley}, M.~S., \& {Sharp}, C.~M. 2000, \apj, 531, 438

\bibitem[{{Burrows} {et~al.}(1995){Burrows}, {Saumon}, {Guillot}, {Hubbard}, \&
  {Lunine}}]{Burrows1995h}
{Burrows}, A., {Saumon}, D., {Guillot}, T., {Hubbard}, W.~B., \& {Lunine},
  J.~I. 1995, \nat, 375, 299

\bibitem[{{Burrows} \& {Sharp}(1999)}]{Burrows1999i}
{Burrows}, A. \& {Sharp}, C.~M. 1999, \apj, 512, 843

\bibitem[{{Burrows} {et~al.}(2006){Burrows}, {Sudarsky}, \&
  {Hubeny}}]{Burrows2006w}
{Burrows}, A., {Sudarsky}, D., \& {Hubeny}, I. 2006, \apj, 640, 1063

\bibitem[{{Burrows} \& {Volobuyev}(2003)}]{Burrows2003m}
{Burrows}, A. \& {Volobuyev}, M. 2003, \apj, 583, 985

\bibitem[{{Campargue} {et~al.}(2012){Campargue}, {Wang}, {Mondelain}, {Kassi},
  {B{\'e}zard}, {Lellouch}, {Coustenis}, {Bergh}, {Hirtzig}, \&
  {Drossart}}]{Campargue2012}
{Campargue}, A., {Wang}, L., {Mondelain}, D., {et~al.} 2012, \icarus, 219, 110

\bibitem[{{Chabrier} {et~al.}(2000){Chabrier}, {Baraffe}, {Allard}, \&
  {Hauschildt}}]{Chabrier2000b}
{Chabrier}, G., {Baraffe}, I., {Allard}, F., \& {Hauschildt}, P. 2000, \apj,
  542, 464

\bibitem[{{Chase}(1998)}]{chase1998monograph}
{Chase}, Jr., M.~W. 1998, NIST-JANAF Thermochemical Tables, Fourth Edition,
  Monograph No. 9

\bibitem[{{Chauvin} {et~al.}(2004){Chauvin}, {Lagrange}, {Dumas}, {Zuckerman},
  {Mouillet}, {Song}, {Beuzit}, \& {Lowrance}}]{Chauvin2004b}
{Chauvin}, G., {Lagrange}, A.-M., {Dumas}, C., {et~al.} 2004, \aap, 425, L29

\bibitem[{{Chilcote} {et~al.}(2015){Chilcote}, {Barman}, {Fitzgerald},
  {Graham}, {Larkin}, {Macintosh}, {Bauman}, {Burrows}, {Cardwell}, {De Rosa},
  {Dillon}, {Doyon}, {Dunn}, {Erikson}, {Gavel}, {Goodsell}, {Hartung},
  {Hibon}, {Ingraham}, {Kalas}, {Konopacky}, {Maire}, {Marchis}, {Marley},
  {Marois}, {Millar-Blanchaer}, {Morzinski}, {Norton}, {Oppenheimer}, {Palmer},
  {Patience}, {Perrin}, {Poyneer}, {Pueyo}, {Rantakyr{\"o}}, {Sadakuni},
  {Saddlemyer}, {Savransky}, {Serio}, {Sivaramakrishnan}, {Song}, {Soummer},
  {Thomas}, {Wallace}, {Wiktorowicz}, \& {Wolff}}]{Chilcote2015}
{Chilcote}, J., {Barman}, T., {Fitzgerald}, M.~P., {et~al.} 2015, \apjl, 798,
  L3

\bibitem[{{Claudi} {et~al.}(2008){Claudi}, {Turatto}, {Gratton}, {Antichi},
  {Bonavita}, {Bruno}, {Cascone}, {De Caprio}, {Desidera}, {Giro}, {Mesa},
  {Scuderi}, {Dohlen}, {Beuzit}, \& {Puget}}]{Claudi2008a}
{Claudi}, R.~U., {Turatto}, M., {Gratton}, R.~G., {et~al.} 2008, in \procspie,
  Vol. 7014, 3

\bibitem[{{Close} {et~al.}(2012){Close}, {Males}, {Kopon}, {Gasho}, {Follette},
  {Hinz}, {Morzinski}, {Uomoto}, {Hare}, {Riccardi}, {Esposito}, {Puglisi},
  {Pinna}, {Busoni}, {Arcidiacono}, {Xompero}, {Briguglio}, {Quiros-Pacheco},
  \& {Argomedo}}]{Close2012g}
{Close}, L.~M., {Males}, J.~R., {Kopon}, D.~A., {et~al.} 2012, in \procspie,
  Vol. 8447, 16

\bibitem[{{Conrath} {et~al.}(1998){Conrath}, {Gierasch}, \&
  {Ustinov}}]{Conrath1998a}
{Conrath}, B.~J., {Gierasch}, P.~J., \& {Ustinov}, E.~A. 1998, \icarus, 135,
  501

\bibitem[{{Currie} {et~al.}(2013){Currie}, {Burrows}, {Madhusudhan},
  {Fukagawa}, {Girard}, {Dawson}, {Murray-Clay}, {Kenyon}, {Kuchner},
  {Matsumura}, {Jayawardhana}, {Chambers}, \& {Bromley}}]{Currie2013}
{Currie}, T., {Burrows}, A., {Madhusudhan}, N., {et~al.} 2013, \apj, 776, 15

\bibitem[{{Daumont} {et~al.}(2013){Daumont}, {Nikitin}, {Thomas},
  {R{\'e}galia}, {Von der Heyden}, {Tyuterev}, {Rey}, {Boudon}, {Wenger},
  {Lo{\"e}te}, \& {Brown}}]{Daumont2013}
{Daumont}, L., {Nikitin}, A.~V., {Thomas}, X., {et~al.} 2013, \jqsrt, 116, 101

\bibitem[{{Galicher} {et~al.}(2014){Galicher}, {Rameau}, {Bonnefoy}, {Baudino},
  {Currie}, {Boccaletti}, {Chauvin}, {Lagrange}, \& {Marois}}]{Galicher2014a}
{Galicher}, R., {Rameau}, J., {Bonnefoy}, M., {et~al.} 2014, \aap, 565, L4

\bibitem[{{Goody} \& {Yung}(1989)}]{Goody1989}
{Goody}, R.~M. \& {Yung}, Y.~L. 1989, {Atmospheric radiation : theoretical
  basis}

\bibitem[{{Hanot} {et~al.}(2010){Hanot}, {Absil}, {Surdej}, {Boccaletti}, \&
  {V{\'e}rinaud}}]{Hanot2010}
{Hanot}, C., {Absil}, O., {Surdej}, J., {Boccaletti}, A., \& {V{\'e}rinaud}, C.
  2010, in \procspie, Vol. 7731, 3

\bibitem[{{Hartmann} {et~al.}(2002){Hartmann}, {Boulet}, {Brodbeck}, {van
  Thanh}, {Fouchet}, \& {Drossart}}]{Hartmann2002al}
{Hartmann}, J.-M., {Boulet}, C., {Brodbeck}, C., {et~al.} 2002, \jqsrt, 72, 117

\bibitem[{{Helling} {et~al.}(2008){Helling}, {Ackerman}, {Allard}, {Dehn},
  {Hauschildt}, {Homeier}, {Lodders}, {Marley}, {Rietmeijer}, {Tsuji}, \&
  {Woitke}}]{Helling2008h}
{Helling}, C., {Ackerman}, A., {Allard}, F., {et~al.} 2008, \mnras, 391, 1854

\bibitem[{{J{\"a}ger} {et~al.}(2003){J{\"a}ger}, {Dorschner}, {Mutschke},
  {Posch}, \& {Henning}}]{Jager2003a}
{J{\"a}ger}, C., {Dorschner}, J., {Mutschke}, H., {Posch}, T., \& {Henning}, T.
  2003, \aap, 408, 193

\bibitem[{{Janson} {et~al.}(2010){Janson}, {Bergfors}, {Goto}, {Brandner}, \&
  {Lafreni{\`e}re}}]{Janson2010c}
{Janson}, M., {Bergfors}, C., {Goto}, M., {Brandner}, W., \& {Lafreni{\`e}re},
  D. 2010, \apjl, 710, L35

\bibitem[{{Konopacky} {et~al.}(2013){Konopacky}, {Barman}, {Macintosh}, \&
  {Marois}}]{Konopacky2013}
{Konopacky}, Q.~M., {Barman}, T.~S., {Macintosh}, B.~A., \& {Marois}, C. 2013,
  Science, 339, 1398

\bibitem[{Kramida {et~al.}(2014)Kramida, {Yu.~Ralchenko}, Reader, \& {and NIST
  ASD Team}}]{NIST_ASD}
Kramida, A., {Yu.~Ralchenko}, Reader, J., \& {and NIST ASD Team}. 2014, {NIST
  Atomic Spectra Database (ver. 5.2), [Online]. Available:
  {\tt{http://physics.nist.gov/asd}} [2015, March 17]. National Institute of
  Standards and Technology, Gaithersburg, MD.}

\bibitem[{{Lagrange} {et~al.}(2012{\natexlab{a}}){Lagrange}, {Boccaletti},
  {Milli}, {Chauvin}, {Bonnefoy}, {Mouillet}, {Augereau}, {Girard}, {Lacour},
  \& {Apai}}]{Lagrange2012a}
{Lagrange}, A.-M., {Boccaletti}, A., {Milli}, J., {et~al.} 2012{\natexlab{a}},
  \aap, 542, A40

\bibitem[{{Lagrange} {et~al.}(2010){Lagrange}, {Bonnefoy}, {Chauvin}, {Apai},
  {Ehrenreich}, {Boccaletti}, {Gratadour}, {Rouan}, {Mouillet}, {Lacour}, \&
  {Kasper}}]{Lagrange2010e}
{Lagrange}, A.-M., {Bonnefoy}, M., {Chauvin}, G., {et~al.} 2010, Science, 329,
  57

\bibitem[{{Lagrange} {et~al.}(2012{\natexlab{b}}){Lagrange}, {De Bondt},
  {Meunier}, {Sterzik}, {Beust}, \& {Galland}}]{Lagrange2012d}
{Lagrange}, A.-M., {De Bondt}, K., {Meunier}, N., {et~al.} 2012{\natexlab{b}},
  \aap, 542, A18

\bibitem[{{Lagrange} {et~al.}(2014){Lagrange}, {Gilardy}, {Beust}, {Chauvin},
  {Rameau}, {Boccaletti}, {Girard}, \& {Bonnefoy}}]{Lagrange2014a}
{Lagrange}, A.-M., {Gilardy}, H., {Beust}, H., {et~al.} 2014, in Proc. of the
  IAU, IAUS, Vol. 299, Exploring the Formation and Evolution of Planetary
  Systems, ed. M.~{Booth}, B.~C. {Matthews}, \& J.~R. {Graham}, 299

\bibitem[{{Lagrange} {et~al.}(2009){Lagrange}, {Gratadour}, {Chauvin}, {Fusco},
  {Ehrenreich}, {Mouillet}, {Rousset}, {Rouan}, {Allard}, {Gendron}, {Charton},
  {Mugnier}, {Rabou}, {Montri}, \& {Lacombe}}]{Lagrange2009}
{Lagrange}, A.-M., {Gratadour}, D., {Chauvin}, G., {et~al.} 2009, \aap, 493,
  L21

\bibitem[{{Lagrange} {et~al.}(2013){Lagrange}, {Meunier}, {Chauvin}, {Sterzik},
  {Galland}, {Lo Curto}, {Rameau}, \& {Sosnowska}}]{Lagrange2013d}
{Lagrange}, A.-M., {Meunier}, N., {Chauvin}, G., {et~al.} 2013, \aap, 559, A83

\bibitem[{{Langlois} {et~al.}(2010){Langlois}, {Dohlen}, {Augereau},
  {Mouillet}, {Boccaletti}, \& {Schmid}}]{Langlois2010g}
{Langlois}, M., {Dohlen}, K., {Augereau}, J.-C., {et~al.} 2010, in \procspie,
  Vol. 7735, 2

\bibitem[{{Lecavelier Des Etangs} {et~al.}(1997){Lecavelier Des Etangs},
  {Vidal-Madjar}, {Burki}, {Lamers}, {Ferlet}, {Nitschelm}, \&
  {Sevre}}]{LecavelierDesEtangs1997b}
{Lecavelier Des Etangs}, A., {Vidal-Madjar}, A., {Burki}, G., {et~al.} 1997,
  \aap, 328, 311

\bibitem[{{Lenzen} {et~al.}(2003){Lenzen}, {Hartung}, {Brandner}, {Finger},
  {Hubin}, {Lacombe}, {Lagrange}, {Lehnert}, {Moorwood}, \&
  {Mouillet}}]{Lenzen2003}
{Lenzen}, R., {Hartung}, M., {Brandner}, W., {et~al.} 2003, in \procspie, Vol.
  4841, Instrument Design and Performance for Optical/Infrared Ground-based
  Telescopes, ed. M.~{Iye} \& A.~F.~M. {Moorwood}, 944

\bibitem[{{Lin} \& {Ida}(1997)}]{Lin1997}
{Lin}, D.~N.~C. \& {Ida}, S. 1997, \apj, 477, 781

\bibitem[{{Lodders}(2010)}]{Lodders2010}
{Lodders}, K. 2010, in Principles and Perspectives in Cosmochemistry, ed.
  A.~{Goswami} \& B.~E. {Reddy}, 379

\bibitem[{{Lodders} \& {Fegley}(2006)}]{Lodders2006}
{Lodders}, K. \& {Fegley}, Jr., B. 2006, {Chemistry of Low Mass Substellar
  Objects}, ed. J.~W. {Mason}, 1

\bibitem[{{Lunine} {et~al.}(1989){Lunine}, {Hubbard}, {Burrows}, {Wang}, \&
  {Garlow}}]{Lunine1989e}
{Lunine}, J.~I., {Hubbard}, W.~B., {Burrows}, A., {Wang}, Y.-P., \& {Garlow},
  K. 1989, \apj, 338, 314

\bibitem[{{Macintosh} {et~al.}(2014){Macintosh}, {Graham}, {Ingraham},
  {Konopacky}, {Marois}, {Perrin}, {Poyneer}, {Bauman}, {Barman}, {Burrows},
  {Cardwell}, {Chilcote}, {De Rosa}, {Dillon}, {Doyon}, {Dunn}, {Erikson},
  {Fitzgerald}, {Gavel}, {Goodsell}, {Hartung}, {Hibon}, {Kalas}, {Larkin},
  {Maire}, {Marchis}, {Marley}, {McBride}, {Millar-Blanchaer}, {Morzinski},
  {Norton}, {Oppenheimer}, {Palmer}, {Patience}, {Pueyo}, {Rantakyro},
  {Sadakuni}, {Saddlemyer}, {Savransky}, {Serio}, {Soummer},
  {Sivaramakrishnan}, {Song}, {Thomas}, {Wallace}, {Wiktorowicz}, \&
  {Wolff}}]{Macintosh2014c}
{Macintosh}, B., {Graham}, J.~R., {Ingraham}, P., {et~al.} 2014, Proc. of the
  National Academy of Science, 111, 12661

\bibitem[{{Males} {et~al.}(2014){Males}, {Close}, {Morzinski}, {Wahhaj}, {Liu},
  {Skemer}, {Kopon}, {Follette}, {Puglisi}, {Esposito}, {Riccardi}, {Pinna},
  {Xompero}, {Briguglio}, {Biller}, {Nielsen}, {Hinz}, {Rodigas}, {Hayward},
  {Chun}, {Ftaclas}, {Toomey}, \& {Wu}}]{Males2014c}
{Males}, J.~R., {Close}, L.~M., {Morzinski}, K.~M., {et~al.} 2014, \apj, 786,
  32

\bibitem[{{Marleau} \& {Cumming}(2014)}]{Marleau2014}
{Marleau}, G.-D. \& {Cumming}, A. 2014, \mnras, 437, 1378

\bibitem[{{Marley} {et~al.}(2000){Marley}, {Ackerman}, \&
  {Seager}}]{2000AAS...19712705M}
{Marley}, M., {Ackerman}, A., \& {Seager}, S. 2000, in Bulletin of the American
  Astronomical Society, Vol.~32, American Astronomical Society Meeting
  Abstracts, 127.05

\bibitem[{{Marley} {et~al.}(2012){Marley}, {Saumon}, {Cushing}, {Ackerman},
  {Fortney}, \& {Freedman}}]{Marley2012a}
{Marley}, M.~S., {Saumon}, D., {Cushing}, M., {et~al.} 2012, \apj, 754, 135

\bibitem[{{Marois} {et~al.}(2008){Marois}, {Macintosh}, {Barman}, {Zuckerman},
  {Song}, {Patience}, {Lafreni{\`e}re}, \& {Doyon}}]{Marois2008c}
{Marois}, C., {Macintosh}, B., {Barman}, T., {et~al.} 2008, Science, 322, 1348

\bibitem[{{Mayor} \& {Queloz}(1995)}]{Mayor1995b}
{Mayor}, M. \& {Queloz}, D. 1995, \nat, 378, 355

\bibitem[{{Mordasini} {et~al.}(2012){Mordasini}, {Alibert}, {Georgy},
  {Dittkrist}, {Klahr}, \& {Henning}}]{Mordasini2012d}
{Mordasini}, C., {Alibert}, Y., {Georgy}, C., {et~al.} 2012, \aap, 547, A112

\bibitem[{{Morley} {et~al.}(2012){Morley}, {Fortney}, {Marley}, {Visscher},
  {Saumon}, \& {Leggett}}]{Morley2012}
{Morley}, C.~V., {Fortney}, J.~J., {Marley}, M.~S., {et~al.} 2012, \apj, 756,
  172

\bibitem[{{Mouillet} {et~al.}(1997){Mouillet}, {Larwood}, {Papaloizou}, \&
  {Lagrange}}]{Mouillet1997b}
{Mouillet}, D., {Larwood}, J.~D., {Papaloizou}, J.~C.~B., \& {Lagrange}, A.~M.
  1997, \mnras, 292, 896

\bibitem[{{Neuh{\"a}user} {et~al.}(2005){Neuh{\"a}user}, {Guenther},
  {Wuchterl}, {Mugrauer}, {Bedalov}, \& {Hauschildt}}]{Neuhauser2005}
{Neuh{\"a}user}, R., {Guenther}, E.~W., {Wuchterl}, G., {et~al.} 2005, \aap,
  435, L13

\bibitem[{{Nikitin} {et~al.}(2002){Nikitin}, {Brown}, {F{\'e}jard}, {Champion},
  \& {Tyuterev}}]{Nikitin2002k}
{Nikitin}, A., {Brown}, L.~R., {F{\'e}jard}, L., {Champion}, J.~P., \&
  {Tyuterev}, V.~G. 2002, Journal of Molecular Spectroscopy, 216, 225

\bibitem[{{Nikitin} {et~al.}(2013){Nikitin}, {Brown}, {Sung}, {Rey},
  {Tyuterev}, {Smith}, \& {Mantz}}]{Nikitin2013d}
{Nikitin}, A.~V., {Brown}, L.~R., {Sung}, K., {et~al.} 2013, \jqsrt, 114, 1

\bibitem[{{Nikitin} {et~al.}(2006){Nikitin}, {Champion}, \&
  {Brown}}]{Nikitin2006o}
{Nikitin}, A.~V., {Champion}, J.-P., \& {Brown}, L.~R. 2006, Journal of
  Molecular Spectroscopy, 240, 14

\bibitem[{{Ordal} {et~al.}(1988){Ordal}, {Bell}, {Alexander}, {Newquist}, \&
  {Querry}}]{Ordal1988}
{Ordal}, M.~A., {Bell}, R.~J., {Alexander}, Jr., R.~W., {Newquist}, L.~A., \&
  {Querry}, M.~R. 1988, \ao, 27, 1203

\bibitem[{{Plez}(1998)}]{Plez1998}
{Plez}, B. 1998, \aap, 337, 495

\bibitem[{{Quanz} {et~al.}(2010){Quanz}, {Meyer}, {Kenworthy}, {Girard},
  {Kasper}, {Lagrange}, {Apai}, {Boccaletti}, {Bonnefoy}, {Chauvin}, {Hinz}, \&
  {Lenzen}}]{Quanz2010b}
{Quanz}, S.~P., {Meyer}, M.~R., {Kenworthy}, M.~A., {et~al.} 2010, \apjl, 722,
  L49

\bibitem[{{Rameau} {et~al.}(2013){Rameau}, {Chauvin}, {Lagrange}, {Boccaletti},
  {Quanz}, {Bonnefoy}, {Girard}, {Delorme}, {Desidera}, {Klahr}, {Mordasini},
  {Dumas}, \& {Bonavita}}]{Rameau2013d}
{Rameau}, J., {Chauvin}, G., {Lagrange}, A.-M., {et~al.} 2013, \apjl, 772, L15

\bibitem[{{Rothman} {et~al.}(2010){Rothman}, {Gordon}, {Barber}, {Dothe},
  {Gamache}, {Goldman}, {Perevalov}, {Tashkun}, \& {Tennyson}}]{Rothman2010h}
{Rothman}, L.~S., {Gordon}, I.~E., {Barber}, R.~J., {et~al.} 2010, \jqsrt, 111,
  2139

\bibitem[{{Rousset} {et~al.}(2003){Rousset}, {Lacombe}, {Puget}, {Hubin},
  {Gendron}, {Fusco}, {Arsenault}, {Charton}, {Feautrier}, {Gigan}, {Kern},
  {Lagrange}, {Madec}, {Mouillet}, {Rabaud}, {Rabou}, {Stadler}, \&
  {Zins}}]{Rousset2003}
{Rousset}, G., {Lacombe}, F., {Puget}, P., {et~al.} 2003, in \procspie, Vol.
  4839, Adaptive Optical System Technologies II, ed. P.~L. {Wizinowich} \&
  D.~{Bonaccini}, 140

\bibitem[{Rowe {et~al.}(2015)Rowe, Coughlin, Antoci, Barclay, Batalha, Borucki,
  Burke, Bryson, Caldwell, Campbell, Catanzarite, Christiansen, Cochran,
  Gilliland, Girouard, Haas, Hełminiak, Henze, Hoffman, Howell, Huber, Hunter,
  Jang-Condell, Jenkins, Klaus, Latham, Li, Lissauer, McCauliff, Morris,
  Mullally, Ofir, Quarles, Quintana, Sabale, Seader, Shporer, Smith, Steffen,
  Still, Tenenbaum, Thompson, Twicken, Laerhoven, Wolfgang, \&
  Zamudio}]{Rowe2015a}
Rowe, J.~F., Coughlin, J.~L., Antoci, V., {et~al.} 2015, \apjs, 217, 16

\bibitem[{{Schmid} {et~al.}(2010){Schmid}, {Beuzit}, {Mouillet}, {Waters},
  {Buenzli}, {Boccaletti}, {Dohlen}, {Feldt}, \& {SPHERE
  Consortium}}]{Schmid2010v}
{Schmid}, H.~M., {Beuzit}, J.~L., {Mouillet}, D., {et~al.} 2010, in In the
  Spirit of Lyot 2010, 49

\bibitem[{{Smith} \& {Terrile}(1987)}]{Smith1987ed}
{Smith}, B.~A. \& {Terrile}, R.~J. 1987, in \baas, Vol.~19, 829

\bibitem[{{Spiegel} \& {Burrows}(2012)}]{Spiegel2012a}
{Spiegel}, D.~S. \& {Burrows}, A. 2012, \apj, 745, 174

\bibitem[{{Tsuji}(2002)}]{Tsuji2002g}
{Tsuji}, T. 2002, \apj, 575, 264

\bibitem[{{Tsuji} {et~al.}(1996){Tsuji}, {Ohnaka}, \& {Aoki}}]{Tsuji1996}
{Tsuji}, T., {Ohnaka}, K., \& {Aoki}, W. 1996, \aap, 305, L1

\bibitem[{{van Leeuwen}(2007)}]{2007h}
{van Leeuwen}, F. 2007, \aap, 474, 653

\bibitem[{{Vigan} {et~al.}(2010){Vigan}, {Moutou}, {Langlois}, {Allard},
  {Boccaletti}, {Carbillet}, {Mouillet}, \& {Smith}}]{Vigan2010c}
{Vigan}, A., {Moutou}, C., {Langlois}, M., {et~al.} 2010, \mnras, 407, 71

\bibitem[{{Vinatier} {et~al.}(2007){Vinatier}, {B{\'e}zard}, {Fouchet},
  {Teanby}, {de Kok}, {Irwin}, {Conrath}, {Nixon}, {Romani}, {Flasar}, \&
  {Coustenis}}]{Vinatier2007}
{Vinatier}, S., {B{\'e}zard}, B., {Fouchet}, T., {et~al.} 2007, \icarus, 188,
  120

\bibitem[{{Yurchenko} {et~al.}(2011){Yurchenko}, {Barber}, \&
  {Tennyson}}]{Yurchenko2011}
{Yurchenko}, S.~N., {Barber}, R.~J., \& {Tennyson}, J. 2011, \mnras, 413, 1828

\bibitem[{{Yurchenko} \& {Tennyson}(2014)}]{Yurchenko2014a}
{Yurchenko}, S.~N. \& {Tennyson}, J. 2014, \mnras, 440, 1649

\end{thebibliography}

\begin{appendix} 
   \section{Radiative-convective equilibrium model, numerical method}
      In a one-dimensional radiative-convective equilibrium model, the net flux 
      (radiative + convective) is assumed to be constant as a function of pressure level.
      This net flux $\pi F$ is equal to         
      \begin{equation}
         \pi F = \sigma T_\mathrm{eff}^4
         \label{piFluxTeff}
      ,\end{equation}
      where $T_\mathrm{eff}$ is the effective temperature of the planet. We first solve
      for purely radiative equilibrium and neglect heating from the parent star. This is
      justified as long as we are interested in hot young giant exoplanets relatively far 
      from their parent star. Assuming a planet with $T_\mathrm{eff}$ = 700 K at a distance 
      of 7 AU from the star, the stellar flux absorbed by the planet would amount to less 
      than 0.1\% of the planet's thermal emission. 

      Discarding scattering, the net flux at pressure level $p$, in a plane parallel 
      geometry, is given by
      \begin{multline}
         \pi F(p) = 2\pi \int\limits_0^\infty d \sigma
         \Biggl[ \int\limits_{\tau_\mathrm{\sigma}(p)}^\infty
         B_\mathrm\sigma(\tau_\mathrm{\sigma}')
         E_2(\tau_\mathrm{\sigma}'-  \tau_\mathrm{\sigma})d\tau_\mathrm{\sigma}'- \\
         \int\limits_0^{\tau_\mathrm{\sigma}(p)} B_\mathrm\sigma(\tau_\mathrm{\sigma}')
         E_2(\tau_\mathrm{\sigma}-\tau_\mathrm{\sigma}')d\tau_\mathrm{\sigma}' \Biggr] ,
      \end{multline}
      where $\tau_\mathrm{\sigma}(p)$ is the optical depth at pressure level $p$ and 
      wavenumber $\sigma$, $B_\mathrm\sigma(\tau_\mathrm{\sigma}')$ is the Planck function 
      at the temperature of level of optical depth $\tau_\mathrm{\sigma}'$ and wavenumber 
      $\sigma$, and $E_2$ is the second-order exponential integral.\\
         
      The integral over wavenumber is calculated over the range [$\sigma_\mathrm{min}$, 
      $\sigma_\mathrm{max}$], with $\sigma_\mathrm{min}= 20$ cm$^{-1}$  and 
      $\sigma_\mathrm{max}= 16000$ cm$^{-1}$. This range is sliced into $n_\mathrm{\sigma}$ 
      intervals of width $\delta \sigma = 20$ cm$^{-1}$, over which the Planck function is 
      taken as constant. The radiative transfer integral over each interval of width $\delta 
      \sigma$ is calculated through a correlated-$k$ distribution method with $n_\mathrm{k}$ 
      quadrature points \citep{Goody1989}. The atmospheric grid consists of $n_\mathrm{p}$ 
      atmospheric levels equally spaced in $ln(p)$ between pressure levels $p_\mathrm{max}$ 
      at the bottom of the grid ($j=1$) and $n_\mathrm{p}=64$). Assuming a linear variation 
      of the Planck function $B$ with optical depth $\tau$ within any layer 
      [$p_\mathrm{j-1}$, $p_\mathrm{j}$], i.e.
      \begin{equation}
          B_\mathrm{\sigma} (\tau_\mathrm{\sigma}) = B_\mathrm{\sigma} 
          (\tau_\mathrm{\sigma,j-1})
          \frac{\tau_\mathrm{\sigma}-\tau_\mathrm{\sigma,j}}
          {\tau_\mathrm{\sigma,j-1}-\tau_\mathrm{\sigma,j}}+
          B_{\sigma} (\tau_\mathrm{\sigma,j})
          \frac{\tau_\mathrm{\sigma,j-1}-\tau_\mathrm{\sigma}}
          {\tau_\mathrm{\sigma,j-1}-\tau_\mathrm{\sigma,j}}, 
          \label{plank}
      \end{equation}
      the contribution of this layer to the flux at wavenumber $\sigma$ can be analytically 
      calculated and expressed as a linear combination of the Planck functions at pressure 
      levels $p_\mathrm{j}$ and $p_\mathrm{j+1}$. We also add a contribution from below the 
      atmospheric grid ($p > p_\mathrm{max}$) assuming a semi-infinite layer with the same 
      variation of the Planck function as in Eq. ( \ref{plank} ) for the first layer. 
      Summing over all layers and spectral intervals, the net flux at level $p_\mathrm{j}$ 
      can then be expressed as
      \begin{equation}
         \pi F(p_\mathrm{j}) = 2\pi \delta \sigma 
         \sum\limits_\mathrm{i=1}^\mathrm{{n_\mathrm{\sigma}}}\sum\limits_\mathrm{j'=1}^\mathrm{n_\mathrm{p}}  
         B_{\sigma_{\mathrm{i}}}(T_\mathrm{j'}) 
         \sum\limits_\mathrm{l=1}^\mathrm{n_\mathrm{k}} \varpi_\mathrm{l} 
         A(\sigma_\mathrm{i},p_\mathrm{j},p_\mathrm{j'},l) ,
         \label{piflux}
      \end{equation}
      where $B_{\sigma_{\mathrm{i}}}(T_\mathrm{j'})$  is the Planck function at temperature 
      of the ${j'}^\mathrm{th}$ pressure level of the grid and wavenumber 
      $\sigma_\mathrm{i}$ at the middle of the $i^\mathrm{th}$ spectral interval of width 
      $\delta \sigma$. The parameter $\varpi_\mathrm{l}$ is the weight applied to the $l^\mathrm{th}$ set 
      of $k$-correlated coefficients used in the quadrature for the spectral integration 
      over any spectral interval $(\sum\limits_\mathrm{l=1}^\mathrm{n_\mathrm{k}} 
      \varpi_\mathrm{l} = 1)$. The parameter $A(\sigma_\mathrm{i},p_\mathrm{j},p_\mathrm{j'},l)$ is a 
      dimensionless factor that couples pressure levels $p_\mathrm{j}$ and $p_\mathrm{j'}$ 
      and only depends on the grid of optical depths for the $l^\mathrm{th}$ set of $k$-
      correlated coefficients of the $i^\mathrm{th}$ spectral interval.
         
         We then search for the temperature profile that ensures radiative equilibrium, 
         i.e., 
         \begin{equation}
          \pi F( p_j ) = \sigma T_\mathrm{eff}^4
         \end{equation}
         for $j$ varying from $2$ to $n_\mathrm{p}$. We do not use the flux at the first, 
         deepest level as a constraint because the variation of the Planck function at 
         deeper levels is fixed arbitrarily in the model to that of the first layer. To 
         solve this system of $n_\mathrm{p-1}$ equations, we use a constrained linear 
         inversion method described in \cite{Vinatier2007} and based on \cite{Conrath1998a}. 
         The algorithm minimizes the quadratic difference ($\chi^2$) between desired 
         ($\sigma T_\mathrm{eff}^4$) and calculated fluxes with the additional constraint 
         that the solution temperature profile lies close to the reference profile. Starting 
         from an initial guess profile $T_0$, an approximate solution $T_1$ is derived from 
         the equation
         \begin{equation}
          T_{\mathrm{n}} = T_{\mathrm{n-1}} + \alpha SK^{\mathrm{T}} C^{-1}\Delta F
          \label{profil}
         ,\end{equation}
         with $n=1$, where $\Delta F$ is the difference vector between the desired and 
         calculated fluxes ($\sigma {T_{\mathrm{eff}}}^4-\pi F( p_j ) $), $K$ is the kernel 
         matrix with $K_\mathrm{jj'}$ equal to the derivative of the flux at level 
         $p_\mathrm{j}$ with respect to the temperature at level $p_\mathrm{j'}$, $S$ is a 
         normalized two-point Gaussian correlation matrix that provides a vertical filtering 
         of the solution needed to avoid numerical instabilities, and $\alpha$ a scalar 
         parameter that controls the emphasis placed on the proximity of the solution $T_1$ 
         to the reference profile $T_0$. We used a correlation length of 0.4 pressure scale 
         height. The kernel matrix is calculated from Eq. (\ref{piflux}), neglecting the 
         dependence of $A$ with temperature, which  is generally much weaker than that 
         of the Planck function, i.e.,
         \begin{equation}
          K_\mathrm{jj'} = 2\pi \delta \sigma \sum\limits_\mathrm{i=1}^{n_\mathrm{\sigma}}
          \frac{\partial B_\mathrm\sigma(i)(T_\mathrm{j'})}{\partial T_\mathrm{j'}}  
           \sum\limits_\mathrm{l=1}^{n_\mathrm{k}} \varpi_\mathrm{l} 
           A(\sigma_\mathrm{i},p_\mathrm{j},p_\mathrm{j'},l) .
           \label{kernel}
         \end{equation}
         Matrix $C$ is equal to
         \begin{equation}
          C = \alpha KSK^T + E
         ,\end{equation}         
         where $E$ is a diagonal matrix with $E_\mathrm{jj'}$ equal to the square of the 
         flux error acceptable at the $j^\mathrm{th}$ pressure level, usually set 
         to $0.1\%$ of $\sigma T_\mathrm{eff}^4$.
         
         The nonlinearity of the problem requires an iterative process in which
         $T_\mathrm{n}$ is obtained from Eq. (\ref{profil}) after updating the reference 
         profile to $T_\mathrm{n-1}$ and recalculating the kernel matrix $K$ for profile  
         $T_\mathrm{n-1}$. The iteration process is pursued until $\chi^2$ is less than 1 and
         no longer significantly decreases. The $\alpha$ parameter in Eq. (\ref{kernel}) is 
         chosen to be small enough to ensure convergence and large enough to reduce the number of 
         iterations needed. Typically ten iterations are needed. Note that the final solution 
         does not depend on the initial profile $T_0$ or on the choice
         of $\alpha$.   For $T_0$, we used one of the three temperature profiles calculated 
         by \cite{Allard2003j} for $T_\mathrm{eff}=$ 900, 1300, and 1700 K. We choose that 
         having $T_{\mathrm{eff}}$ closest to the input value to ensure rapid convergence.      
         
         In a second step, the solution profile is checked against convective instability by 
         comparing the model lapse rate $\bigtriangledown = \frac{ln 
         (T_\mathrm{j}/T_\mathrm{j-1} )}{ln ( p_\mathrm{j}/p_\mathrm{j-1} )} $ with the 
         adiabatic value $\bigtriangledown_\mathrm{ad} = R/C_\mathrm{p}$, where $R$ is the 
         gas constant and $C_\mathrm{p}$ the temperature-dependent molar heat capacity for 
         the H$_2$-He atmosphere. Regions where the lapse rate exceeds the adiabatic value 
         are unstable against convection. They are found in the bottom of the pressure grid 
         $p > p_\mathrm{ad}$, with $p_\mathrm{ad}$ being the level where the lapse rate 
         equals the adiabatic value. In that case, convective heat transfer occurs setting 
         back the lapse rate to the adiabatic value. In Exo-REM, we do not solve 
         explicitly for convection. We add a convective flux to the radiative flux in Eq.
         (\ref{piflux}) through an analytical function that is essentially zero when 
         $\bigtriangledown\leq\bigtriangledown_\mathrm{ad}$ and rapidly gets very large when 
         $\bigtriangledown>\bigtriangledown_\mathrm{ad}$. We found that the following 
         function:         
         \begin{equation}
          F_\mathrm{conv} ( p_\mathrm{j} ) = 10^{-3} \sigma T_\mathrm{eff}^4 \text{e}^{ 200
          [ \bigtriangledown/\bigtriangledown_\mathrm{ad}-1]}\end{equation}
         is adequate to ensure negligible superadiabaticity in the final solution profile. 
         We then set the lapse rate of the purely radiative solution to the adiabatic value, 
         plus a small amount (0.015), at levels $p > p_\mathrm{ad}$ . The kernel matrix $K$ 
         is calculated adding         
         \begin{multline}
          \frac{\partial F_\mathrm{conv} ( p_\mathrm{j} )}{\partial T_\mathrm{j}}  = 0.2 
          \sigma T_\mathrm{eff}^4
          \text{e}^{ 200 [ \bigtriangledown/\bigtriangledown_\mathrm{ad}-1]}/
          (T_\mathrm{j}\bigtriangledown_\mathrm{ad} 
          ln(p_\mathrm{j}/p_\mathrm{j-1}))~~(a)\\~to~K_\mathrm{jj}~and\\
          \frac{\partial F_\mathrm{conv} ( p_\mathrm{j} )}{\partial T_\mathrm{j-1}}  = -0.2 
          \sigma T_\mathrm{eff}^4
          \text{e}^{ 200 [ \bigtriangledown/\bigtriangledown_\mathrm{ad}-1]}/
          (T_\mathrm{j-1}\bigtriangledown_\mathrm{ad} ln(p_\mathrm{j}/p_\mathrm{j-1}))~~(b),
         \end{multline}         
         to $K_\mathrm{j j-1}$ in Eq.(\ref{kernel}). The iterative process is finally 
         restarted with the modified flux and kernel until convergence is achieved. Typically 
         another set of ten iterations is needed. Our model in this paper has 64 pressure 
         levels equally spaced in $\ln (p)$ between 50 bar and 0.01 mbar.
        
\end{appendix}
 
\end{document}